\documentclass[conference,a4paper]{IEEEtran}
\usepackage[table]{xcolor}
\usepackage{graphicx}
\usepackage{float}
\usepackage{longtable}
\usepackage{mdwmath}
\usepackage{eqparbox}
\usepackage{url}
\usepackage{multirow}
\usepackage{lineno,hyperref}
\usepackage{multirow}
\usepackage{bigstrut}
\usepackage{hhline}
\usepackage{tabularx}

\usepackage{booktabs}
\usepackage{rotating}
\usepackage{adjustbox} 
\usepackage{blindtext}
\usepackage{makecell}
\usepackage{longtable,multirow,array}
\usepackage[utf8]{inputenc}
\usepackage[table]{xcolor}
\usepackage{color, colortbl}
\usepackage{threeparttable,booktabs,multirow,lscape}

\graphicspath{ {./Figures/} }
\usepackage{rotating,booktabs}

\usepackage[verbose]{placeins}
\usepackage{blindtext}
\usepackage{longtable} 

\definecolor{orange}{rgb}{ 1,  .6,  0}
\definecolor{beige}{rgb}{1.0, 0.75, 0.0}
\definecolor{lightorange}{rgb}{1.0, 1.0, 0.19}
\definecolor{lightgreen}{rgb}{0.0, 0.8, 0.6}
\definecolor{lightgreend}{rgb}{1,1,1}

\definecolor{red}{rgb}{1,1,1}
\definecolor{green}{rgb}{1,1,1}
\definecolor{yellow}{rgb}{1,1,1}
\definecolor{lightgreen2}{rgb}{1,1,1}
\definecolor{maron}{rgb}{1,1,1}
\definecolor{NormalRow}{rgb}{0.8,0.9,1}
\definecolor{TopRow}{HTML}{87CEEB}
\definecolor{applayer}{HTML}{E1D5E7}
\definecolor{execlayer}{HTML}{F8CECC}
\definecolor{conslayer}{HTML}{DAE8FC}
\definecolor{datalayer}{HTML}{FFF2CC}
\definecolor{mycolor}{HTML}{fbfbeb}
\DeclareRobustCommand{\IEEEauthorrefmark}[1]{\raisebox{0pt}[0pt][0pt]{\textsuperscript{\footnotesize #1}}}

\begin{document}
\title{Drawing the boundaries between
Blockchain and Blockchain-like systems: A Comprehensive Survey on Distributed Ledger Technologies}
\author{
\IEEEauthorblockN{
Badr Bellaj \IEEEauthorrefmark{* \$},
Aafaf Ouaddah\IEEEauthorrefmark{*},
Emmanuel Bertin \IEEEauthorrefmark{\S}(IEEE Senior), 
Noel Crespi\IEEEauthorrefmark{\$}(IEEE Senior),
Abdellatif Mezrioui\IEEEauthorrefmark{*}
}
\IEEEauthorblockA{
\IEEEauthorrefmark{\$}Telecom SudParis, Paris, France,
\IEEEauthorrefmark{*}Institut National des Postes et Télécommunications, Rabat, Morocco,
\IEEEauthorrefmark{\S}Orange, France
}
\IEEEauthorblockA{Corresponding e-mail: bellaj.badr@mchain.uk }
}
\maketitle
\begin{abstract}
Bitcoin's success as a global cryptocurrency has paved the way for the emergence of blockchain, a revolutionary category of distributed systems. However, the growing popularity of blockchain has led to a significant divergence from its core principles in many systems labeled as "blockchain" This divergence has introduced complexity into the blockchain ecosystem, exacerbated by a lack of comprehensive reviews on blockchain and its variants. Consequently, gaining a clear and updated understanding of the diverse spectrum of current blockchain and blockchain-like systems has become challenging. This situation underscores the necessity for an extensive literature review and the development of thematic taxonomies.

This survey seeks to offer a comprehensive and current assessment of existing blockchains and their variations, while delineating the boundaries between blockchain and blockchain-like systems. To achieve this objective, we propose a holistic reference model for conceptualizing and analyzing these systems. Our layer-wise framework envisions all distributed ledger technologies (DLT) as composed of four principal layers: data, consensus, execution, and application. Additionally, we introduce a new taxonomy that enhances the classification of blockchain and blockchain-like systems, offering a more useful perspective than existing works.

Furthermore, we conduct a state-of-the-art review from a layered perspective, employing $23$ evaluative criteria predefined by our framework. We perform a qualitative and quantitative comparative analysis of $44$ DLT solutions and $26$ consensus mechanism, while discussing differences and boundaries between blockchain and blockchain-like systems. We emphasize the significant challenges and trade-offs encountered by distributed ledger designers, decision-makers, and project managers during the design or adoption of a DLT solution. Finally, we outline crucial research challenges and directions in the field of DLTs.

\end{abstract}

{\smallskip \keywords Blockchain, DLT, Consensus, Blockchain-like.}

\IEEEpeerreviewmaketitle

\vspace{5pt}
\section{Introduction}

\IEEEPARstart{T}{he} blockchain space has rapidly evolved, starting with the introduction of Bitcoin \cite{nakamoto} a decade ago, and progressing to the development of modern enterprise versions of DLT. Bitcoin or Bitcoin-based project such as Litecoin \cite{team} and Peercoin \cite{King2012} are often acknowledged as blockchain $1.0$. The introduction of Ethereum in 2015, along with projects like IOTA \cite{Popov2017}, Hyperledger \cite{Foundation}, and Solana \cite{Hedra}, ushered in the era of blockchain $2.0$. This phase witnessed significant deviations from Nakamoto's original blockchain design, giving rise to a distinct category of technologies.
The emergence of this new category, deeply influenced by blockchain principles but not confined to them, has led the industry to market a more inclusive term: DLT when describing this particular category. Despite this shift, a lack of rigorously defined terminologies and a universally accepted taxonomy has resulted in confusion. Terms such as "blockchain," "DLT", or even "distributed database" have been subject to misunderstanding, misuse, and misinterpretation. Many projects and enterprises extensively employ the term "blockchain" as mere marketing jargon, further complicating the landscape.
Despite several proposals aiming to standardize blockchain (ITU \cite{ITU}, ISO \cite{Dawson}-\cite{ISO/TR}, IEEE \cite{Association}), there is currently no standardized recognized definition of blockchain or DLT. Consequently, varying opinions may arise regarding the extent to which a system qualifies as a blockchain. Additionally, a noticeable pattern is the rising use of unclear and inconsistent language in diverse projects. This has led to instances where the same term signifies different concepts. This linguistic discordance poses a potential hindrance to the development and widespread adoption of the DLT sector. Another ramification of this lack of consensus and standardization manifests as a deficiency in interoperability between DLT networks.
The DLT ecosystem lacks interoperability as the DLT community, driven by intense competition and commercial pressures, focuses on the introduction of new systems that emphasize improved performance.
Yet, the absence of a standard technical reference model makes it challenging to evaluate and compare these systems.
In light of these challenges, our goal here is to establish clearer distinctions between different categories within the DLT ecosystem. We propose a new taxonomy designed to highlight the unique differences between different DLT groups. Additionally, we adopt a systematic and holistic approach to conceptualize and scrutinize DLTs as functional systems, emphasizing key layers across four levels of analysis. This comprehensive effort aims to provide a structured framework for understanding the diverse technologies within the evolving DLT landscape.

\color{black}
\subsection{ Motivation, aims and impact of the survey }
\subsubsection{  Problem statement and motivations}
At the time of writing this survey, the evolution of blockchain technology initiates an era marked by inconsistency and intense technical variations. In fact, there is a big number of blockchain-based projects under development. Some of which are simple replications of well-known projects, such as Bitcoin or Ethereum, whereas others propose entirely new functionalities and architectures. The current variations in blockchain systems pose a number of concerns from different perspectives, particularly concerning heterogeneity, coupled with the lack of interoperability, which may hinder the adoption of blockchains in our techno- and socio-economic systems. Furthermore, the diverse designs of DLT and their adaptable configurations pose a challenge for software architectures and developers when it comes to making informed decisions for constructing genuinely decentralized systems.Furthermore, like other technologies, \textit{the lack of standards} harms privacy, security, governance, and more importantly, interoperability. \\
Several potential issues may arise, including but not limited to:
\begin{enumerate}
\item Hindering consistency in the formulation of of DLT regulatory laws and policies.
\item Causing confusion in the application of consumer protection laws and regulations.
\item Diminishing the precision of academic research aimed at exploring the foundational concepts essential for the development of innovative applications and solutions, thereby potentially impeding advancements in various fields.

\item Hindering the widespread adoption of the technology and its interoperability and integration with existing standardized technologies, thus hindering the utilization of blockchains/DLT and their potential applications.

\end{enumerate}



In fact, any DLT or Cryptocurrency regulation or law begins by providing a correct legal definition of these elements. A DLT taxonomy helps establish that basis and assists regulators in crafting a flexible and granular DLT regulatory framework that considers the differences between existing projects rather than treating them as a single entity. For certain regulations such as eIDAS (electronic Identification, Authentication and Trust Services), the GDPR (General Data Protection Regulation), or Mica(markets in crypto-assets), the scope of applicability can differ from one category to another, depending on the specifications of a given DLT project (e.g., Data shareability). Thus, in the absence of a standard that helps regulators differentiate between different types of DLTs, confusion may arise regarding when to apply consumer protection laws and regulations. 
Furthermore, the visibility provided by the proposed framework makes it easier to integrate blockchain-based registries into other legislation. 
     
\subsubsection{Our proposed approach and methodology}


A comprehensive solution to the aforementioned challenges involves proposing a reference model that delineates standardized structures, elements, and relationships. Such a model would serve as a framework for distinguishing authentic blockchain systems from those that exhibit blockchain-like characteristics. Establishing clear boundaries between these categories is imperative to establish a fair and equitable environment, facilitating the design and adoption of blockchain-based products or services by industry participants and community members.
Drawing parallels with the Internet, where various standardization bodies (such as IETF in collaboration with W3C, ISO/IEC, ITU) continually define standards to enhance interoperability among systems and technologies, a similar approach is essential for the blockchain domain. By instituting blockchain standards, we can encourage the development and widespread adoption of interoperable blockchain and blockchain-like applications. Analogous to how standards for the World Wide Web contribute to interoperability, accessibility, and usability of web pages, blockchain standards would play a pivotal role in fostering the flourishing and widespread use of blockchain technologies.
To achieve this goal, we conduct the following steps:

$(1)$ Definition and Vocabulary Framework: An initial step involves scrutinizing the vocabulary and terms to clarify any ambiguities and resolve discrepancies. Conducting a literature review of existing technologies serves as the foundation for streamlining complexity and structuring information systematically. This process culminates in the development of a comprehensive vocabulary encompassing key blockchain terms. This established vocabulary serves as the groundwork for readers, enabling a better understanding of the subsequent classification and taxonomy presented in the analysis.

$(2)$ Framework Setting \& Component and Property Identification: A review of the DLT literature by adopting a common multi-layered approach that inspects each key design separately. In fact, layering is a basic structuring technique used in different models such as TCP/IP, Open Systems Interconnection (OSI) \cite{Wittbrodt}, and Linux Systems. This layerization divides the studied systems into distinct layers to make it easier to understand and analyze. The structure of the different defined layers helps us to determine to which taxon a system belongs.


$(3)$ Blockchain Classification: Finally, by examining the components and properties identified in each layer, we introduce and compare two categories. However, due to the continuous evolution of technology, these categories are expanding over time. Consequently, for the purpose of simplicity, our study focuses on two primary configurations for each component and property.

\subsubsection{ Results and impact of the survey}
The outcome of the analysis at the component level yields a comprehensive blockchain taxonomy and a layered framework. This framework organizes major components hierarchically, elucidating their functional relationships and potential design patterns. This conceptual framework (DCEA) serves as a tool that assists DLT designers and decision-makers in analyzing the state of DLTs and their interactions for a comprehensive understanding of the DLT landscape and different proposed solutions. Moreover, the proposed framework helps DLT designers and DLT adopters build a structured vision of the proposed solutions in different DLTs and thus opt for the suitable design choice.
In general, assessing the quality of a taxonomy or ontology proves challenging, particularly in dynamic domains such as blockchains. Taxonomies and ontologies are typically crafted to manage complexity and structure information, each serving distinct purposes and undergoing evolutionary changes over time (as seen, for instance, in the evolution of the renowned Linnaean taxonomy in biology).
Our taxonomy seeks to lay the groundwork for classifying diverse blockchain components, acknowledging that it does not claim to be the definitive structure. Nevertheless, the proposed taxonomy holds practical significance in various scenarios. It can:
\begin{itemize}
\item[(1)] Support software architectures in exploring different system designs, evaluating, and comparing diverse design options;
\item[(2)] Serve as preparatory work for the development of blockchain standards, aiming to enhance the widespread adoption of blockchain-based solutions and services;
\item[(3)] Facilitate research into architectural frameworks for blockchain-based systems, fostering the adoption of blockchain-based systems through increased interoperability and compatibility.
\end{itemize}

\color{black}

\subsection{Contributions}
\noindent
This paper significantly extends our previous works \cite{bellaj2022sok}, \cite{bellaj2022untangling}, contributing in the following key ways:

\begin{enumerate}
    \item Reference Model Establishment: The paper introduces a reference model that offers a comprehensive perspective on Distributed Ledger Technology (DLT) systems over time. This model categorizes current systems across four distinct layers: data, consensus, execution, and application, providing a thorough exploration of the state of the art.

    \item DLT Taxonomy: A novel taxonomy of DLTs is proposed, challenging the distinctions between blockchain and blockchain-like systems. This taxonomy is based on different architectural configurations across the four layers defined in the reference model.

    \item Comprehensive Comparison: The paper provides the first comprehensive comparison of a broad spectrum of DLT technologies, encompassing 44 projects and $26$ consensus mechanism, offering valuable insights into their unique features and characteristics.

    \item Consensus Mechanism Evaluation: A qualitative and quantitative comparison of various consensus mechanisms, including recent contributions, is conducted through the established framework. This evaluation enhances understanding and aids in decision-making for system designers.

    \item Academic Achievements Summary: The paper summarizes practical academic achievements that have positively impacted the design and performance of DLTs. This inclusion provides a succinct overview of advancements in the field.

    \item Exploration of New Trends: Emerging blockchain trends such as Blockchain modularity, zKvms, accounts abstractions, and others are presented and discussed. This forward-looking analysis identifies and explores evolving aspects within the blockchain landscape.
\end{enumerate}

The paper will serve as a valuable guide for blockchain system designers, aiding them in making informed design choices for new DLT implementations. By encompassing a broad spectrum of DLT technologies, offering a novel taxonomy, and addressing current and future trends, the paper provides a comprehensive resource for both researchers and practitioners in the blockchain domain.


\color{black}
\subsection{Paper’s Organization}

The structure of the survey is outlined as follows. Section \ref{sect 2} provides the necessary background, defines the adopted terminology, and offers a synopsis of the contextualized history of the Distributed Ledger Technology (DLT) evolution. In Section \ref{sect 3}, the proposed DCEA framework is introduced, outlining its layers and components. The section also elaborates on how this framework is employed to classify Distributed Ledger Technologies (DLTs). Sections \ref{sect 4}, \ref{sect 5}, \ref{sect 6}, and \ref{sect 7} individually delve into the DCEA layers, addressing the data, consensus, execution, and application layers, respectively. Within each section, we provide an overview of the key components and properties of the examined layer, accompanied by an exploration of the latest developments in the field. Section \ref{sect 8} provides an in-depth comparative assessment and critical analysis of 44 diverse DLTs in academic and industrial settings. This evaluation employs the proposed referential framework and encompasses the examination of 26 consensus protocols. Section \ref{sect 9} discusses the lessons learned from the reviewed literature. Section \ref{sect 11} focuses on the open challenges and research directions. Finally, we conclude with a summary in Section \ref{sect:conclusion}.

\subsection{Comparison with existing surveys and tutorials}	
In the past few years, many studies have been conducted on reviewing and taxonomizing DLT technologies, covering multiple topics. Generally, we observe three types of surveys as shown in Table \ref{tab:taxonomy_survey}. First, there are components-oriented surveys focusing on specific parts of DLTs, e.g., consensus mechanisms. Second, application-oriented surveys covering DLT applications in different domains; and third, surveys providing conceptual and holistic views of the DLT landscape. Our work belongs to the last category, even though none of the prior work has the scope or the width we adopt in our survey. A comparative summary of earlier similar works is shown in Table \ref{tab:surveys-table}.

Here, we provide an overview of some recent holistic research. \cite{Tschorsch2015a} attempts to reflect the state of the art in the area of fully distributed digital currencies. However, it predominantly focuses on the Bitcoin protocol, its building blocks, and its applications, with a very short introduction to other Bitcoin-inspired projects. \cite{Xu2017} proposes a taxonomy that captures a limited number of architectural characteristics of blockchains, namely: the level of decentralization, computation and client storage, as well as blockchain configuration. Besides, the authors provide only a brief surface-level review of concepts and notions needed to understand or classify DLTs. In \cite{Butijn2020}, a good academic literature review on Blockchain technology is presented, with a focus on providing a non-exhaustive list of architecturally-relevant characteristics and quality attributes. However, the authors define the blockchain as a distributed ledger technology without setting a clear distinction between the two.

Interestingly, \cite{Ballandies2018} proposes a conceptual architecture, a taxonomy that distinguishes cryptoeconomic design (CED) from DLT, and a classification of 29 distributed ledger systems (DLT). However, the paper does not present a detailed overview of the underlying concepts of DLTs. Besides, the proposed taxonomy focuses more on CED and considers the blockchain as a mere data structure without a clear and systematic definition. Similarly, \cite{Tasca2019} introduced a comprehensive taxonomy of DLTs with extensive coverage of multiple relevant concepts that revolve around CED. The proposed taxonomy lacks a conceptual architecture that defines what a blockchain or a DLT is. More particularly, this taxonomy does not explicitly differentiate between CED and DLT or between different types of distributed ledgers.

\begin{table*}[ht]
\caption{Three broad types of survey studies reported in the blockchain literature: Component-oriented, application-oriented, and holistic surveys.}
\label{tab:taxonomy_survey}
\begin{adjustbox}{max width=\linewidth}
\begin{tabular}{|l|l|l|l|l|}
\hline
Survey Type &
  Subcategory &
  Paper &
  Year &
  Targeted topic \\ \hline
\multirow{17}{*}{\begin{sideways}Component-oriented  \end{sideways}} &
  \multirow{7}{*}{Consensus protocols} &
\cite{Vukolic2016} & 2021
   &
  A survey and taxonomy of consensus protocols for blockchains
 \\ \cline{3-5} 
 &
   &
  \cite{Cachin2017} & 2022
   &
  A survey and taxonomy of consensus protocols for blockchains \\ \cline{3-5} 
 &
   &
  \begin{tabular}[c]{@{}l@{}}\cite{Dinh2017}\\
  \end{tabular} & 2023
   &
  A Survey Paper on Blockchain Technology and Consensus Algorithms \\ \cline{3-5} 
 &
   &
  \cite{Bano2017} & 2017
   &
  A Taxonomization and evaluation of consensus protocols \\ \cline{3-5} 
 &
   &
  \cite{Wang2019} &2019
   &
  \begin{tabular}[c]{@{}l@{}}A comprehensive survey \\ of permissionless blockchain consensus protocols focusing on the underpinning\\ incentive mechanism.\end{tabular} \\ \cline{3-5} 
 &
   &
  \cite{Xiao} & 2019
   &
  A self-complete tutorial on different types of distributed consensus protocols. \\ \cline{3-5} 
 &
   &
  \cite{Ferdous} & 2020
   &
  \begin{tabular}[c]{@{}l@{}}A survey taxonomizing blockchain consensus\\ algorithms based on incentivization.\end{tabular} \\ \cline{2-5} 
 &
  \multirow{4}{*}{Smart contract} &
  \cite{Chen2020} &
  2020 &
  A survey on vulnerabilities and Attacks on Ethereum smart contracts \\ \cline{3-5} 
 &
   &
  \cite{Zheng2020} &
  2019 &
  A survey on security, application and performance of smart contracts. \\ \cline{3-5} 
 &
  &
 \cite{Zheng2020} & 
 2020
 &
  A survey on challenges, advances and platforms of Smart Contracts \\ \cline{3-5} 
 &
   &
  \cite{Almasoud2020} &
  2020 &
  A Survey of Smart contracts for blockchain-based reputation systems. \\ \cline{2-5} 
 &
  \multirow{3}{*}{Security and privacy} &
 \cite{Saad2020} &
  2020 &
  A Survey on the attacks targeting the public blockchain \\ \cline{3-5} 
 &
   &
  \cite{Li2020} &
  2020 &
    A survey examining of the security aspects within blockchain systems \\ \cline{3-5} 
 &
   &
\cite{islam2023survey}&
  2023 &
  A Survey on Blockchain Security and Its Impact Analysis
 \\ \cline{2-5} 
 &
  \multirow{3}{*}{Scalability} &
  \cite{Zhou2020} &
  2020 &
  A survey on the Scalability of blockchain \\ \cline{3-5} 
 &
   &
  \cite{Hafid2020} &
  2020 &
  A survey on sharding in blockchains \\ \cline{3-5} 
 &
   &
  \cite{Zhou2020} &
  2020 &
  A survey on the scalability solution in blockchains \\ \hline
\multirow{10}{*}{ \begin{sideways}Application-oriented  \end{sideways}} &
  \multirow{4}{*}{IOT} &
  \cite{Dai2019} &
  2019 &
  A survey on Blockchains in Internet of Things \\ \cline{3-5} 
 &
   &
  \cite{Ali2019} &
  2018 &
  A survey on the application of Blockchain in Internet of Things \\ \cline{3-5} 
 &
   &
  \cite{Wang2019b}&
  2019 &
  A survey on Blockchain for Internet of Things

   \\ \cline{3-5} 
 &
   &
  \cite{Lao2020} &
  2020 &
  A Applications in Blockchain Systems: Architecture, Consensus, and Traffic Modeling \\ \cline{2-5} 
 &
  Different domains &
  \cite{Chen2018a} &
  2018 &
  A Survey on Blockchain Applications in Different Domains \\ \cline{2-5} 
 &
  Smart cities &
  \cite{Xie2019}&
  2019 &
  A Survey on the application of Blockchain Technology in Smart Cities \\ \cline{2-5} 
 &
 Telecoms &
  \cite{Nguyen2020} &
  2020 &
  A state of the art survey on the application of Blockchain in 5G networks \\ \cline{2-5} 
 &
 Cloud and Edge computing
   &
  \cite{Yang2019} &
  2019 &
  A survey on the application of Blockchain on Edge Computing Systems \\ \cline{2-5} 
 &
  Blockchain and machine learning &
  \cite{Liu2020} &
  2020 &
  A survey on the application of Blockchain in machine learning \\ \cline{2-5} 
 &
  Blockchain and artificial intelligence &
  \cite{Salah2019} &
  2019 &
  A survey on the application of Blockchain in AI \\ \hline  
  
  \vtop{\hbox{\strut Holistic and }\hbox{\strut conceptual}}
 &
\multicolumn{4}{c|}{\vtop{\hbox{\strut  }\hbox{\strut This category is presented separately in Table II}}}

 \\ \hline

\end{tabular}
\end{adjustbox}
\end{table*}

\cite{Kolb2020} is a tutorial that explains the fundamental elements of blockchains using Ethereum as a case study. The paper considers comparing, at a high level, blockchains to traditional distributed systems. However, the paper compares both categories in a manner that hardly differentiates between a traditional distributed system and a permissioned blockchain—at least for many settings in which the private blockchain can be configured.

\cite{Belchior2020} proposes an exhaustive literature review on blockchain interoperability. The paper presents the necessary background and highlights definitions tailored for both industry and academia. It categorizes blockchain solutions into three categories considering solely the interoperability aspect.

\cite{Monrat2019} presents a comparative study of the applications and trade-offs of blockchain. The paper explains the architecture of the blockchain adopted by Bitcoin and Ethereum without considering other variants, e.g., IOTA or Hyperledger. Besides, most of the reviewed DLTs are briefly introduced without discussing their inner working mechanisms.

\cite{Belotti2019} provides a conceptual framework with a multi-layer abstraction of a blockchain. The authors propose a vademecum containing the information necessary to understand blockchain from a technical perspective, then introduce a decision model on which, when, and how to use blockchain technology. However, this paper covers a limited number of DLTs ($7$) and consensus mechanisms, making it an incomplete guide for readers looking to understand new DLT technologies operating in different modes than conventional blockchains.

In summary, the current state of the art on DLT system taxonomies suffers from a few limitations. First, the number of evaluated DLTs or consensus mechanisms across the papers varies significantly, from $0$ to $29$ for the DLTs and from $0$ to $15$ for the consensus mechanisms, and the provided evaluations are often superficial. Second, most papers adopt the same vision that often distinguishes real-world DLT systems based on their operation modes—private or public—or based on their access modes—permissioned or permissionless—which are insufficient characteristics to define whether a system is a blockchain or not. Third, none of the papers, except one, proposes a conceptual framework for DLTs and blockchain systems; none offers a systematic definition for fundamental notions like DLTs, Blockchain, DAGs (Directed Acyclic Graph), and more.

Table \ref{tab:taxo} presents a comparison between our proposed taxonomy and other existing taxonomies. Guided by the adopted criterion, our taxonomy surpasses the simplistic classifications found in the literature, which typically distinguish DLTs based on permissibility (permissioned or permissionless) or their public or private nature. Our taxonomy introduces a new perspective that aims to define what constitutes a blockchain and differentiates it from similar systems.

The framework used for taxonomizing DLTs has a positive impact on the DLT design process by facilitating decision-making through a systematic comparison of the capabilities of different design options. Furthermore, it illustrates the impact of these design choices on various quality attributes. The trade-off analysis of these quality attributes forms the foundational basis for effective comparisons.

\color{black}

\begin{table*}[ht]
\caption{Comparison of Proposed Frameworks for Understanding and Utilizing Blockchain Technology}
\centering
\label{tab:taxo}
 \colorbox{mycolor}{%
\begin{adjustbox}{max width=\linewidth}
\begin{tabular}{|p{2cm}|p{2.5cm}|p{2.5cm}|p{2.5cm}|p{2.5cm}|p{2.5cm}|p{2.5cm}|p{2cm}|}
\hline
\textbf{Taxonomy} & \textbf{Focus} & \textbf{Classification Dimensions} & \textbf{Strengths} & \textbf{Limitations} & \textbf{State-of-the-art Review} & \textbf{Technical review} & \textbf{Solutions covered} \\ \hline

\textbf{A Taxonomy of Blockchain Consensus Protocols \cite{bouraga2021taxonomy}} & Consensus protocols & Fault tolerance model, Block creation method, Leader selection mechanism, Scalability & Technical precision and depth of analysis & Limited to consensus protocols & Partially - Focuses on consensus protocols & Yes & $28$ consensus protocols \\ \hline
\textbf{A Taxonomy of Centralization in Public Blockchain Systems \cite{sai2021taxonomy}} & Centralization levels in public blockchains & Application, Contract, Consensus, Incentive, Network and Data layer & Captures various aspects of centralization and evaluate multiple research papers & Emphasizes the assessment of research papers rather than concentrating on Blockchain projects.  focuses on evaluating research papers rather than Blockchain projects  & Partially - Focuses on centralization aspects & No & $2$ blockchain projects \\ \hline

\textbf{Taxonomy of Blockchain Technologies. Principles of Identification and Classification \cite{tasca2017taxonomy}} & Standardization of blockchain architectures & consensus, sharing and rewarding systems, Transaction capabilities, Native currency/Tokenization, Extensibility, Security and Privacy, Code Base and  Identity Management & Bottom-up approach, Component-based taxonomy & Early stage analysis, Potential complexity, Focus on software architecture, Limited discussion of existing projects & No - Engages with research and standardization discussions, but doesn't explicitly compare specific frameworks or solutions & No& None \\ \hline

\textbf{A Taxonomy For Governance Mechanisms Oftowards A Taxonomy For Governance Mechanisms Of
Blockchain-Based Platforms
 \cite{werner2020towards}} & Blockchain governance structures & Centralization level, Decision-making process, Stakeholder participation & Analyzes different governance models & Limited to blockchain governance   & Partially - Focuses on formal governance structures & No& None\\ \hline
\textbf{A Taxonomy for Characterizing Blockchain Systems \cite{alzhrani2022taxonomy}} & Composition of blockchain systems & Platform, P2P Network,
Distributed Ledger,
Smart Contract,
Consensus Protocol,
Digital Wallet, Token and
Network Node  & Comprehensiveness, Hierarchy, Software Development guidance, Case Studies & Complexity, limited discussion and comparison, The layer-based taxonomy may result in overlapping classifications, complicating information organization & Partially - Offers a restricted overview of the state-of-the-art, focusing on whether a layer or component exists in a solution without providing detailed explanations. & Yes - Provides a high-level technical review and doesn't cover consensus mechanisms.& $10$ blockchain projects and $11$ consensus mechanism \\ \hline

\textbf{A Vademecum on Blockchain Technologies \cite{Belotti2019}} & Practical guide to blockchain decision-making & Decision factors, Use cases, Platform selection, Development & Accessibility to a broad audience & Focuses primarily on the decision-making process for adopting blockchain technology, not a comprehensive analysis of the technology itself. The provided analysis focuses on only seven projects & No - Offers practical insights, not exhaustive review & Yes & $7$ blockchain projects  \\ \hline

\textbf{A Taxonomy of DLTs (Our Work)} & Reference model for DLTs (blockchain and blockchain-like) & Data, Consensus, Execution, Application layers & Comprehensiveness, Component-based taxonomy, Inclusion of Blockchain and Blockchain-Like Systems, A 360° overview of the technical and practical dimensions of DLTs, Technical review of each layer of the DLT & Analysis limited to $44$ selected DLT projects & Yes - Comprehensive examination of the state-of-the-art, considering both technical and design perspectives. & Delve into the technical details of different blockchain platforms or specific implementation challenges.& $44$ projects blockchain and $26$ consensus mechanism.
\\ \hline
\end{tabular}
\end{adjustbox} 
}
\end{table*}

In contrast to existing literature, our contribution entails a comprehensive, standalone tutorial offering a thorough exploration of the architecture underlying blockchain technology. Moreover, we propose a new taxonomy and classification that is highly comprehensive with a robust view of the DLT landscape based on a rigorous and systematic definition of what blockchain is and is not. Furthermore, we evaluate the largest number of DLTs ($44$) and consensus mechanisms ($26$) compared to other holistic surveys with a focus on covering the whole DLT landscape. Both the taxonomy and the evaluation are intended to help decision-makers architecting new blockchain-based systems or choosing among the existing solutions through enabling a systematic comparison between the capabilities of different design choices.

\begin{table*}[ht]
\caption{ Comparison of related comprehensive and holistic surveys}
\label{tab:surveys-table}
\begin{adjustbox}{max width=\linewidth}
\begin{tabular}{|l|l|l|l|l|l|l|l|l|l|l|l|l|l|l|l|l|}
\hline
 { Paper}&
 
{ \vtop{\hbox{\strut Publication}\hbox{\strut year}}}  &
 
 { \vtop{\hbox{\strut Conceptual}\hbox{\strut framework}}}   &
 { \vtop{\hbox{\strut Taxonomy}\hbox{\strut of DLTs}}}  &
 {  \vtop{\hbox{\strut Number of}\hbox{\strut DLTs attributes}}}  &
 {  \vtop{\hbox{\strut Evaluated}\hbox{\strut DLTs}}}  &
 {  \vtop{\hbox{\strut Evaluated}\hbox{\strut consensus}}}   &
 { Limitations}  &
 {  \vtop{\hbox{\strut Solutions }}}   &
 {  \vtop{\hbox{\strut Surveyed}\hbox{\strut layers*}}}  &
 { DLTs categories}  &
 { \vtop{\hbox{\strut Applica-}\hbox{\strut tions}}}  \\ \hline
\textbf{\cite{Belchior2020}}  &
  2020 &
  No &
  No &
  18 &
  21 &
  0 &
  Yes  &
  Yes &
  C,E &
  No &
  Yes \\
\textbf{\cite{Kolb2020}} &
  2020 &
  No &
  No &
  17 &
  9 &
  5 &
  Yes &
  Yes &
  C,E &
  Yes (Distributed Database, Blockchain) &
  No \\
\textbf{\cite{Belotti2019}} &
  2019 &
  No &
  Yes &
  19 &
  7 &
  15 &
  Yes &
  Yes &
  D,C,E &
  No &
  Yes \\
\textbf{\cite{Monrat2019}} &
  2019 &
  No &
  Yes &
  11 &
  7 &
  5 &
  Yes &
  No &
  D,C &
  No &
  Yes \\
\textbf{\cite{Ballandies2018}} &
  2018 &
  Yes &
  Yes &
  10 &
  29 &
  0 &
  No &
  No &
  D,C,A &
  Yes (Cryptoeconomic design, distributed ledger) &
  No \\
\textbf{\cite{Butijn2020}} &
  2018 &
  No &
  No &
  13 &
  0 &
  7 &
  Yes &
  Yes &
  D,C,E &
  No &
  Yes \\
\textbf{\cite{Xu2017}}&
  2017 &
  No &
  Yes &
  4 &
  0 &
  4 &
  Yes &
  No &
  D,C &
  No &
  No \\
\textbf{\cite{Tasca2019}} &
  2017 &
  No &
  Yes &
  25 &
  0 &
  6 &
  No &
  No &
  D,C,E &
  Yes (Cryptoeconomic design, distributed ledger) &
  No \\
\textbf{\cite{Tschorsch2015a}} &
  2016 &
  No &
  No &
  12 &
  0 &
  5 &
  Yes &
  Yes &
  D,C,E &
  No &
  Yes \\ \hline
Our work &
  2020 &
  Yes &
  Yes &
  25 &
  44 &
  26 &
  Yes &
  Yes &
  D,C,E,A &
  Yes (Blockchain, Blockchain-like) &
  Yes
  
\\ \hline
\end{tabular}
\end{adjustbox}
\begin{tablenotes}
        \scriptsize
         \item[1]  \emph{* D} : Data \emph{C}: Consensus \emph{E} : Execution \emph{A}: Application.

    \end{tablenotes}
\end{table*}

\subsection{Survey’s approach and methodology}

To conduct our survey, we followed the five-step process outlined by Biolchini et al. \cite{Biolchini2005} for elaborating a systematic review, as explained below:

\textbf{Problem Formulation:}
\begin{itemize}
  \item RQ1: How can blockchain technology be defined within a unified layered-wise reference model?
  \item RQ2: What are the properties and components distinguishing each layer?
  \item RQ3: How can we distinguish a blockchain system from a blockchain-like one?
  \item RQ4: What obstacles and challenges does blockchain technology currently face?
  \item RQ5: What are the existing research gaps in the realm of blockchain technology?
\end{itemize}

\begin{figure*}[ht]
\centering
\includegraphics[width=18cm]{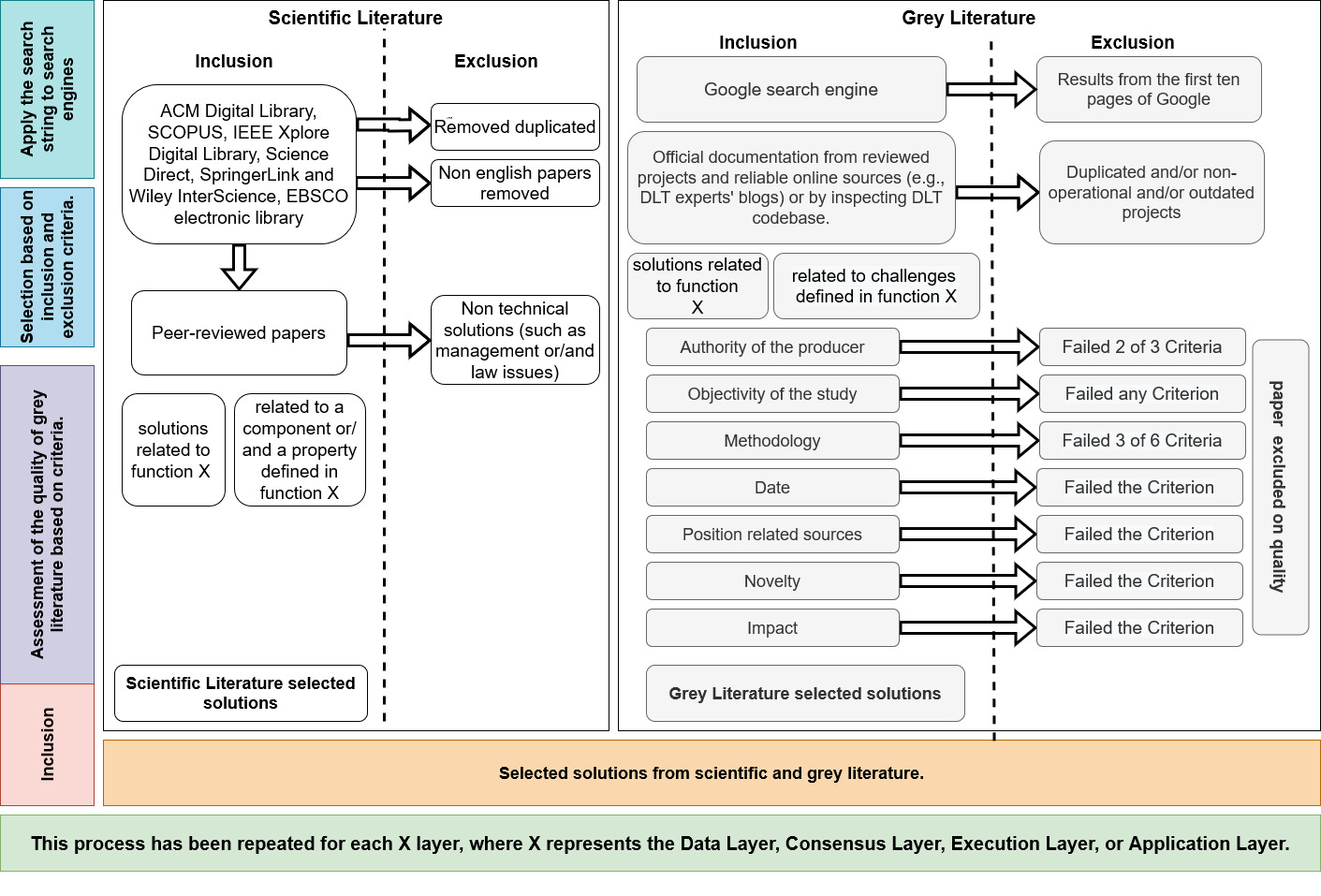}
\caption{Search and selection strategy, depicting a process model for a single layer.}
\label{fig:methodology}
\end{figure*}

\textbf{Data Collection:}
After defining the layered-wise reference model, we gathered information for each layer from scientific literature, official documentation of reviewed projects, and reliable online sources (e.g., DLT experts' blogs), known as Grey Literature. This collection and selection were carried out using the systematic review protocol suggested by Kitchenham \cite{Kitchenham2004}, involving three stages: (1) elaborating the search string; (2) applying the string on chosen search engines; (3) filtering out and extracting primary papers based on pre-established exclusion criteria from search results. 
The implementation of these steps is illustrated in a streamlined process model representing the methodology employed in this research, as shown in Figure \ref{fig:methodology}.
The reviewed projects from both the scientific and Grey literature are selected based on their notoriety and scientific significance. We assessed the quality and relevance of the sources from the grey literature within the exclusion criteria suggested by Garousi et al. \cite{Garousi2017}. These criteria represent $7$ quality categories, ranging from the credibility of the producer to the objectivity of the study, as outlined in Table \ref{tab:CRITERIA} under the assessment of quality grey literature.

\begin{table}[ht]
 \caption{Quality criteria Grey literature}\label{tab:CRITERIA}
  \includegraphics[width= 8cm]{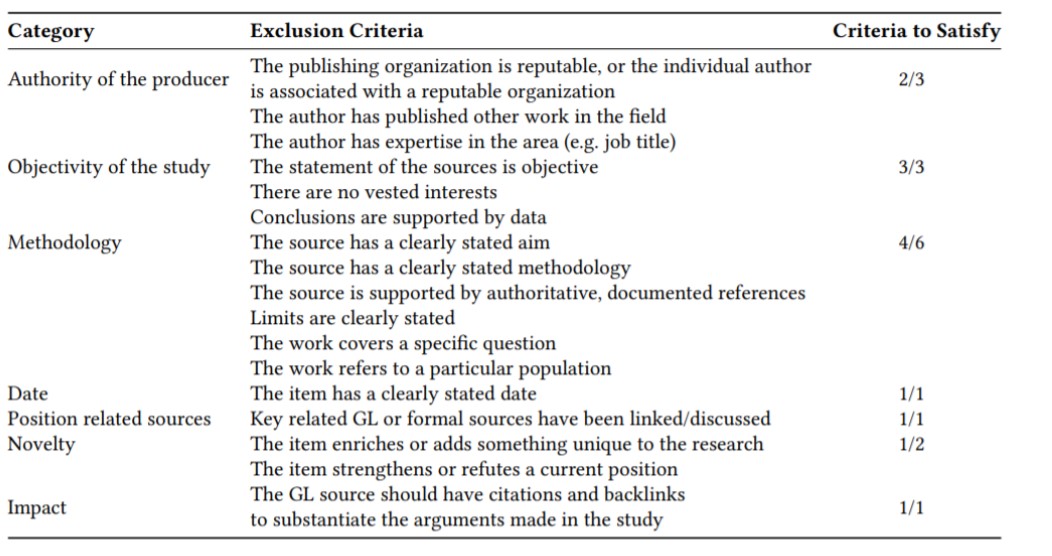}
\end{table}

The choice of the $44$ studied solutions was driven by the need to have good representativity of different components defined by the DCEA framework. For instance, we aimed to select DLTs that represent all different consensus mechanisms ($26$ different consensus mechanisms) and other DLTs representing other settings defined by the DCEA framework at the four layers.
\begin{itemize}
    \item Data Evaluation: We systematically evaluate the DLT literature, highlight the pros and cons of each solution, and discuss how each solution matches the components and properties defined in each layer.\\
    \item Analysis and Interpretation Process: Within the same multi-layered approach, we conduct a qualitative and quantitative analysis of the surveyed solutions based on the $23$ criteria (components and properties) defined in the first step.\\
    \item Conclusion and Presentation: Finally, we discuss the main remaining issues and draw future research directions and insights for the four defined layers.
\end{itemize}

\section{HISTORY AND BACKGROUND} \label{sect 2}
\subsection{Contextualized History}	

The earliest identified occurrences of the concept of a ‘blockchain’ can be traced back to a 1991 research paper \cite{Haber1990} entitled “How to Time-Stamp a Digital Document” by Haber and Stornetta, introducing the notion of a cryptographically secured chain of blocks. They proposed a timestamping system where the server would link a document to a previous document to avoid tampering with data. In 1992, they upgraded their design by introducing Merkle trees and collecting document certificates in blocks. Although their project has a different structure than that of the current blockchain system, they laid the basis for the modern blockchain.

Around the same time, cryptocurrencies were already emerging. In fact, David Chaum proposed “untraceable payments” in 1983 \cite{Chaum1983}, in which he conceptualized an anonymous cryptographic system for processing digital money that could be transferred in the form of blindly signed coins. The proposed blind signatures provide a cryptographic means to prevent linking users to coins and to enforce fungibility, while still allowing a central server to ensure protection against double-spending. Later, Chaum proposed DigiCash \cite{Chaum1994} in 1990, as the first global electronic cash system. DigiCash gained some interest, but its centralized nature led to its failure and stoppage in 1998.

In 1997, the Hashcash \cite{Back2002} project proposed a cryptographic hash-based algorithm called proof-of-work to protect against denial of service, a concept that will be used later in Bitcoin \cite{nakamoto} as a fraud countermeasure. In 1998, Wei Dai proposed "b-money" \cite{Dai1998}, characterized as a pseudo-anonymous, distributed electronic cash system with a replicated, transparent, and public ledger. Several years later, in 2005, Nick Szabo described Bit Gold; a decentralized cryptocurrency inspired by Hashcash. Bit Gold was notable for using an underpinning system similar to the current blockchain. It included many of the basic components of modern cryptocurrencies, such as decentralization, anonymity protections, public registry, a proof-of-work mechanism, chains of hashes, and timestamping. However, Bit Gold was never implemented, and its design suffers from the double-spending problem, wherein users can spend a coin twice in the network. Nick Szabo is also renowned for a major contribution in the blockchain field. He proposed (in 1997) the concept of smart contracts, which enables counterparties to formally codify a cryptographically enforceable agreement without the need for any third party — a concept that can be considered a predecessor of Bitcoin’s scripting mechanism.

Despite the vast advancement in cryptographic-based currencies, the problem of proposing a decentralized network resistant to the double-spending problem remained unsolved until 2008, when Bitcoin was introduced by the anonymous person or persons, Satoshi Nakamoto, in “Bitcoin: A Peer-to-Peer Electronic Cash System” \cite{nakamoto}. One year later, Bitcoin network was widely deployed, with the genesis block mined on or around January 3, 2009. Bitcoin presented a convincing solution for preventing double spending in a global decentralized system without depending on trust in any third party, using a mix of techniques developed in earlier projects. Since then, multiple alternative cryptocurrencies with different features have been created, such as Ethereum \cite{Wood2014} and IOTA \cite{Popov2017}.
    
In 2014, motivated by the decentralized automation of Bitcoin, the blockchain technology was separated from the currency, mostly by the media and the industry, in an attempt to harness the power of Bitcoin’s underlying technology and its potential for other purposes. One year later, global financial companies formed R3, a consortium of 42 institutions \cite{R3} with an agenda of exploring and harnessing blockchain and DLT technology for financial, inter-organizational transactions with a blockchain-like product called Corda \cite{Hearn2016}. Around the same time, the Linux Foundation launched Hyperledger \cite{Dhillon2017} as an umbrella project of open-source DLT solutions to encourage the use of blockchain to support global business transactions.

\subsection{Terminology and Background}	

In the blockchain literature, there is significant overlap between the terms DLT and blockchain. The situation is made even worse by the absence of a universal standard defining a blockchain and its boundaries, which results in conflicting use of the terminology and different meanings being assigned to the same terms. The precise definition of multiple notions presented hereafter may be controversial, but for a better understanding and categorization of blockchain and similar systems, we provide here the terminology we have chosen to adopt throughout this paper.
\begin{itemize}
  \item \textbf{Nodes:} Nodes are entities, maintained by individual participants of the network, that issue and validate transactions in a DLT network.  
  \item \textbf{Ledger:} The term ledger refers to the single storage endpoint hosted by a node in a blockchain or DLT network. It represents a local append-only log of ordered transactions validated by a distributed network. 
  \item \textbf{DLT:} An umbrella term for distributed ledger technologies, representing technologies operating as replicated, shared and synchronized ledgers throughout a distributed network, whereby parties who do not fully trust each other maintain consensus about shared information. A DLT, as a data structure, can adopt some of the design principles of blockchain technology whilst dropping others and might be governed in a partially or fully centralized manner. Although blockchain is initially a distributed ledger, the term DLT is now used frequently to separately identify technologies which do not embrace the full Nakamoto vision. To the best of our knowledge, the DLT term was  initially adopted, in the industry, for commercial purposes to differentiate \cite{Hearn2016} some solutions (e.g. Corda or Hyperledger Fabric) from traditional blockchain systems like Bitcoin. 
  \item \textbf{Blockchain:} A distributed system that shares fundamental architectural principles laid out by Nakamoto in Bitcoin’s initial paper. That is, we can define a blockchain as a replicated database, managed by a consensus mechanism, solving the double spending problem, in a peer-to-peer fashion and shared by non-trusting parties. The inner database presents special key properties such as a chain of blocks, data structure, data immutability, global data shareability, the use of cryptography to secure data and ensure direct ownership, and interaction through transactions. The core fundamental characteristic of a blockchain is its fully or mostly decentralized and censorship-resistant nature. Initially, the term blockchain (as a single word) did not appear in Bitcoin's initial paper, but instead the term "chain of blocks" was used.
  \item \textbf{Blockchain-like:} A general term encompassing distributed ledger systems that are constructed in a similar manner to the conventional blockchain system (Bitcoin), while not necessarily conforming to Nakamoto’s vision (Bitcoin’s white paper \cite{nakamoto}).
  \item \textbf{State:} While state can have different meanings depending on the context, the use of state within the blockchain and distributed ledger environment refers to the atomic unit of data stored in the ledger, where it is prone to change under the execution of validated transactions.  As described by Gavin Wood in the Yellow Paper  \cite{Wood2014}, a blockchain is a cryptographically secure, transactional singleton machine with shared state. 
  \item \textbf{P2P gossip:} DLT networks are built upon P2P networks that transport all data needed to support the DLT protocol.  In a blockchain network, nodes are equally privileged, equipotent participants in the application, whereas a Blockchain-like system is usually a hybrid combination of a peer-to-peer network and centralized entities where some nodes are more privileged than others. Blockchain’s P2P networks are structured \cite{Buterin2014a} or unstructured \cite{nakamoto} networks where nodes randomly connect to other nodes. Data is exchanged directly over the underlying TCP/IP network implemented as multiple TCP unicasts between connected nodes, but at the application level, nodes are able to communicate with each other directly, via the virtual overlay connections built upon the underlying physical network. These virtual overlay networks facilitate node discovery and indexing, besides they make the P2P system independent from the physical topology. In order to converge on an eventually consistent sequential log of transactions. Many blockchain and blockchain-like systems adopt gossip protocol \cite{He2019}\cite{Berendea2020} whereby nodes exchange notifications about new data (a transaction or a block) \cite{Decker2013}. At a high level, a node first discovers and learns about a set of peers (a list of nodes it knows), then it constantly advertises each new data it has received and validated, to his peers announcing its availability. Afterward, the peers who do not have the advertised data, in their turn, request the missing data avoiding thus broadcasting data directly to the nodes \cite{Decker2013} \cite{KyunKim2018}. The gossip process continues until all reachable nodes receive the new data.
  \item \textbf{On-chain/off-chain:} We consider the terms ‘off-chain’ or ‘off-ledger’ to refer to actions that occur outside the formal boundaries of the transactional ledger distributed between the nodes. Conversely, ‘on-chain’ or ‘on-ledger’ refers to actions that occur within the boundaries of the distributed ledger. 
\end{itemize}

Blockchain and DLT are often mistakenly conflated: a blockchain is a sub-class of DLTs, but a distributed ledger does not always comprise a blockchain. Thus, in the rest of this paper, we use the term DLT as a unifying term covering blockchains and blockchain-like systems.

\begin{figure}[ht]
\hspace*{-0.4cm}
\includegraphics[width=8cm, height=5cm]{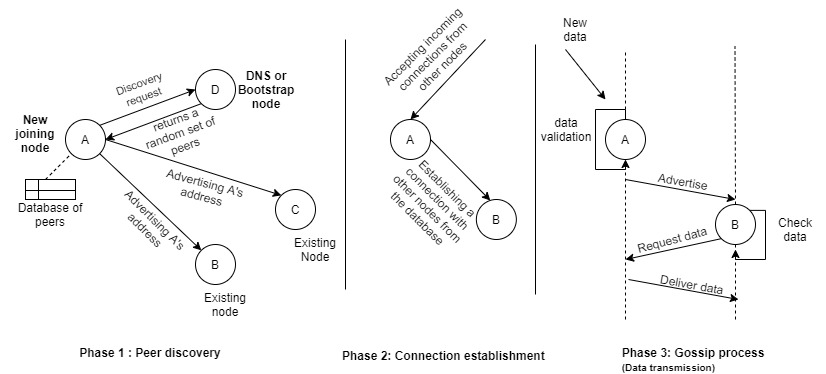}
\caption{Communication phases in a blockchain P2P network (Bitcoin)}\label{fig: p2p}
\textbf{\centering}
\end{figure}
  
\vspace{2pt}
\section{DCEA Framework: A Taxonomy-Oriented Approach to Conceptualize and Explore Distributed Ledger Technologies (DLTs)} \label{sect 3}

We present the DCEA framework, a structured model defining a layered and diverse stack tailored for the intricate realm of Distributed Ledger Technologies (DLTs). It's application organizes DLTs into $4$ fundamental and distinct layers: data, distributed consensus protocols, execution, and application layers. Each layer serves a precise and integral function within the overarching DLT architecture.

\begin{figure}[b]
\centering
\includegraphics[width=7cm, height=4cm]{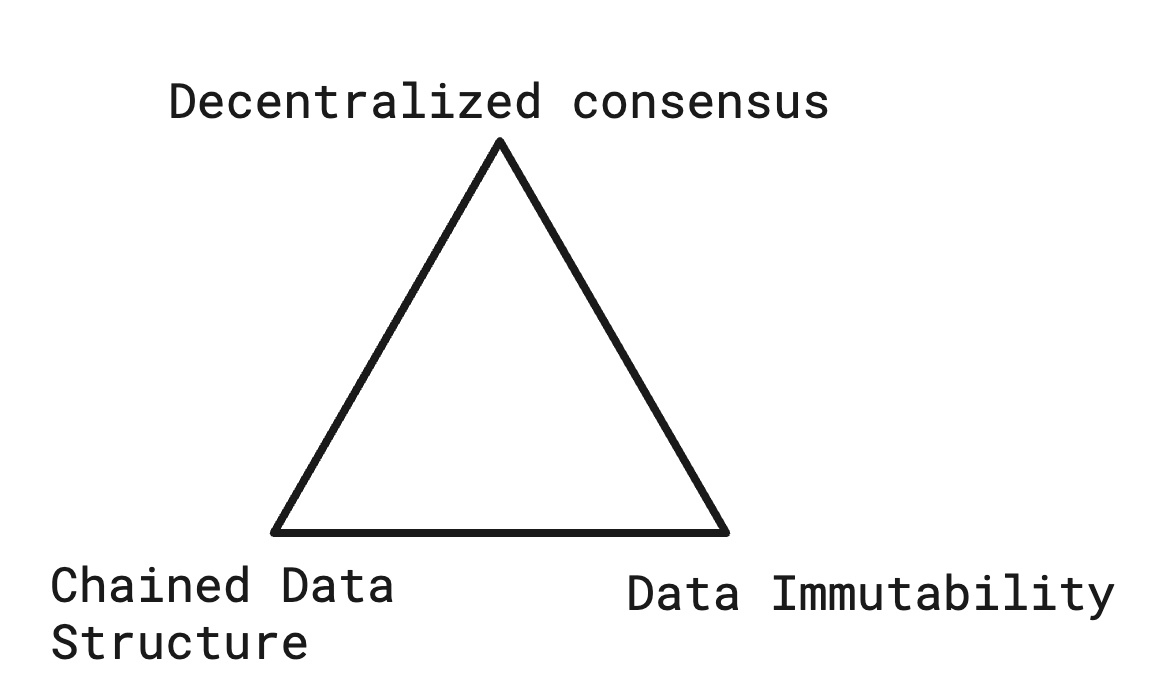}
\caption{According to our definition, a system resembling a blockchain can only possess two of these specified traits.}\label{fig: triangle}
\end{figure}

Utilizing the DCEA framework results in a nuanced dual-level classification of Distributed Ledger Technologies (DLTs). The primary classification is rooted in the varied configurations of key designs, while the secondary classification involves a comprehensive two-dimensional taxonomy, distinguishing between blockchain and blockchain-like technologies. This taxonomy evaluates the impact of different DCEA settings at 3 levels (refer to Figure \ref{fig: triangle}). The results of the exhaustive review and taxonomy, outlined in Sections \ref{sect 4}, \ref{sect 5}, \ref{sect 6}, and \ref{sect 7}, are applied in Section \ref{sect 8} to scrutinize and categorize a diverse spectrum of DLTs into the two overarching categories.

\begin{table*}[ht]
\caption{Layers and components of DCEA framework}
\label{tab:my-table4}
\centering
\includegraphics[width=\textwidth, height=3cm] {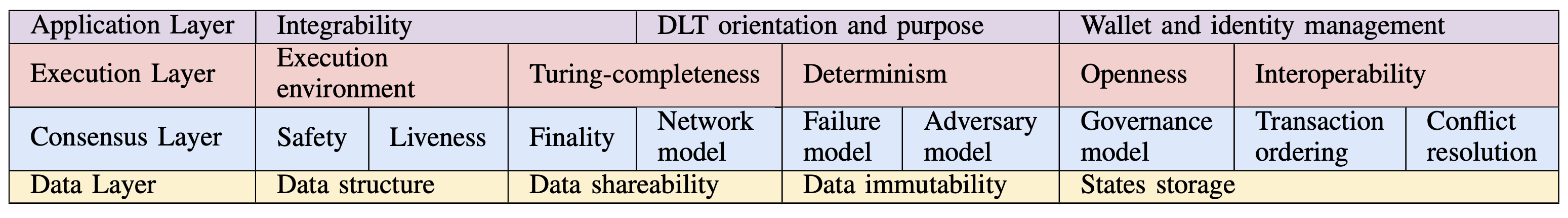}
\end{table*}

\subsection{Introduction to the DCEA Framework}

The DCEA framework encompasses the components of DLTs along with their key properties, as illustrated in Table \ref{tab:my-table4}. The components are logically organized, aligning with the modular architecture of DLTs. This layered structure enhances the comprehension of DLTs and establishes a foundational basis for comparative analyses across various DLT variants.
Below, we present the four layers constituting the DLT stack.

\begin{itemize}
\item Data Layer: encompasses the representation, storage, and flow of data within the distributed network. This layer deals with the transactions and states recorded in the ledger, forming the core information infrastructure of the blockchain. The Data Layer ensures that entries in the ledger are recorded under consensus, meaning all participants in the network agree on the validity of the data. These records can be elements built-in the underlying DLT protocols (e.g., cryptocurrency or Tokens or smart contracts states) or data sourced from external environments (e.g., IoT data). This layer encompasses both on-chain storage (stored directly on the blockchain) and off-chain storage (utilizing a distributed database).

\item Consensus Layer: Establishes a set of rules through software definition to reach consensus between network participants, ensuring alignment on a unique ledger (source of truth). This layer defines governance mechanisms that govern the validation, addition, and verification of transactions. The critical role of the Consensus Layer merits detailed exploration, which is extensively covered in Subsection \ref{subsection consensus analysis}.

\item Execution Layer: Serves as the computational environment where distributed programs, including smart contracts, are executed. These programs encode specific logic (e.g., business logic) into programmatic instructions which are executed to manipulate the states recorded in the ledger. In essence, the Execution Layer acts as the engine that interprets and processes these instructions, facilitating the decentralized execution of logic encoded in smart contracts. It plays a pivotal role in ensuring the integrity and automation of transactions within the distributed ledger, providing a secure and transparent framework for the execution of programmable business logic across the network. 

\item Application Layer: Refers to the topmost layer of the technology stack, serving as an abstraction layer that provides protocols and APIs (Application Programming Interfaces). This layer is designed to enable the development and execution of distributed applications, commonly known as DApps (Distributed Applications). It acts as a bridge between external entities, such as end-users or other applications, and the underlying code and data stored on the blockchain ledger.

\end{itemize}

\begin{figure*}[ht]
\hspace*{-0.3cm}
\includegraphics[width=\textwidth ]{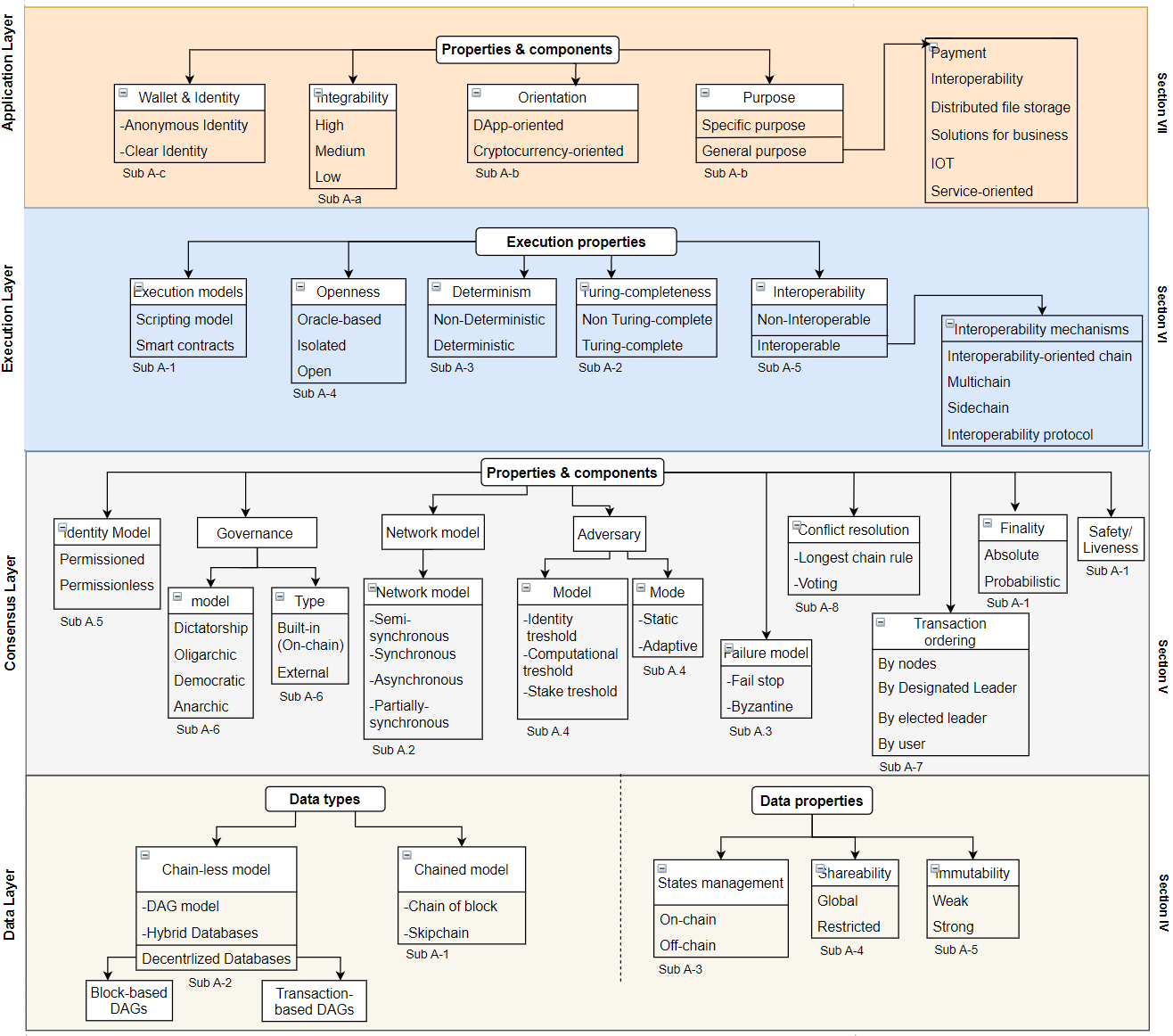}
\caption{ Layered taxonomy of DLT systems (With references to the corresponding sections and subsection)} \label{fig:taxon}
\centering
\end{figure*}

Built upon the layered architecture discussed earlier, we introduce a comprehensive four-layered taxonomy (Figure \ref{fig:taxon}) designed to categorize DLT systems. This taxonomy serves a dual purpose:
\begin{itemize}
\item To classify the various DLT systems presented in academic and industrial domains.
\item To assess the inherent strengths and weaknesses of existing systems, pinpointing deficiencies within the current landscape of DLTs.
\end{itemize}

At each layer within the DCEA framework, DLTs encompass diverse configurations of DCEA properties detailed in Table \ref{tab:my-table4}. Through the examination of these property combinations across the four layers, distinct classes of DLTs can be characterized. For example, distinctions at the data layer arise between Directed Acyclic Graph (DAG) and chain based DLTs. Also, the consensus layer facilitates categorizations such as permissioned and permissionless DLTs, contingent on the identity model of the consensus mechanism. The execution layer enables the differentiation between Smart-contract and script based DLTs. Meanwhile, at the application layer, classifications emerge, including DApps-oriented DLTs and Cryptocurrency-oriented DLTs. This comprehensive analysis results in the classification of DLT systems into two major categories: blockchain and blockchain-like systems.

\begin{table*}[hbt!]
\label{tab:blockchain_like}
\caption{Configurations of both blockchain and blockchain-like systems in the DCEA framework}
\includegraphics[width= \textwidth,height= 7cm]{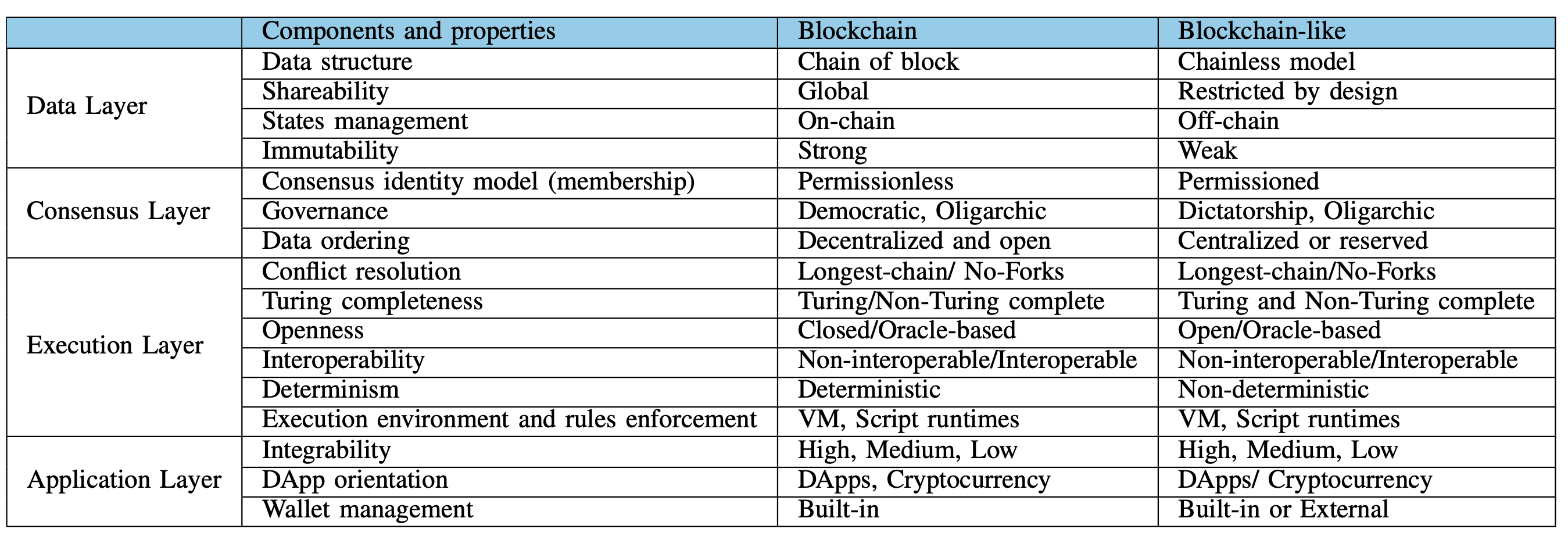}
\end{table*}

\subsection{Distinguishing Between Blockchain and Blockchain-Like Systems}

In scrutinizing DLT systems through the lens of the DCEA framework and discerning the disparities between the two overarching taxonomic categories, blockchain and blockchain-like, we identify shared commonalities alongside distinctive traits (refer to Table \ref{tab:blockchain_like}).
Taking a broader perspective, we assert that a system falls under the blockchain-like category, rather than being a blockchain, if it exhibits at least two of the following attributes, as depicted in Figure \ref{fig: triangle}. First, a noticeable lack of robust decentralization—several DLTs may opt to compromise decentralization for reasons such as enhanced scalability, streamlined governance, or suitability for deployment in contexts where full decentralization is unnecessary. Second, a blockchain-like system permits data tampering and provides weak immutability for either states or transactions. Third, its data structure does not rely on the chaining of blocks of transactions to store data.


\textbf{Caption:} Table \ref{tab:blockchain_like} summarizes distinctive settings within each category, aiding the differentiation between blockchain and blockchain-like DLTs. Despite not being entirely separate, these categories often exhibit significant overlap. It is possible to find a system with blockchain-like features, encompassing all properties except one or two. Moreover, our distinction is not contingent on operational settings—whether public or private, permissioned or permissionless—as the demarcation between blockchain and blockchain-like is not bound by these factors. For instance, a project initially designed for public and permissionless settings (e.g., Ethereum) can seamlessly transition to private and permissioned settings, and vice versa. In Table \ref{tab:my-table9}, we present a classification of various DLT technologies as either blockchain-like or blockchain, based on our established criteria.

\begin{figure*}[ht]
\centering
\includegraphics[width= \textwidth]{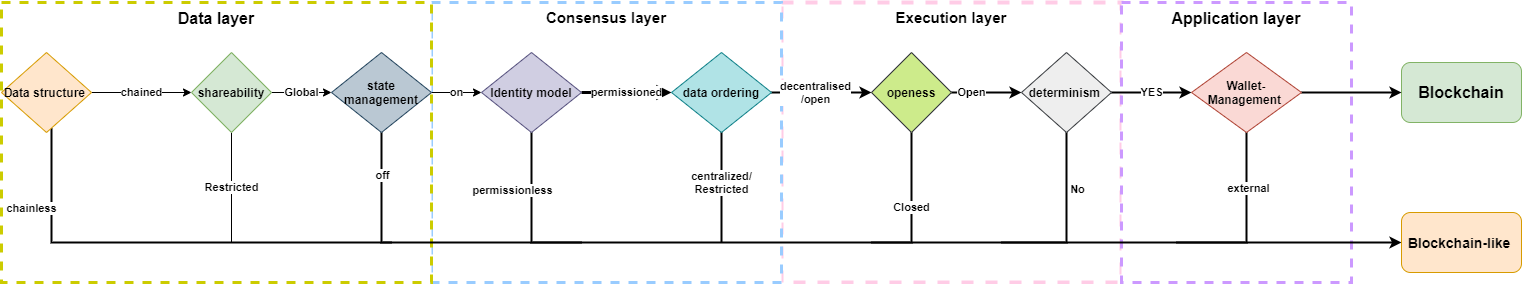}
\caption{A graphical dichotomous key for distinguishing between blockchain and blockchain-like systems}
\label{fig:taxo}
\end{figure*}

In Figure \ref{fig:taxo}, we present a graphical dichotomous key aimed at facilitating the differentiation between the two categories of DLTs.
In Figure \ref{fig:taxo}, we present a graphical dichotomous key to facilitate the differentiation between the two categories of DLTs.

The methodology adopted first deconstructs any DLT (blockchain or blockchain-like system) project into four layers that encompass different components. Second, it evaluates the different components to define the design choices adopted at the four levels. Third, based on the distinctive settings defined for the different components at each layer, as presented in Table IV, we classify a system as either a blockchain or a blockchain-like system. For instance, Bitcoin is classified as a Blockchain since it exhibits all the settings (Table IV) used to define a system as such. Hyperledger Fabric, on the other hand, is classified as a blockchain-like system since it displays some distinctive settings adopted for defining a system as such. Specifically, at the data layer, Hyperledger Fabric does not adopt a chainless model, and it does not provide strong immutability. Moreover, at the consensus layer, it adopts a centralized governance with a leader-based consensus mechanism.

Following the comprehensive analysis of DLTs across the four primary layers in sections \ref{sect 4}, \ref{sect 5}, \ref{sect 6}, and \ref{sect 8}, we present and discuss the proposed taxonomy by highlighting distinguishing properties. Additionally, we offer an in-depth exploration of the current state of the art at the data, consensus, execution, and application layers, respectively.

\vspace{5pt}

\section{Data layer} \label{sect 4}
In this section, we delineate the fundamental components and their respective characteristics that form the data layer, as outlined in Table \ref{tab:my-table4}. Additionally, we provide an overview of the current state of the art regarding various data structures employed by prominent DLT solutions. The section concludes with an examination of key challenges and trade-offs inherent in the design of the data layer.

\subsection{Components and properties}	
The ledger in DLTs essentially functions as a distributed "source of truth", with data is replicated and synchronized across multiple nodes. The macroscopic organization of data structures differs among DLT technologies, primarily falling into two main models: the linear chain of blocks and chain-less models.
\subsubsection{Data-Structure}

\begin{figure}[t]
\hspace*{-0.3cm}
\includegraphics[width=9cm, height=5cm]{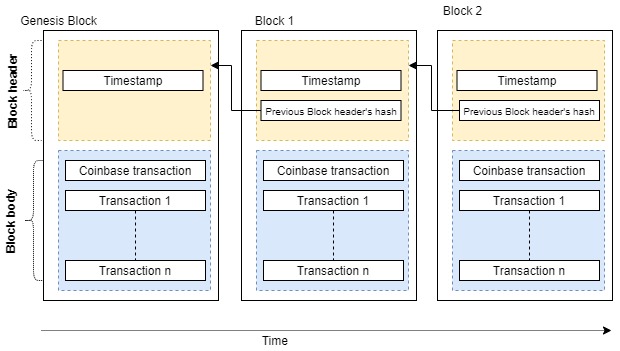}
\caption{Basic chain of blocks data structure}\label{fig:blockchain2}
\centering
\end{figure}

\paragraph\textbf{Linear Chain Model}\\
\textbf{Chain of Blocks}
In the chain of blocks model, transactions are bundled into blocks that are chained together in chronological order, creating an unalterable history of data. Each block is cryptographically linked to its predecessors, ensuring the integrity of the entire chain and preventing unauthorized modifications. A block comprises a header and a transaction record, as illustrated in Figure \ref{fig:blockchain2}.The header's inclusion of a timestamp and a cryptographic reference to the previous block empowers all network participants to independently verify the block's authenticity and its rightful place in the chain. This decentralized validation process eliminates the need for a central authority, fostering trust and transparency. At the transaction level, data is conveyed and stored in the ledger, with transactions representing the fundamental data type. Transactions are organized and hashed into a Merkle tree at the block level, and the hash root is embedded in the corresponding block's header. This architecture guarantees a cryptographically sealed and tamper-resistant data repository, providing an immutable record of transactions. The cryptographic links between blocks, enforced by the hash pointers makes it computationally infeasible to alter any part of the data without affecting the entire chain, thus safeguarding against unauthorized modifications. This robust design contributes to the trustworthiness and reliability of the blockchain, fostering transparency and accountability in data management.

\textbf{Skipchain} Skipchain derives their data structure inspiration from skip lists \cite{Pugh1990}. Skip lists are characterized by multiple linked lists arranged in layers with varying skip distances. This architectural influence enhances the efficiency of blockchain operations in a Skipchain by selectively engaging a subset of nodes in block validation. This hierarchical structure improves efficiency and scalability in consensus processes, streamlining the validation of new blocks. In addition to its unique consensus mechanism, a Skipchain employs an organized data management approach. The skip list-inspired structure allows for expedited searches and verification, enhancing the overall performance of the blockchain. This adaptation reduces the computational burden on individual nodes, contributing to faster transactions and improved scalability while maintaining the integrity and security of the distributed ledger.
\paragraph{Chainless model} 
Some DLTs break away from the conventional chain of blocks, opting instead for non-linear data structures like DAGs to achieve performance gains.

\textbf{Direct Acyclic Graph} 
The DAG is essentially a graph without cycles, progressing in one direction. It serves as a meticulously woven network of interconnected transactions, where each transaction (or block of transactions) functions as an individual node or vertex in the graph. This structure maintains a chronological growth pattern, resembling the organic expansion of branches from a continually developing tree rooted in the initial node.
The distinctive characteristic of the DAG lies in its cyclical-free and unidirectional growth, allowing for automatic transaction confirmation based on preceding transactions within the network. DAGs are further categorized into two types based on the nature of their nodes:
\begin{itemize}
    \item \textbf{Transaction-based DAGs:} Individual transactions are depicted as nodes within the DAG. Transactions are linked through directed edges, establishing a clear order of execution.
    \item \textbf{Block-based DAGs:} Nodes signify blocks containing multiple transactions, resembling traditional blockchains. Transactions are aggregated into blocks before integration into the DAG.
\end{itemize}

\textbf{Decentralized database model} 
decentralized databases distribute the responsibility of data maintenance across multiple nodes, allowing various parties, even those with mutual distrust, to collaboratively contribute to and synchronize shared records. This model leverages novel technologies to enable a distributed, secure, and transparent framework for managing data. Data is typically stored in a distributed fashion across the network of nodes, employing cryptographic techniques to ensure integrity and security.  
\paragraph {Hybrid data model} 
An Hybrid Data Model, integrates both blockchain and block-less models to govern transactions and states within the network. This hybridization is implemented to harness the strengths of each model, facilitating enhanced scalability and swift transaction validation. In this framework, states are typically stored in external dedicated key-value databases, while the blocks exclusively contain transactions impacting the ledger's states. The use of key-value databases simplifies direct access to the updated value of a state, eliminating the need to traverse transaction trees for calculation. This duality of blockchain and key-value databases represents a pragmatic approach, combining the security of the blockchain with the efficiency of direct state access, thereby optimizing the overall performance of the distributed ledger system.

\textbf{Distributed Hash Table}
A Distributed Hash Table (DHT) is a decentralized, distributed infrastructure that provides a scalable and fault-tolerant method for organizing and retrieving key-value pairs in a network. In a DHT, the keys and associated values are distributed across multiple nodes, allowing for efficient decentralized storage and retrieval of data. Each node in the network is responsible for a specific range of keys, and the DHT algorithms ensure that nodes can efficiently locate the node responsible for a given key. DHTs are commonly used in peer-to-peer (P2P) networks, distributed file systems, and other decentralized applications to enable efficient and resilient data storage and retrieval across a network of interconnected nodes.

\subsubsection{State Management}

DLTs offer various approaches to managing data within their systems. While they all share a distributed ledger for recording transactions, they diverge in how they store the actual data itself. Some DLTs, like traditional blockchains, keep data directly on the shared ledger (\textit{on-chain/on-ledger}), while others opt for off-chain storage in separate databases (\textit{off-chain/off-ledger}).

When examining the management of general states, such as user balances, in existing DLTs, three prevalent models come to the forefront:

\begin{enumerate}
  \item  UTXO Model:  In the UTXO (Unspent Transaction Output) model, transactions are constructed as a web of interconnected links, where each new transaction (UTXO) (\textit{new UTXO}) explicitly references the previous transactions that it draws upon as inputs (\textit{inputs}). This structure creates a traceable path of ownership and spending history and ensures that each UTXO can only be spent once, preventing fraudulent activities.

  \item Account Model: The account model resembles a traditional bank ledger, maintaining a central record of individual account balances. Transactions update these balances directly, offering a familiar and straightforward approach for managing user-specific data.

  \item World-State Model: Represents a design where the current state of the ledger is stored and managed as a single, comprehensive snapshot. This model separates the current state from the transaction log, allowing for efficient querying and retrieval of the most recent data.
\end{enumerate}
A complete review of the UTXO model can be found in \cite{Delgado-Segura2019}.

\vtop{}
\subsubsection{Data Shareability in DLT Networks}

Within a DLT network, the exchange of transactions among all participating nodes is central to achieving consensus. However, the concept of data shareability varies across different DLT systems, often influenced by privacy considerations. Two predominant visions of data shareability emerge within this landscape.

 Global Shareability:
In certain DLT systems, a philosophy of complete shareability is embraced. This approach, which we refer to as \textit{global shareability}, advocates for an unrestricted exchange of all data among all nodes in the network. In this model, transparency and accessibility are paramount, with every participant having access to a comprehensive and unfiltered view of the shared data. This promotes a decentralized environment where information is universally accessible, fostering a high degree of transparency and collaboration.

 Restricted Shareability:
Contrastingly, some DLT systems adopt a vision of restricted shareability. In this model, the perimeter of data shareability is delineated, encompassing specific nodes while excluding others. This restricted approach is motivated by privacy concerns, allowing for a more selective dissemination of information. Nodes within this restricted shareability model have limited access to certain data, ensuring a controlled flow of information based on predefined criteria. While this approach may limit universal transparency, it provides a mechanism to tailor data access to specific needs, enhancing privacy and security.

\subsubsection {Data immutability / Atomicity}
There is a common belief is that records in DLT (especially a blockchain) are inherently immutable and impervious to alteration. However, this assumption requires clarification, as the degree of immutability varies across different DLT systems based on their design. Consequently, in some DLTs nodes may retain inconsistent states or witness the rollback of transactions and states after confirmation. For data immutability, we differentiate between:
\begin{itemize}
    \item Strong Immutability: Strong immutability is the property of a system where data, once created and defined, is permanently unalterable. No operation, actor, or event can modify the data after its initial creation. This ensures absolute, tamper-proof preservation of historical records and guarantees data integrity through all time.
    \item Weak Immutability: Weak immutability allows data to be updated under defined and controlled conditions. Modifications follow specific rules and governance mechanisms, ensuring transparency and accountability in the change process. This balance preserves historical records while enabling necessary adaptations.
\end{itemize}
It's noteworthy that in certain systems characterized by strong immutability, updates to their states can occur without compromising this immutability. This is accomplished through the use of tree structures, allowing for the persistent storage of old and new values associated with a specific entry (e.g, Ethereum state structure \ref{fig: eth}).
\subsection{State-of-the-Art in Data Layer}
This subsection aims to provide a comprehensive overview of DLT projects that align with the diverse data structures delineated in our framework. Additionally, it entails an evaluation of the properties associated with these projects.

\subsubsection{Data-structure}

1.1) Chained DLTs

Many DLTs adopt the foundational linear data chain structure initially introduced by Bitcoin. Within this broad category, data is stored in inter-linked blocks; however, various projects within this domain introduce distinct inner block structures to tailor the architecture to their specific needs and objectives.

\paragraph{Bitcoin} In the Bitcoin ecosystem and its derivatives, transactions are assembled within the block's body and are subsequently linked in a Merkle tree. The Merkle root is a unique cryptographic hash created by combining the hashes of all the data blocks in a Merkle tree. Beyond the Merkle root, the block header encapsulates the hash of the previous block. Beyond the Merkle root, the block header encapsulates additional critical information, including the hash of the previous block. This hash of the previous block is crucial for maintaining the chronological order and integrity of the entire blockchain. It creates a cryptographic link between the current block and the one that preceded it, establishing a continuous and tamper-resistant chain of blocks. This interconnection of blocks through their headers ensures the immutability and trustworthiness of the Bitcoin blockchain.
Notably, Bitcoin relies on the UTXO model's distributed tracking of spendable value within the blockchain, enabling the inference of wallet balances through UTXO analysis. Nodes often cache frequently accessed UTXOs in memory for faster retrieval, improving transaction processing efficiency.. This approach contributes to the efficiency and scalability of the Bitcoin blockchain.

\begin{figure*}[ht]
\hspace*{0.09cm}
\includegraphics[width=\textwidth,height=8cm]{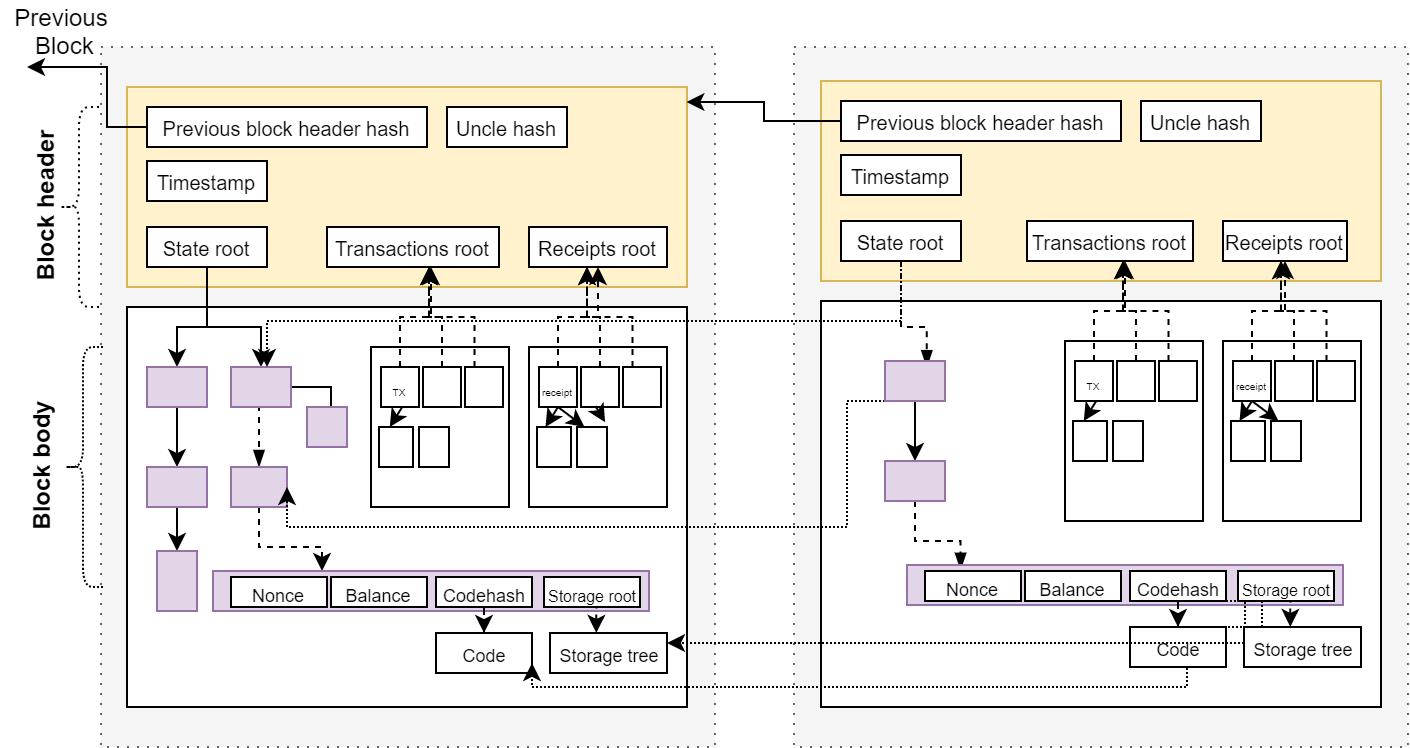}
\caption{ Structure of Ethereum's chain of blocks} \label{fig: eth}
\centering
\end{figure*}

\paragraph{Ethereum} 
Ethereum employs a state tree as a meticulously organized database, mirroring the account model to oversee account balances, smart contract data, and other critical network states. This tree-like structure relies on a specialized data structure known as a Merkle Patricia Tree (MPT), facilitating efficient retrieval and verification of information to maintain synchronization across all nodes. This highly efficient, self-balancing tree structure combines the strengths of Merkle trees and Patricia tree data structures. Each node in the MPT represents a key-value pair, storing the name (key) of a state variable and its corresponding value. Nodes have 16-byte keys and contain either a leaf value (for end nodes) or 16-byte hashes of child nodes (for intermediate nodes). As a state variable changes, a new node is added to the tree to reflect the updated value \ref{fig: eth}. Recalculation of the root hash captures the change, ensuring secure and tamper-proof state tracking.

\begin{figure}[b]
\centering
\includegraphics[width=8cm, height=2.5cm]{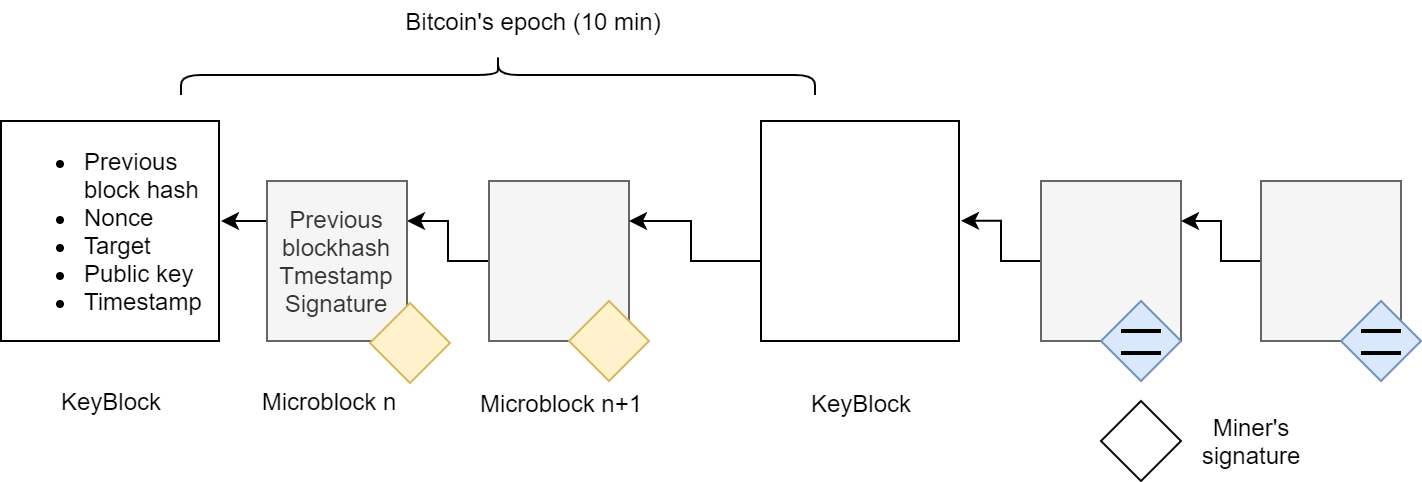}
\caption{Bitcoin-NG Blockchain Data Structure. Keyblocks and Microblocks contain the public key and the signature belonging to the miner.}\label{fig: btng}
\end{figure}

\paragraph{Bitcoin-NG} proposed by Eyal et al. in \cite{Eyal2016}, Bitcoin-NG uses a special data structure inspired from bitcoin's chain of blocks. This structure is consisting of two kind of blocks  key-blocks and Microblocks. The Keyblocks are produced using proof-of-work and share the same structure as Bitcoin's blocks. Their role is to determine the block miner. Between two key-blocks, the selected block miner creates and signs multiple Microblocks, each contains a collection of signed transactions. The miner responsible for creating Microblocks is chosen based on a predefined algorithm, such as a rotating leader schedule or Proof of Stake. As in Bitcoin, Bitcoin-NG blocks form a chain where Microblocks reference previous Microblocks and Keyblocks, as illustrated in Figure \ref{fig: btng}. Notably, Microblocks significantly reduce the overall size of data stored on the blockchain, improving storage efficiency and network bandwidth usage.

\paragraph{ByzCoin} Taking inspiration from Bitcoin-NG, ByzCoin \cite{Jovanovic2016} embraces the concept of decoupling in two blocks. However, rather than adhering to a single chain structure, ByzCoin introduces a novel approach by forming two distinct and parallel chains: Keyblocks and Microblocks. The Keyblocks serve as anchors or reference points, securing the overall structure, while the Microblocks contain a more granular level of transactional data. 
\smallskip

1.2) Skipchain 

\begin{figure}[ht]
\includegraphics[width=8cm, height=3cm]{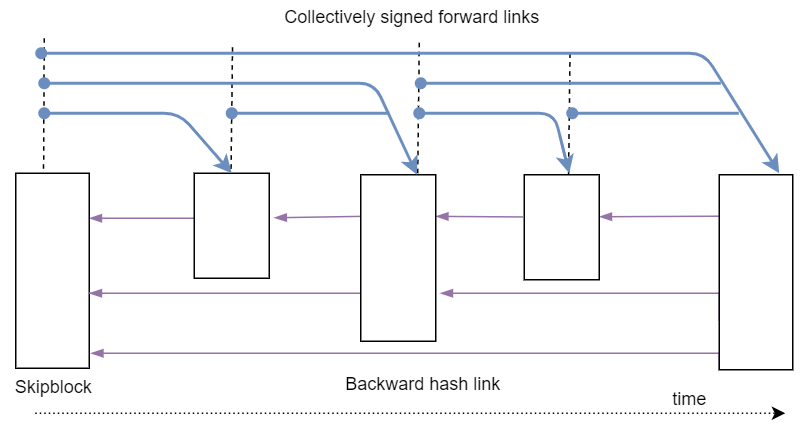}
\caption{Backward and forward links in skipchain} \label{ fig: skipchain}
\centering
\end{figure}

\paragraph{Chainiac} Introduced by Nikitin et al. \cite{Nikitin2017}, addresses challenges related to offline transaction verification. This involves enabling nodes to ascertain whether a transaction has been committed without requiring a complete copy of the ledger. The innovative solution proposed by Chainiac involves introducing traversability forward in time through the use of a Skipchain, as depicted in Fig. \ref{ fig: skipchain}. Back-pointers are represented by cryptographic hashes, while forward-pointers are realized through collective signatures. By incorporating long-distance forward links and employing collective signatures, Chainiac facilitates efficient transaction verification for a client or node at any point in time. 

1-3) Chainless DLT
\paragraph{DAG based chains}
The concept of utilizing Directed Acyclic Graphs (DAGs) as the underlying data structure has garnered significant attention from DLT designers across various projects. Notable examples include Byteball \cite{Churyumov2016}, DagCoin \cite{Paper}, IOTA \cite{Popov2017}, Nano \cite{Churyumov2016}, Phantom \cite{Dwivedi2018}, and Hedera \cite{Hedra}. 
Some studies have explored the integration of DAGs instead chain of blocks. For example, the GHOST protocol \cite{Kiayias2019}, A DAG-Based Modification for Bitcoin, aims to reduce confirmation time and to secure the network \cite{Sompolinsky2015}.

\begin{figure}[b]
\includegraphics[width=9cm, height=4cm]{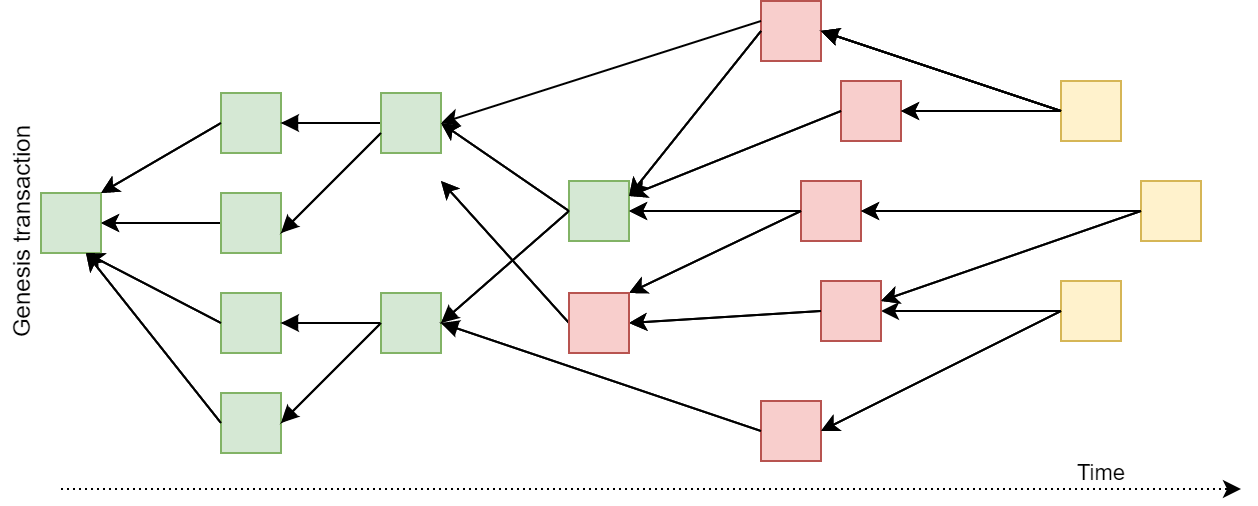}
\caption{  IOTA Tangle’s DAG structure.  Green blocks represent final validated transactions; Red blocks represent uncertain validated transactions; Yellow blocks represent transactions awaiting validation.}\label{fig: iotaa}
\centering
\end{figure}

\paragraph{IOTA}\ IOTAcite{Popov2017} is a DLT designed to provide a scalable and fee-less environment for machine-to-machine transactions. IOTA is among the early adopters of a DAG data structure known as Tangle. The tangle is a block-less DAG where transactions are directly interlinked without the need for traditional blocks. Unlike traditional blockchain systems, IOTA's Tangle eliminates the need for miners and introduces a decentralized validation model, where each participant contributes to the network's security by validating two previous transactions for every transaction they make. Transactions in the Tangle are stored as DAG edges, each referencing two immediate predecessors and forming a directed acyclic network with confirmed, unconfirmed, and potentially conflicting subsets (Fig. \ref{fig: iotaa}).

IOTA’s graph begins with a root transaction known as the Genesis Transaction. This transaction serves as the starting point, initiating the creation of the Tangle. In the creation of the Tangle, by design, a single address holds all the created tokens (IOTA tokens); The Genesis Transaction then orchestrates the distribution of tokens to various accounts within the network. This distribution sets the stage for a decentralized ownership structure as tokens move from the centralized address to numerous accounts.

\paragraph{Hashgraph} \cite{Baird2016} deviates from traditional blockchain technology by employing a unique, chainless architecture to achieve significantly higher throughput. Instead of relying on linear chains of blocks, each node within the Hashgraph network maintains its own DAG. The vertices of this DAG are referred to as "events", which are analogous to blocks in traditional blockchains. Similar to blocks, events contain a single or several transactions alongside other critical data. This data includes:
\begin{itemize}
    \item Event parent hashes: These are the hashes of two previous events – one created by the gossip receiver and the other by the sender.
    \item A timestamp.
    \item A signature from the node that created the event.
\end{itemize}

\begin{figure}[ht]
\centering
\includegraphics[width=5cm, height=5.5cm]{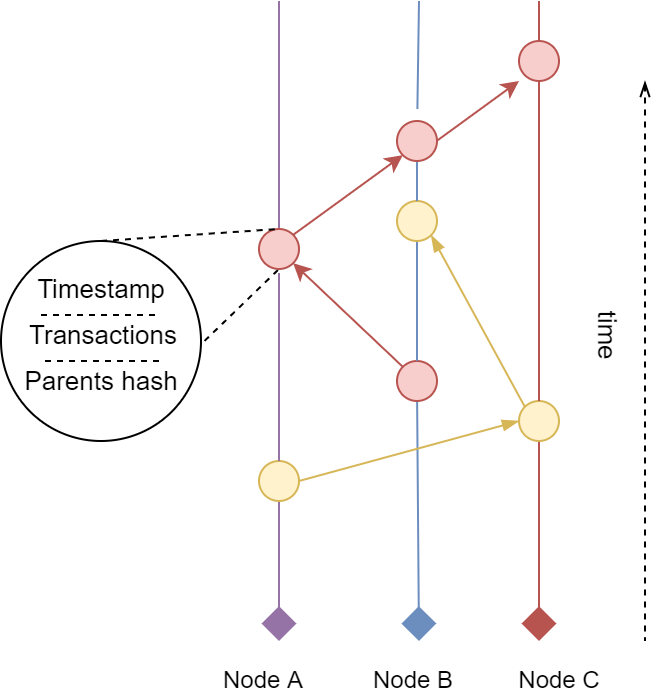}
\caption{ Evolution of the Hashgraph} \label{fig: hashgraph}
\end{figure}

Figure \ref{fig: hashgraph} depicts the structure of an event within the Hashgraph network.

\paragraph{Byteball} Described by Churyumov in \cite{Churyumov2016}, Byteball is a transaction-based Directed Acyclic Graph (DAG) system that differs from Bitcoin-NG. Instead of utilizing a hybrid structure of key-blocks and microblocks, Byteball relies solely on a DAG where each transaction directly references one or multiple previous transactions. Transactions establish a clear lineage through direct references to their parent transactions, ensuring data integrity. The DAG experiences dynamic growth as new transactions can attach to any existing transaction, enabling parallel processing and efficient network expansion. Initially employing proof-of-work for consensus, Byteball utilizes miners who compete to solve cryptographic puzzles, similar to Bitcoin. Additionally, Byteball employs a weighting system, where transactions are assigned weights based on age and lineage depth within the DAG, prioritizing older and more deeply connected transactions to promote network stability.

\begin{figure}[b]
\centering
\includegraphics[width=5cm, height=5cm]{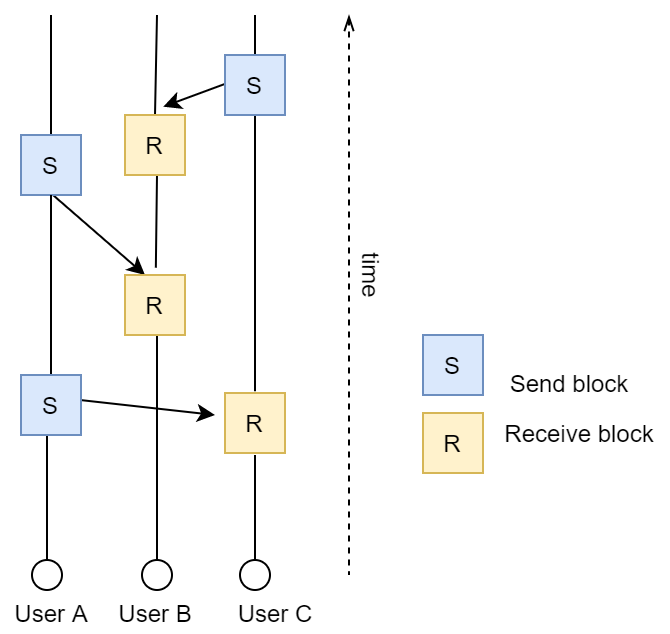}
\caption{ Structure of DAG in Blocklattice}\label{fig:latice}
\end{figure}

\paragraph{Nano} formerly known as RaiBlocks \cite{LeMahieu2017}, employs a unique DAG data structure called Blocklattice. In the Nano network, each account possesses its dedicated chain of blocks, as illustrated in Figure \ref{fig:latice}. These individual chains are replicated across all network peers, enabling the simultaneous growth of multiple single chains. Significantly, only the wallet owner has the authority to make changes to their specific chain, allowing for asynchronous updates to each wallet. In contrast, Dexon \cite{Chen2018} integrates a Blocklattice single chain into a globally-ordered chain without requiring additional communication. In the work by Sompolinsky et al. \cite{Sompolinsky2016,Sompolinsky2020}, a DAG-based DLT is considered, where the nodes of the DAG represent blocks rather than individual transactions. Referred to as BlockDAG, this structure involves blocks referencing multiple other blocks, with newly added blocks pointing to recent blocks that are not yet referenced. This innovative paradigm serves as the foundation for emerging consensus protocols such as Inclusive \cite{Lewenberg2015}, SPECTRE \cite{Sompolinsky2016}, and PHANTOM \cite{Sompolinsky2020}.

\subsubsection{Decentralized Databases}
\paragraph{Corda R3 \cite{Hearn2016}} 
In Corda, each node has its own private, secure RDBMS (Relational Database Management System) called a "vault". The vault efficiently stores time-stamped records of transactions, contracts, and other relevant data for that node. Corda's design differs from traditional blockchains where all nodes share a single, replicated ledger. The vault includes tables dedicated to various services, such as ledger management, transactions (NODE\textunderscore TRANSACTIONS), states (NODE\textunderscore STATES), participants (NODE\textunderscore IDENTITY), and additional metadata.

While the metadata associated with a state is stored in the vault tables, the actual state data itself is stored in a separate binary format \cite{R3a}. Attachments, which are files linked to states, are stored in a distinct location on the node's file system. This separation enhances scalability and performance, particularly for managing large attachments.

In tandem with its storage approach, Corda adopts a UTXO model for handling state data. This model dictates that a transaction can consume existing states and may or may not produce new states, providing a clear representation of the state changes within the distributed ledger.

\subsubsection{Hybrid DLTs}
\paragraph{Hathor} is a DLT that combines DAG and traditional blockchain architectures. Unlike traditional blockchains, Hathor utilizes a DAG structure for transaction processing, allowing for high scalability and throughput. The chain of blocks structure ensures security when the number of transactions per second is small, whereas the DAG prevails when the number increases significantly.

\paragraph{Hyperledger Fabric \cite{Androulaki2018} }

\begin{figure}[b]
\centering
\includegraphics[width=8cm, height=4cm]{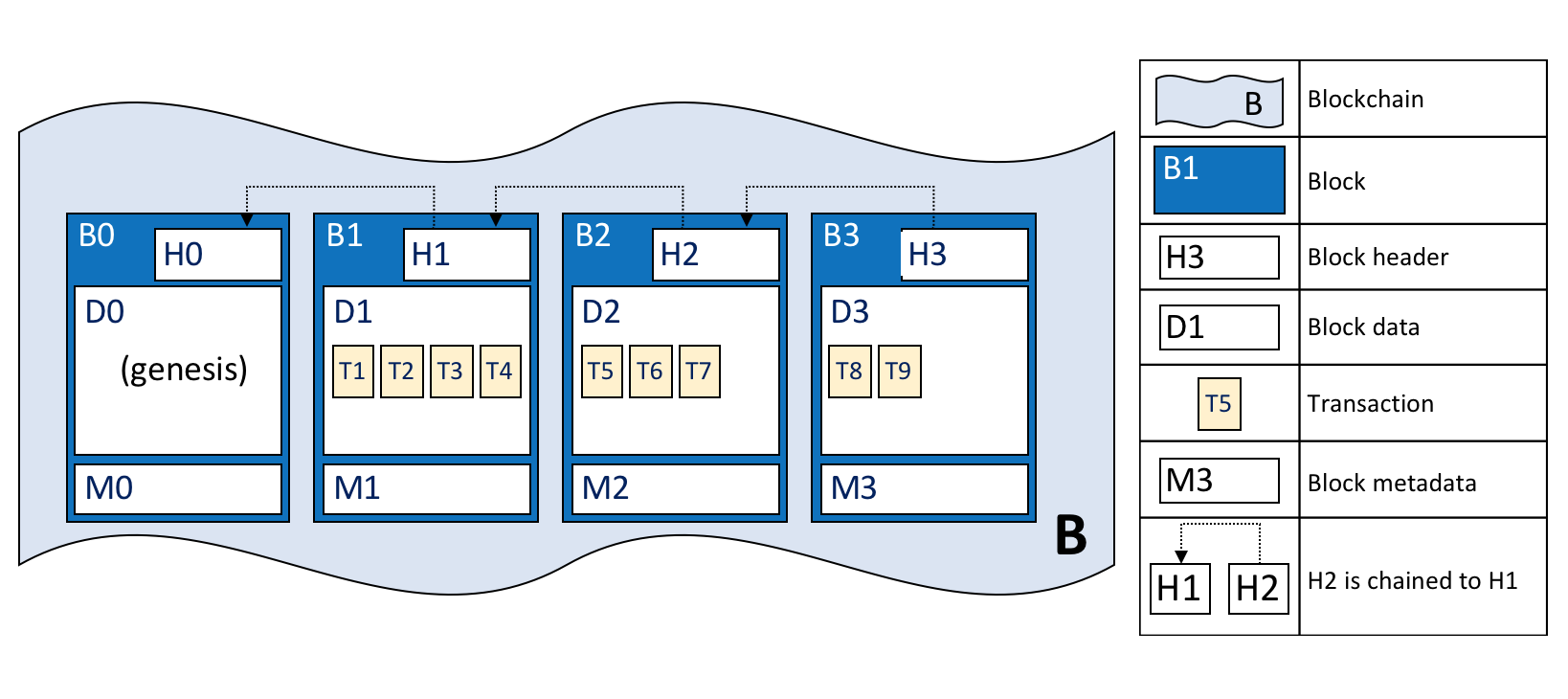}
\caption{Structure of Hyperledger fabric’s chain of blocks (Source: Hyperledger official documentation) }\label{fig: hyperledger}

\end{figure}

Hyperledger Fabric incorporates both a chain of blocks for storing validated transactions and a classical key-value database for managing the system's states, as depicted in Figure \ref{fig: hyperledger}. While the ledger captures transaction metadata and state hashes, the actual state data—encompassing details like asset ownership and financial information—is deliberately stored off-chain. This strategic off-chain storage is designed to optimize efficiency and uphold privacy considerations. Hyperledger Fabric employs three distinct stores:
\begin{itemize}
    \item Ledger: Stores transaction metadata and state hashes.
    \item Peer State Database: Manages state data associated with transactions validated by the peer.
    \item Private Data Storage: Houses private state data accessible only to authorized nodes.
\end{itemize}

Within the Fabric chain, the block structure resembles that of a traditional blockchain, yet it incorporates an additional segment: block metadata. This supplementary section includes a the certificate, public key, timestamp and signature of the block writer. In contrast to certain other blockchain structures, the Fabric block header is straightforward, and transactions within the block body are ordered without Merklization.

\begin{figure}[t]
\centering
\includegraphics[width=8cm, height=4cm]{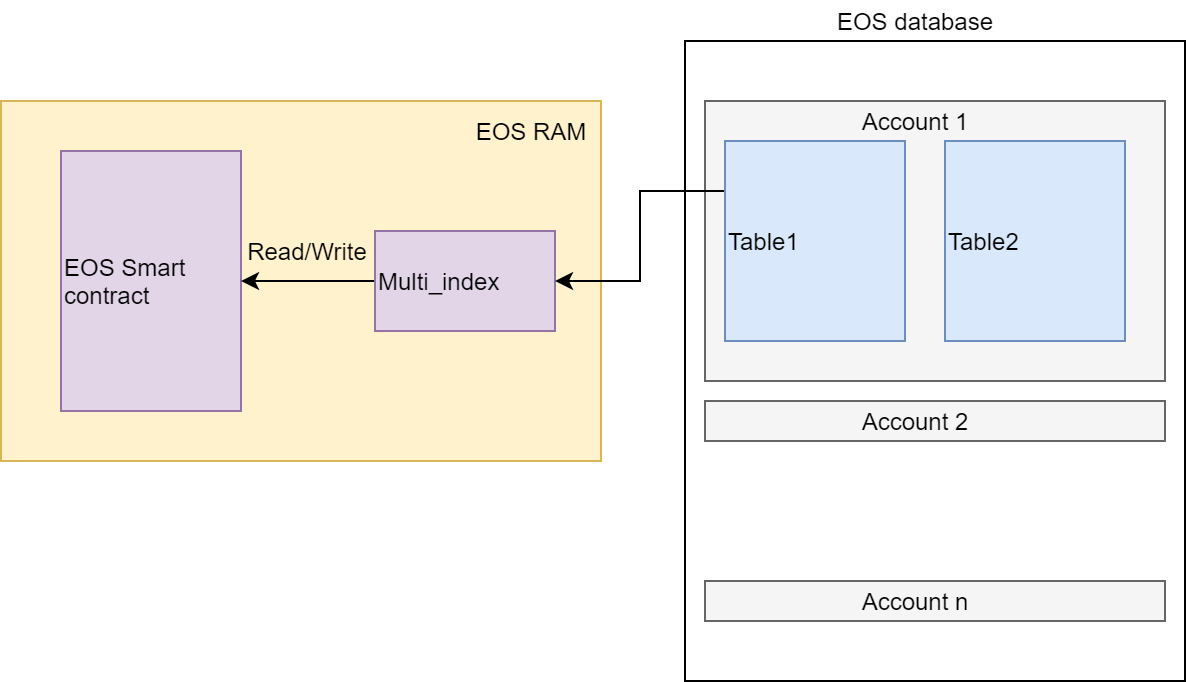}
\caption{EOS database structure and its interactions with smart contracts}\label{fig:eos}
\end{figure}

\paragraph{EOS \cite{Grigg2017}} 
EOSIO employs a multi-layered approach to data storage, seamlessly integrating the strengths of on-chain, off-chain, and RAM storage (Fig. \ref{fig:eos}). This layered architecture ensures efficient data management, catering to diverse needs while maintaining security and performance.
On-chain storage forms the bedrock of the system, acting as an immutable ledger for transactions and key state data. Replicated across the network, this data remains permanently accessible and tamper-proof. Within smart contracts, multi-index tables provide structured storage, allowing developers to organize and efficiently retrieve specific data.
For data that requires scalability and privacy, off-chain storage comes into play. EOSIO offers various options to meet these needs, including the specialized Chainbase database for fast access and management of crucial data like user accounts, token balances, and smart contract states. Other options like MongoDB, SQLite, and RocksDB cater to specific data types and access requirements.
To further optimize performance, RAM storage serves as a temporary caching layer. Frequently accessed data is stored in the memory of nodes, significantly reducing retrieval times and enhancing user experience. However, RAM resources are limited and require purchase or lease, encouraging efficient data management.
 
\paragraph{BigchainDB} BigchainDB's architectur \cite{McConaghy2016} revolves around two key databases: the "Backlog," a staging area for pending transactions, and the "Chain of Blocks," a repository for validated transactions, ensuring efficient processing and secure record-keeping. It bridges the gap between NoSQL databases and blockchains, offering a hybrid approach that blends the flexibility of MongoDB with the immutability and security of blockchain technology. BigchainDB stores all its data in Json documents. 
These documents are then organized into three main categories:
\begin{itemize}
    \item Transactions: Represent actions like creating assets, transferring ownership, or voting. Each transaction includes details like sender, receiver, asset information, and timestamp.
    \item Blocks: Combine multiple transactions and additional metadata to form a chain of data. This chain ensures data integrity and immutability.
    \item Votes: Nodes use votes to confirm the validity of transactions and blocks, maintaining consensus within the network.
\end{itemize}

\paragraph{Ripple}: \cite{Schwartz2014}
 utilizes the XRP Ledger (XRPL), a hybrid of a database and a chain of blocks, offering unique advantages for its global payment network. The XRPL employs a multi-layered data storage architecture, combining centralized and decentralized databases for optimal efficiency. Primary data, decentralized and managed by NuDB or Cassandra, provides redundancy. Secondary data, using PostgreSQL, can be centralized or distributed. Chainbase, for user accounts and smart contracts, operates centrally on dedicated servers. Off-chain databases like MongoDB, SQLite, and RocksDB are flexible, optimizing for specific data types. The XRPL features a Ledger Object Model (LOM) for structured data storage, efficient data retrieval with unique identifiers (LODIDs), state pruning for storage optimization, and a chain of blocks ensuring an immutable record of transactions. Distributed consensus via the Ripple Consensus Protocol (RLCP) and robust security features, including cryptographic hashing and digital signatures, enhance the authenticity and integrity of XRPL data.

\subsubsection{Data shareability}

Many DLTs designed for global cryptocurrency platforms inherently embrace a model of global shareability for transactions. For instance, Bitcoin and Ethereum operate in a relay mode, where nodes propagate transactions throughout the entire network without restrictions. Hashgraph, another DLT, employs a different approach where senders deliver transactions to a select set of nodes responsible for incorporating them into their DAG and sharing them through Gossiping.
In addition to these, various other blockchain and decentralized ledger projects contribute to the landscape of data shareability. The Interledger Protocol (ILP)\cite{Thomas2015a} facilitates interoperability, enabling different ledgers to share data and value seamlessly between network peers. Polkadot's heterogeneous multi-chain framework\cite{Wood2016} allows connected blockchains to share data through its relay chain, promoting interoperability while controlling data access. Cosmos, aiming for an "Internet of Blockchains," enables different blockchains to selectively share data through the Cosmos Hub, emphasizing interoperability.

Furthermore, IOTA's Tangle structure fosters global shareability, with all network peers having access to efficiently interlinked transactions. Filecoin's decentralized storage network enables global data shareability, allowing peers to access shared data stored across the network. Nervos Network's two-layer architecture supports global shareability of state and assets, with all peers having access to the shared data on the Nervos network. Dfinity's decentralized cloud platform enables global data shareability by providing secure and scalable computing resources, accessible to network peers.

In contrast, DLTs tailored for business applications, such as Corda and Hyperledger Fabric, prioritize restricted shareability of transactions to address privacy concerns. Corda adopts a node-centric approach, where each node maintains a distinct database containing only relevant data, limiting the visibility of the ledger to individual peers. Similarly, Hyperledger Fabric introduces the concept of channels \cite{Brakeville2016} to restrict data shareability. Each transaction occurs on a private subnetwork (channel), accessible only to authenticated and authorized parties, resulting in a distinct ledger for each channel. Other projects like Quorum, built on Ethereum, ensure privacy by facilitating private transactions among network participants through constellations\cite{Jpmorganchase}  or Tessera \cite{Tessera}. Ripple, with its XRP Ledger (XRPL), employs a hybrid model that combines elements of both global and restricted data shareability. The XRP Ledger serves as a decentralized global payment network, where transactions are recorded in a chain of blocks. This chain ensures immutability and transparency of transactions, contributing to global shareability.
 
\subsubsection{Data immutability}
The pinnacle of strong immutability is exemplified in Bitcoin, where robust cryptographic hashing and chained blocks forge an unalterable record, rendering it ideal for applications requiring unwavering security and transparency. However, this rigidity may be less suitable for projects seeking adaptability and evolution. Other protocols, such as Ethereum or Avalanche, prioritize strong immutability, ensuring that once data is written to the ledger, it remains unalterable or erasable.

Business-oriented DLTs like Hyperledger Fabric illustrate the possibilities of weak immutability. By leveraging tree structures and state updates, they permit data modifications under specific conditions while preserving historical values, offering greater flexibility for projects to adapt to changing needs without compromising overall ledger integrity.

Projects like Multichain provide customizable data sharing with varying degrees of immutability, while Hyperledger Sawtooth enables the implementation of different consensus mechanisms and diverse levels of immutability.

Tezos introduces a unique feature known as the "rollback-enabled model," offering weak immutability. This model allows for the reversal of specific on-chain transactions in exceptional circumstances, such as critical bugs or vulnerabilities.

\subsection{Data Layer: Discussion}

The design choice of a data structure and its properties is complex, involving the assessment of numerous challenges and tradeoffs, as described below:

\subsubsection{\textbf{Tradeoff: Data Integrity (Block Size) versus Transaction Throughput (Performance)}}

The sequential nature of the chain of blocks ensures a high level of security and data integrity. However, it limits the number of transactions accepted by the network due to the block size (in megabytes or gas limit). This limitation penalizes system throughput. For instance, Bitcoin introduced a block size limit (1 MB) to prevent DOS attacks caused by huge blocks, but this decision hampers network performance (4 transactions per second). Increasing the block size cannot be an effective solution, as it risks breaking the network's decentralization, making full nodes more expensive to operate, increasing orphan block rates, and delaying propagation speed. To address this dilemma, Bitcoin adopted the Segregated Witness (SegWit) technique \cite{Segwit} \cite{Lombrozo2015}, which separates signature data from Bitcoin transactions and allows its pruning from the ledger, safely increasing the block size limit to 2-4 MB. Additionally, to enhance throughput without increasing orphan blocks, the Bitcoin network reduced block propagation time by incorporating rapid relay networks \cite{Network}-\cite{BASU}-\cite{Klarman}. In Ethereum, miners set the gas limit for each block. The gas limit is a parameter that defines the maximum amount of gas units that can be consumed in a block. Gas is the computational unit used to measure the computational work performed by the Ethereum network, and it is separate from the cryptocurrency ether. That limit helps prevent malicious users from creating transactions that require excessive computational resources, potentially disrupting the network. 
In contrast to the chain of blocks, the parallel nature of DAGs enables the processing of transactions in parallel, contributing to higher throughput. However, a DAG structure comes with its own challenges. A reduction in the volume of transactions may expose it to attacks (e.g., parasite attack \cite{Cullen2019}). Therefore, DAG-based DLTs often resort to centralized mechanisms (e.g., coordinators in IOTA) to mitigate this risk.

\subsubsection{\textbf{The Challenge of Fast-Growing Ledger Size}}

Addressing the issue of fast-growing ledger sizes is a significant concern across all DLTs. Various solutions have been devised to handle this challenge at the data level. For example, projects like Bitcoin employ straightforward methods like ledger pruning. In this approach, nodes download blocks, validate them, and then discard irrelevant data to conserve disk space. On the other hand, more complex solutions, such as sharding techniques, are utilized by projects like Zilliqa \cite{Team2017} and Ethereum 2.0. Sharding involves running multiple parallel subsets of nodes, known as shards, where each shard maintains its sub-ledger. This enables parallel transaction processing and distributes data storage among multiple nodes.

Additionally, there's an active development of off-chain processing techniques to alleviate the network's load, whether in terms of storage or computation. Examples include Bitcoin's Lightning Network \cite{Poon2016} and Raiden \cite{Est.}. In the realm of business-oriented implementations, cloud providers offer Blockchain as a Service (BAAS) solutions, such as IBM BAAS \cite{IBMa} and Azure BAAS \cite{Microsoft}. These solutions involve deploying a ready-to-use DLT solution on the cloud, providing the necessary storage space and network bandwidth. For more in-depth information on current BAAS solutions, readers are encouraged to refer to \cite{Chen2019}.

 \subsubsection{\textbf{The transparent aspect of the blockchain versus privacy of shared data}}
The transparency of the DLTs often raises concerns regarding data privacy, particularly in DLTs with global shareability. While most public blockchains employ pseudo-anonymization to introduce an initial layer of privacy, the persistent risk of deanonymization, as demonstrated by \cite{Koshy2014} \cite{Mastan2017}, remains. This risk is heightened when users neglect to employ distinct single-use addresses for each operation, a practice supported by hierarchical deterministic wallets \cite{Deterministicwallet}. To enhance anonymity and facilitate confidential transactions in the blockchain sphere, solutions such as Monero \cite{Noether2015} and Zcash \cite{Zcash} have been proposed. Zcash introduced zkSNARKs \cite{Gennaro2013}, a zero-knowledge proof system, while Monero utilizes ring signatures \cite{Noether2015} to conceal transaction senders and recipients. An additional strategy to bolster the anonymity of existing public blockchains (e.g., Bitcoin or Ethereum) involves mixing schemes like CoinJoin \cite{Barcelo2007}, Coinshuffle \cite{Ruffing2014}, MimbleWimble \cite{Poelstra2016}, and Grin \cite{Grin}. These solutions employ transaction shuffling mechanisms to safeguard anonymity. Additionally, initiatives like Zether \cite{Bunz2019}, WaterCarver \cite{Xin}, and Aztec \cite{Williamson2018} aim to leverage zkSNARKs, introducing an enclave of privacy to public Ethereum for enabling private transactions. It's worth mentionning that entities such as StarkNet, which currently lack native support for the Ethereum Virtual Machine (EVM), are actively investigating transpilers to overcome the programming language gap. The future is in favour of zkEVMs, as they compete for supremacy not only within the realm of zk-rollups but also against their optimistic counterparts.

\section{Consensus Layer} \label{sect 5}
DLTs have sparked a renewed interest in the development of innovative distributed consensus protocols. In fact, the literature has witnessed the proposal of numerous consensus algorithms tailored for DLTs, each presenting distinct properties and functionalities. This section introduces the properties and features incorporated into the DCEA framework, essential for the study and differentiation of these protocols. Additionally, the second subsection provides an overview of the current state of the art in this domain. Section \ref{sect 8} delves into the presentation and discussion of the results derived from a comparative analysis of the examined protocols.

\subsection{Components and Properties}

\subsubsection{Basic Properties}
The concepts of safety and liveness properties are fundamental to understanding and evaluating consensus algorithms in distributed systems. These properties were originally introduced by Leslie Lamport in 1977 in his seminal work "Proving the Correctness of Multiprocess Programs" \cite{Lamport1977}. Since then, they have become essential tools for analyzing and designing consensus protocols in various distributed computing scenarios. 

\paragraph{Safety} In the context of DLT networks, safety denotes the assurance that correct nodes will not simultaneously validate conflicting outputs or make conflicting decisions, such as chain forks. The safety property plays a crucial role in ensuring availability, ensuring that transactions submitted by honest users are efficiently integrated into the ledger \cite{Pass2017}. Additionally, safety encompasses preventing undesirable phenomena like reorganizations (reorgs), where previously confirmed transaction history undergoes revisions, often due to factors like network forks or changes in consensus rules.

\paragraph{Liveness} 
a consensus protocol ensures liveness in a DLT if it guarantees eventual progress, agreement even in the presence of faults or delays.

\paragraph{Finality} we define the finality as the affirmation and guarantee that once a transaction is recorded on the ledger, it cannot be altered or reversed, guaranteeing data integrity and providing a strong foundation for trust in the system. Finality can be divided into two categories:
\begin{itemize}
    \item Probabilistic finality, where the certainty of a transaction being irreversible increases with time after its inclusion in the ledger. This means that while the transaction may theoretically be reversed, the probability of such an event occurring diminishes significantly as more blocks are added to the chain
    \item Absolute finality guarantees that a transaction is irreversible once it has been validated by a majority of honest nodes in the network.
\end{itemize}
 
\subsubsection{Network Models} \label{subsect network model}
In the literature of distributed systems and DLT consensus protocols, we adhere to the message passing model where nodes communicate by exchanging messages over the network, operating with different assumptions of network synchrony. In this survey, we adopt the taxonomy defined by \cite{Dwork1988}, which categorizes network synchrony assumptions.

\begin{itemize}
    \item \textbf{Synchronous:} Assumes a known upper bound on message delay, ensuring messages are consistently delivered within a specified time after being sent.
    \item \textbf{Partially-synchronous:} This assumption is based on a Global Stabilization Time (GST), indicating that messages sent will be received by their recipients within a predetermined time frame. Before reaching the GST, messages may encounter unpredictable delays.
    \item \textbf{Asynchronous:} Messages sent by parties are eventually delivered, with arbitrary delays and no assumed bound on the delivery time.
\end{itemize}

\subsubsection{Failure Models}
Different failure models have been considered in the literature; we list here two major types.

\begin{itemize}
    \item \textbf{Fail-stop failure (Also known as benign or crash faults):} Nodes go offline because of a hardware or software crash. Fail-stop failures can be detected relatively easily, as the node simply becomes unreachable.

    \item \textbf{Byzantine faults:} This category of faults was introduced and characterized by Leslie Lamport in the Byzantine Generals Problem \cite{Lamport1982} to represent nodes behaving arbitrarily due to software bugs or a malicious compromise. In a Byzantine fault, a node behaves arbitrarily, potentially due to software bugs, security breaches, or malicious intent, allowing the node to send conflicting messages to different nodes, forge false information, and collude with other Byzantine nodes to mislead the system.
\end{itemize}
Therefore, we consider a protocol as fault-tolerant if it can gracefully continue operating without interruption in the presence of failing nodes.

\subsubsection{Adversary Models}
Distributed systems, including blockchain networks, operate in environments where malicious actors may attempt to disrupt their operation or exploit vulnerabilities for their own benefit. To design secure and resilient systems, it is crucial to understand different adversary models and how they can impact the network.

\begin{itemize}
    \item Threshold Adversary Model (Hirt and Maurer) \cite{Hirt2000}: This model, commonly used in traditional distributed computing literature, assumes that the Byzantine adversary can corrupt up to any \(f\) nodes among a fixed set of \(n\) nodes. This model typically assumes a closed membership where permission is required to join the network. The consensus protocol should ensure consensus in the presence of Byzantine nodes as long as their numbers is below a specific threshold.
    
    \item Computational Threshold Adversary: A model introduced by Bitcoin (and applicable to other decentralized systems), where the adversary's control over the network is bounded by computational power not by the number of nodes they can control. This model typically assumes, the membership is open to multiple parties, and the bounding computation involves a brute force calculation.
    
    \item Stake Threshold Adversary \cite{Abraham2017}: In this model, the adversary's control is bound by the control over a specific resource: the system's native token or currency. The Stake Threshold Adversary is the one who hoards enough of these shares (tokens) to gain a significant say in the system's operation.
\end{itemize}

\subsubsection{Adversary Modes}
Consensus protocols consider various types of adversaries based on their capabilities and the time required to compromise a node.

\begin{itemize}
    \item \textbf{Static adversary:} Represents a Byzantine entity with the ability to corrupt a predetermined number of nodes in advance, exerting complete control over them. However, the static adversary exhibits limited adaptability, as it cannot modify the selection of nodes it has corrupted after the initial attack. Furthermore, the adversary achieves instantaneous control over the compromised nodes, without any delay or gradual process.
    
    \item \textbf{Adaptive adversary:} Characterizes a Byzantine entity that possesses the ability to dynamically control nodes within the network, making adjustments based on changing circumstances. This adversary can flexibly change the nodes under its control, enhancing its overall influence over time.

    \item \textbf{Mildly adaptive adversary:} Describes a Byzantine entity capable of an corrupting nodes based on past messages.It can observe and learn from past messages exchanged within the system. This allows it to tailor its attacks based on the acquired information. Importantly, this adversary lacks the ability to alter messages that have already been sent. While the adversary can corrupt entire groups of nodes, this action comes with a time penalty. Corrupting a group takes longer than the group's normal operating phase.
    
    \item \textbf{Strongly adaptive adversary:} Represents a Byzantine entity that possesses the capability to gain knowledge of all messages sent by honest parties. This adversary leverages this information to make decisions, determining whether to corrupt a party by altering its message or delaying message delivery. With a high level of adaptability, the strongly adaptive adversary strategically utilizes its comprehensive knowledge to compromise the system.
\end{itemize}

\subsubsection{Identity Model}
Distributed ledger consensus protocols adopt different approaches to managing nodes' membership, generally falling into two contrasting categories:

\begin{itemize}
    \item \textbf{Permissionless:} This model embraces an open membership system, enabling any node to freely join the network and participate in the validation of new entries.
    \item \textbf{Permissioned:} In contrast, the permissioned model restricts membership, allowing only a predefined set of approved members to validate new entries.
\end{itemize}

In the context of Distributed Ledger Technology (DLT), the identity model is intricately tied to the network's openness, which may manifest as private, public, or a consortium. A comprehensive exploration of these identity types can be found in \cite{BitFuryGroup2015a} and \cite{Nadir2019}.




\subsubsection{Governance Model}
The governance model pertains to the decision-making framework embraced by a DLT network for determining protocol rules and their updates and upgrades. Given that system governance is inherently a social concept, various governance models can be identified:
\begin{itemize}
    \item \textbf{Anarchic:} Protocol upgrade proposals undergo approval by every participant in the network. Each participant has the autonomy to accept or reject a given proposal, potentially leading to network splits.
    \item \textbf{Democratic:} Participants engage in voting on new rules and protocol upgrade proposals. Ultimately, all participants are obligated to follow the decision of the majority, even those who opposed it.
    \item \textbf{Oligarchic:} New rules and protocol upgrades are suggested and endorsed by a select group of participants.
\end{itemize}

Considering the shift of most DLTs in handling governance and related matters "on-chain" or "off-chain", we also recognize the distinction between:
\begin{itemize}
    \item Built-in (On-chain governance): The decision-making process is defined as part of the underlying consensus protocol within the network.
    \item External governance (Off-chain governance): The decision-making process relies on procedures independently performed outside the DLT network.
\end{itemize}

\subsubsection{Transactions Ordering}
Maintaining the chronological order of transactions is crucial for preventing fraud and inconsistencies in both linear and non-linear DLTs, including DAGs. To address this requirement, diverse approaches have been developed within consensus protocols to ensure correct transaction ordering. While transaction ordering is typically integrated into the consensus mechanism for efficiency, specific scenarios may necessitate decoupling it from transaction execution and validation. This allows for greater flexibility and optimization based on specific application needs. Two key approaches for decoupling ordering include:
\begin{itemize}
    \item Optimistic ordering: Transactions are speculatively executed and finalized only after their order is confirmed, allowing for faster processing but requiring additional validation steps.
    \item Off-ledger ordering: Transactions are ordered outside the DLT before being submitted for validation and execution, providing increased scalability but adding complexity to the system.
\end{itemize}

\subsubsection{Conflict Resolution Model}
In certain DLT networks, the coexistence of temporary conflicting ledger versions, known as forks, may arise due to factors such as network latency or parallel block validation. To converge towards a single source of truth, networks and consensus mechanisms implement various rules, with the "longest chain rule" from the Bitcoin protocol being one of the most prominent. In case of conflicting orders, the PoW-based networks converge to a single order following the longest chain — the chain with the largest accumulated Proof of Work discarding the others. This rule is adopted by various protocols, each potentially utilizing a different cumulative parameter (witness votes, endorsements, etc.). Table \ref{tab:my-table7} illustrates different rules employed by various DLTs to resolve conflicting orders and mitigate discrepancies. In Proof of Stake (PoS) blockchains, the authoritative chain may be the one with the highest support from validators, chosen based on their stake in the network. The chain with the highest level of validator support is considered the valid chain.

\subsection{Consensus Layer: State of the Art}
\begin{figure}[b]
\hspace*{-0.3cm}
\centering
\includegraphics[width=8cm, height=2.5cm]{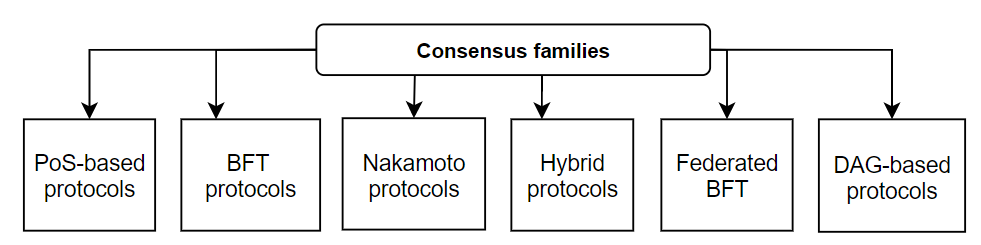}
\caption{Our taxonomy for consensus protocols}\label{fig: families}

\end{figure}

In this subsection, we introduce various consensus mechanisms and their properties. While it is beyond the scope of this paper to provide an exhaustive taxonomy of existing protocols (see fig. \ref{fig: families}), we categorize the reviewed protocols into six groups. This classification forms the basis for categorizing DLTs.

\subsubsection{BFT Consensus Family (PBFT-like)}

\smallskip

This family encompasses classical consensus mechanisms derived from traditional distributed computing literature and their recent variations. The BFT family is distinguished by its characteristic of conducting all-to-all voting rounds \cite{Antoine2019}. Nodes within the network possess known identities, and the participant count is limited. Given the multitude of protocols within this family, our paper's review primarily concentrates on the most widely employed algorithms in the context of DLT, namely PBFT, Raft, IBFT, DBFT, POA (AURA, Clique), HoneyBadgerBFT, and HotStuff.

\paragraph{Practical Byzantine Fault Tolerance (PBFT)}

The PBFT algorithm, introduced by Castro and Liskov \cite{Castro1999}, stands out as a well-established consensus mechanism often synonymous with BFT consensus. Designed for practical, asynchronous environments with a Byzantine minority of \emph{f} out of a total of \emph{3f+1} nodes, PBFT relies on a leader-based approach. In each view (leader election term), a leader is elected to append log entries (blocks in the case of a blockchain).

PBFT ensures asynchronous safety by employing a three-phase protocol: pre-prepare, prepare, and commit. The first two phases (pre-prepare and prepare) facilitate ordering requests sent within views, even when the leader is faulty, while the last phase ensures that committed requests are totally ordered across views. However, PBFT exhibits a weak level of liveness, aiming to prevent faulty nodes from forcing frequent view changes and consistently promoting a faulty node to the primary role.

While PBFT enables high throughput and low latency, it does have certain drawbacks, including the costliness of recoveries in case of faulty leaders and performance degradation with an increasing number of nodes. As a result, PBFT is considered suitable primarily for a small-to-medium group of known participating nodes.

\paragraph{RAFT \cite{Ongaro}} is a log-replication consensus algorithm that uses a leader-based approach. A single leader is elected, chosen through randomized timer timeouts, to append new entries to the log (append-only) and replicate it across the other nodes in the network. When the leader becomes faulty or slow, at most one of the nodes with the most up-to-date log is chosen as the new leader. Similar to PBFT, all nodes in RAFT must be known and interconnected to exchange messages in epochs.

RAFT guarantees network safety by ensuring that every log is correctly replicated and commands are executed in the same order across all nodes. Additionally, only one leader is ever elected at a time, preventing conflicting updates. Notably, attempts have been made to extend RAFT's tolerance to Byzantine faults, as seen in works by \cite{Clow2017}and \cite{Copeland2016}.

\paragraph{Istanbul BFT (IBFT) \cite{Moniz2020}} is a Byzantine Fault Tolerant (BFT) state machine replication-based consensus algorithm. IBFT inherits from the original PBFT by using a three-phase consensus protocol, where before each round, nodes will elect a leader (called Proposer) who will be responsible for proposing new blocks in the network along with the PRE-PREPARE message. In each stage, validators broadcast the state message and wait for 2f+1 state messages (PREPARE, COMMIT) to commit the current state (insert the block to the chain). IBFT can tolerate at most f faulty nodes in a network with 3f+1 nodes.

Three-Phase Consensus in IBFT:
\begin{itemize}
    \item Pre-prepare: The Proposer broadcasts a PRE-PREPARE message containing the proposed block to other nodes.
    \item Prepare: Upon receiving the PRE-PREPARE message, validators verify the block and then broadcast a PREPARE message if they agree.
    \item Commit: Once a validator receives 2f+1 PREPARE messages for the same block, it broadcasts a COMMIT message and adds the block to its chain.
\end{itemize}

While IBFT demonstrates scalability to a large number of peers, it comes with a trade-off as it demands higher communication overhead compared to other consensus algorithms.



Proof of Authority (PoA)
\paragraph{Proof of authority PoA \cite{DeAngelis2018}} is a Byzantine Fault Tolerance (BFT) leader-based consensus mechanism designed for permissioned blockchain deployments with at least \emph{n/2+1} honest nodes, where \emph{n} is the total number of nodes.

PoA relies on a set of trusted and identifiable nodes called authorities, responsible for creating new blocks and securing the blockchain. PoA protocols operate through a series of steps, each with an authority elected as a validation leader using a rotating process. PoA is a reputation-based consensus protocol where the authority's reputation is at stake instead of financial or computational power.

PoA has two major implementations: Aura \cite{PARITY} and Clique \cite{Consensus}. These implementations differ in their process:
\begin{itemize}
    \item Aura : In the Block Proposal phase, the current leader takes the initiative to propose a new block. Following this, in the Acceptance phase, a subsequent round unfolds where other authorities within the network participate in voting to determine the acceptance of the proposed block. This two-phase approach ensures a systematic and distributed mechanism for reaching consensus on the inclusion of new blocks in the blockchain.

\end{itemize}

\begin{itemize}
    \item Clique: In Clique, the current leader proposes a new block, and block acceptance is immediate as other authorities directly append the proposed block to their chains. This streamlined approach eliminates a separate acceptance phase, enhancing efficiency in the consensus process.
\end{itemize}

Unlike PBFT, PoA requires fewer message exchanges, resulting in better performance \cite{Dinh2017}:
Aura offers lower transaction acceptance latency and more predictable block issuance with steady time intervals. In contrast, Clique achieves faster block creation by eliminating a separate block acceptance round from its consensus process.
However, Deangelis et al. \cite{DeAngelis2018} argue that PoA algorithms might not be the most suitable choice for permissioned blockchains deployed over the internet. They advocate for PBFT as a potentially better alternative for permissioned settings.


\paragraph{Democratic BFT (DBFT)} \cite{Gramoli2017} is a deterministic and partially synchronous consensus algorithm that can tolerate up to \emph{f $<$ n/3} Byzantine nodes, where \emph{n} is the total number of nodes. Unlike practical Byzantine Fault Tolerance (PBFT) protocols that rely on a single, correct leader to finalize the consensus process, DBFT is a leaderless protocol.

DBFT offers several key features:
\begin{itemize}
    \item Multiple Proposers: Instead of relying on a single leader, DBFT allows multiple nodes to propose sets of transactions for inclusion in a block, enriching the pool of potential options and potentially accelerating block creation.
    \item Disjoint Sets: Proposers in DBFT can propose disjoint sets of transactions, meaning they don't necessarily need to agree on the exact contents of a block. This allows for flexibility and can potentially lead to higher throughput.
    \item Asynchronous Rounds: Nodes in DBFT can complete asynchronous rounds, meaning they don't need to wait for all other nodes before proceeding. This can further improve performance, especially in large networks.
    \item Democratic Decision-Making: Each node plays a similar role in the consensus process, contributing to a more "democratic" decision-making mechanism compared to leader-based protocols.
    
\end{itemize}
This combination of features allows DBFT to achieve high throughput and scalability, making it suitable for large-scale blockchain deployments. However, DBFT is comparatively complex among BFT protocols, potentially posing implementation challenges. While it exhibits promising performance, ongoing research focuses on formally verifying its security properties.


\paragraph{HoneyBadgerBFT or HBBFT \cite{Miller2016}} is a Byzantine fault-tolerant consensus algorithm designed for fully asynchronous networks with an honest majority (\(n > 3f\)). Unlike traditional BFT, it eliminates the need for a special leader node to propose transactions; instead, every node assumes the role of a proposer. In each epoch, participating nodes exchange a batch of encrypted transactions and collectively agree on them using a randomized agreement protocol. HBBFT relies on threshold encryption, where transactions are encrypted with a shared public key and can only be decrypted when the elected consensus committee collaborates. This ensures that adversaries cannot determine which transactions are proposed by specific nodes until an agreement is reached. Additionally, HBBFT incorporates optimizations to enhance system performance, including communication-optimal reliable broadcast \cite{Cachin2002}, binary consensus as proposed by \cite{Mostefaoui2000}, and batching \cite{Cachin2002}. It is particularly suitable for networks with a small number of known (permissioned) validator nodes.
However, it's essential to note that HBBFT, like any consensus algorithm, has its drawbacks. One potential drawback is the computational overhead associated with the use of threshold encryption, which may impact the algorithm's overall performance, especially in resource-constrained environments. Additionally, the complexity of the algorithm may pose challenges in terms of implementation and maintenance. Despite these considerations, HBBFT remains a viable choice, particularly for networks with a small number of known (permissioned) validator nodes.

\paragraph{HotStuff \cite{Yin2019}} is a leader-based Byzantine fault-tolerant replication protocol designed for partially synchronous networks, offering both linearity (linear change view) and responsiveness. It re-examines the original BFT designs and aims to significantly reduce the authenticator complexity of PBFT. In cases of a correct leader or view-change, the complexity is brought down from O($n^2$) and O($n^3$) to O($n$), where \emph{n} is the number of nodes. Unlike PBFT, HotStuff adopts a three-round process, rotating the leader every three rounds after a single attempt to commit a command. The leader replacement (view-change) operation is integrated with the normal process, eliminating a separate view change process and reducing overall complexity. HotStuff streamlines PBFT authentication complexity by introducing a new proposer in the first phase carrying only a single commit-certificate, and in the second phase, replicas can reject a proposition conflicting with their highest-level certificate without requiring a leader proof. Additionally, it shifts PBFT's communication model from a mesh to a star topology, relying on the leader for communication between nodes, and utilizes threshold digital signature schemes to further reduce authenticator complexity. Notably, HotStuff ensures optimistic responsiveness, with a leader requiring only \emph{n-f} (where \emph{f} is the number of faulty nodes) votes from other nodes to guarantee progress. It maintains simplicity and modularity by decoupling safety (voting and commit rules) from liveness, guaranteed by the pacemaker mechanism after the global stabilization time (GST). Furthermore, HotStuff enhances BFT scalability by linearizing the algorithm's authenticator complexity, making it suitable for wide networks. However, the introduction of three phases in HotStuff, compared to the two phases in PBFT, induces additional latency, limiting the throughput to a single commit per three phases. For a comprehensive exploration of other Byzantine consensus algorithms with extensive analysis, refer to \cite{Berger2018}.

\subsubsection{Nakamoto consensus family}
The Nakamoto consensus family comprises protocols that utilize a chain of block data structures and employ the longest chain fork choice rule (or variants like GHOST \cite{Sompolinsky2015}) to ensure network safety, all while incorporating economic incentives as a motivating factor. Originally designed to facilitate secure global currency transfers, these protocols offer a simpler alternative to PBFT while tolerating significant corruption (up to \emph{n/2} nodes). Notably, they operate on a permissionless basis, allowing nodes to freely join or leave the network without requiring prior authentication. Here, we explore some of the most prominent protocols within this category: PoW, memory-bound PoW, and BitcoinNG.


\paragraph{proof-of-work (or PoW)} PoW is a consensus mechanism where a leader, commonly known as a miner, is chosen probabilistically based on the computational power contributed. Upon solving a cryptographic puzzle, the miner constructs and submits a block, accompanied by a valid proof-of-work nonce, to the network. Verification by other peers involves computing the hash of the block header and ensuring it meets the condition of being smaller than the current target value. Valid blocks are added to the chain with the largest cumulative difficulty. In PoW, each peer effectively casts a vote on transaction validity using its hashing power. The protocol relies on partial hash collisions to thwart Sybil attacks\cite{Douceur2002}, preventing the system from being tainted by multiple misbehaving nodes. 

A noteworthy aspect of PoW is its resilience to potential forks or reorganizations in the blockchain. In the event of competing blocks being added simultaneously by different miners, the network may experience a temporary fork. However, PoW relies on the principle that the longest chain is considered the valid one. Nodes in the network continuously work on extending the chain, and the longest chain, representing the majority of computational power, ultimately becomes the accepted version. This mechanism ensures a coherent and agreed-upon transaction history.
Additionally, PoW introduces economic measures and incentives, such as transaction fees and mining rewards, to actively discourage denial-of-service attacks and protect the network against spam, contributing to the overall robustness of the protocol.


\paragraph{Memory bound PoW} Originally conceived as an egalitarian process accessible to all, mining in traditional PoW faced concerns over centralization due to the widespread adoption of application-specific integrated circuits (ASICs). These specialized devices provided a significant advantage over conventional hardware, leading to the concentration of mining power among a few entities and the dominance of ASIC manufacturers.

To address this issue, various ASIC-resistant solutions were introduced. Dwork et al. \cite{Dwork2005} \cite{Tromp2015} \cite{Ren2017} proposed a memory-bound proof-of-work that relies on random access to slow memory rather than computational hashing power. This design makes ASIC mining less efficient, as ASICs cannot accommodate the substantial memory requirements. Several PoW implementations, such as Scrypt \cite{Percival2009}, Primecoin \cite{King2013}, Equihash \cite{Biryukov2017}, and CryptoNight \cite{VanSaberhagen2013}, have adopted memory-bound mining processes.

For example, Ethereum 1.0 employed a PoW protocol called Ethash \cite{Wiki}, intentionally designed to be ASIC-resistant through memory-hardness. Miners on the Ethereum network are required to compute a sizable in-memory Directed Acyclic Graph (DAG), which stood at 4 GB as of December 23, 2020 \cite{DAG}, to mine new blocks. Ethereum is further planning to enhance its Ethash algorithm by transitioning to ProgPoW (programmatic proof-of-work), aiming to improve resilience against ASIC mining before transitioning to the proof-of-stake \cite{Ranjan} consensus protocol. This strategic approach seeks to maintain a fair and decentralized mining ecosystem within the Ethereum network.

\paragraph{Bitcoin-NG \cite{Eyal2016}} Bitcoin-NG was proposed to scale PoW-based Nakamoto protocols by decoupling transaction verification from the leader (miner) election process. Similar to Bitcoin, Bitcoin-NG operates in epochs, where each epoch features a single leader chosen via proof-of-work. Once a leader is identified, they are responsible for unilaterally serializing transactions via Microblocks in the succeeding epoch, respecting predefined limits on rate and block size, until a new leader is chosen. The decoupling in Bitcoin-NG reduces delays caused by periodic leader elections and scales Bitcoin in terms of transaction throughput and propagation latency. However, Bitcoin-NG does not ensure strong consistency, as short forks may occur during the leader-switching process.

\subsubsection{Proof of stake and its variants}
\smallskip
Proof-of-Stake (PoS) emerged as an alternative to the resource-intensive Proof of Work (PoW) and was initially proposed in PPCoin \cite{King2012a}. Instead of engaging in a competitive hash-calculation race, participants aspiring to become validators and forge new blocks in a PoS system must lock a specific amount of coins into the network as a financial stake. The likelihood of a node being selected as the next validator is then determined by the size of their stake. Various implementations of PoS have been introduced, each bringing unique features to address specific challenges. Notable representatives include Ethereum PoS, which combines PoW and PoS, Delegated Proof of Stake (DPoS) as seen in EOS, Liquid Proof of Stake (LPoS) allowing users to "lease" their coins to validators, and Ouroboros and its variants. Snow White is another example within the PoS landscape. The evolution of PoS and its diverse variants reflects ongoing efforts to refine and optimize blockchain consensus mechanisms.

\paragraph{Ouroboros} was introduced by Aggelos Kiayias et al \cite{Kiayias2017} in 2017 as a provably secure proof of stake protocol. The protocol operates in epochs, with each epoch divided into slots during which blocks are produced by a randomly chosen slot leader. Within each epoch, a committee of stakeholders employs a secure multiparty implementation of a coin-flipping protocol to generate the required randomness for electing a random list or committee of block producers for the slots in the current epoch. From this list, slot leaders are then elected using a random lottery algorithm called Follow the Satoshi (FTS). Given that Ouroboros is a PoS system, the probability of selecting a block producer is proportional to their stake. While Ouroboros was presented as the first provably secure proof-of-stake in the synchronous setting, it is susceptible to desynchronization attacks. This vulnerability arises as slot leaders in Ouroboros require precise synchrony to utilize their allocated slots accurately, potentially leading to network stalling or hindering liveness.

\paragraph{Ouroboros Praos \cite{David2018}} extends the Ouroboros protocol to accommodate semi-synchronous networks, addressing desynchronization attacks and providing security against fully-adaptive corruption. In contrast to Ouroboros, Praos incorporates a special verifiable random function (VRF) that enables the randomly selected slot leader to anticipate the slots they will lead in advance. Unlike Ouroboros, where stakeholders learn about the slot leader once they publish a block, Praos stakeholders privately verify the corresponding VRF to determine the slot leader. The protocol is underpinned by formal analysis and delivers a robust Proof-of-Stake (PoS) implementation with a security level equivalent to Proof of Work (PoW).

\paragraph{Ouroboros Genesis \cite{Badertscher2018}} represents a variant of the Ouroboros Praos protocol designed to address the synchronization challenge for new nodes joining Proof-of-Stake (PoS) systems. In contrast to Praos, which maintains moving checkpoints, Genesis introduces a distinct chain selection rule, employing the so-called $maxvalid$ procedure \cite{Badertscher2018}. This innovation enables new nodes to safely participate in the protocol without requiring external information beyond the genesis block. As a result, new validators can verify the true longest chain with only knowledge of the genesis block. The security of Ouroboros Genesis has been formally proven in \cite{Badertscher2018} against a fully adaptive adversary controlling less than half of the total stake in a partially synchronous network.

\paragraph{Ouroboros Chronos \cite{Badertscher2019}} extends the Ouroboros Genesis protocol, introducing a significant innovation by eliminating the necessity for a global clock to maintain network synchronization. This design removes the dependency on an external service providing timestamps, such as NTP \cite{NTP}. Ouroboros Chronos achieves this by introducing a novel synchronization mechanism that allows nodes to synchronize their local clocks based solely on knowledge of the genesis block and the assumption that their local clocks, initially desynchronized, advance at approximately the same speed.


\paragraph{Snow White} was introduced by Daian et al. \cite{Daian2019} as a PoS protocol providing end-to-end and formal proofs of security in asynchronous and permission-less settings. This blockchain-style protocol emphasizes the ability of users to freely enter and exit the network, incorporating a modified version of the sleepy consensus protocol introduced by \cite{Pass2017a}.
Similar to Ouroboros, Snow White operates in epochs. In each epoch, a leader is randomly and publicly elected from a committee of stakeholders to append a block to the blockchain. To address security concerns related to committee reconfiguration and random block-proposer selection, particularly adaptive chosen key and randomness-biasing, Snow White introduces a novel “two-lookback” mechanism. In each epoch, a new consensus committee is determined in advance (multiple blocks before) than the randomness seed, making seed prediction challenging for malicious nodes. The previously generated randomness seeds of the previous blocks serve as a source of randomness entropy to seed the new random oracle for electing a slot’s leader. Additionally, akin to PoW, Snow White’s nodes always choose the longest chain in the presence of multiple chains.

\paragraph{Delegated Proof of Stake (DPoS) \cite{Grigg2017}} was originally conceptualized by Daniel Larimer and implemented in the Bitshares blockchain \cite{Schuh2017}. Serving as a variant of the PoS consensus, DPoS relies on a group of delegates, commonly referred to as witnesses, who are elected by token holders (stakeholders) to generate and validate blocks on behalf of other network participants. DPoS networks typically feature an odd number of elected witnesses, ranging from $21$ to $101$, who take turns forming and signing new blocks for approval through a voting system. However, the specific range of witnesses can vary depending on the chosen DPoS implementation and network configuration. Some networks might employ a lower number of witnesses, such as $11$ or $17$, for improved performance or efficiency. Conversely, others might utilize a higher number, exceeding $101$, to accommodate larger network sizes or enhance decentralization.

The specifics of the voting system in DPoS can vary between implementations, but generally, each witness presents a single proposal when soliciting votes from other witnesses. Similar to PoS, a voter's weight is determined by their stake in the network. Moreover, witnesses in DPoS cannot sign arbitrary blocks or produce a block outside their scheduled time slot. Refusal by witnesses to produce blocks results in swift expulsion, with replacement by other elected witnesses.

While DPoS offers enhanced scalability compared to PoS or PoW, it faces criticism for the significant risk of centralization arising from the concentration of power among a limited number of actors (delegates).

\subsubsection{ Hybrid protocols }
\smallskip
Hybrid consensus protocols, drawing strength from established mechanisms like PoW and PoS, combine elements of different approaches to overcome individual limitations and achieve better performance. This hybrid approach offers the potential for enhanced scalability, improved security, and flexibility to adapt to changing network conditions. However, increased complexity, potential conflicts, and finding the right balance between mechanisms remain challenges to be addressed. 

\paragraph{Byzcoin} was introduced by Kokoris-Kogias et al. \cite{Kogias2016} as a solution to enhance Bitcoin’s consistency and performance by leveraging the advantages of Practical Byzantine Fault Tolerance (PBFT) and employing collective signing. In contrast to Bitcoin, Byzcoin forms a BFT consensus group through the collaboration of Proof of Work (PoW) miners. The protocol incorporates a proof-of-membership mechanism based on PoW to establish the consensus group using a fixed-size sliding share window. Membership shares are distributed to successful miners for mining a valid block, and the cumulative shares represent their voting power in the consensus process. As miners mine new blocks, the share window advances, and miners without valid shares exit the consensus group.

Byzcoin employs a collective signing protocol named scalable collective signing (CoSi) \cite{Syta2016} to aggregate thousands of signatures, reducing the PBFT communication complexity from O($n^2$) to O($n$). This scalability enables BFT protocols to accommodate large consensus groups. Moreover, ByzCoin ensures safety and liveness under Byzantine faults, with near-optimal tolerance allowing for up to f faulty group members among \emph{3f + 1} participants. To minimize transaction processing latency, ByzCoin adopts the decoupling of transaction verification from block mining, a concept introduced by Bitcoin-NG.

\paragraph{Solana \cite{pierro2022can} \cite{li2021bitcoin}}
Solana actually utilizes a hybrid consensus mechanism that combines both Proof of History (PoH) and Proof of Stake (PoS). This is one of the unique aspects of Solana's architecture, allowing it to achieve high scalability and transaction throughput.
Proof of History (PoH) operates on the principles of Verifiable Delay Function (VDF) and Cryptographic Hash Functions to establish a reliable and verifiable timeline of events on the Solana blockchain. The VDF serves as a time-sequencing mechanism, designed to be computationally expensive to solve but easy to verify. Validators independently execute VDFs, generating unique outputs that act as proofs of the time spent solving the function. These outputs are then hashed and linked together to form a continuous chain of timestamps.

The PoH process involves the execution of VDFs by validators, generating unique outputs that serve as proofs of time. These outputs are hashed and linked in a chain, creating an immutable and tamper-resistant history of timestamps. Validators share their hash chains with the network, and other validators can easily verify the chain by checking the validity of each VDF output and ensuring the cryptographic integrity of the chain.

Consensus is built based on these verified timestamps, allowing validators to agree on the order of transactions and establish the current state of the blockchain. PoH introduces additional features to enhance its functionality, such as Sealevel, a parallel execution environment that enables multiple validators to work on different parts of the VDF chain simultaneously, thereby improving processing speed and scalability. Tick Verification is another feature that ensures the accuracy of timestamps by allowing validators to periodically compare their internal clocks and make adjustments if necessary.
Solana's Proof of Stake (PoS) and Proof of History (PoH) work in tandem, each playing a distinct role in the blockchain's consensus mechanism. Their contributions can be outlined as follows:
\begin{itemize}
    \item PoH: Primarily responsible for ordering transactions and creating a verifiable timeline of events. This allows validators to quickly and efficiently reach consensus on the state of the network without needing to rely on computationally expensive calculations like in Proof of Work (PoW).
    \item PoS: Primarily responsible for securing the network by incentivizing validators to act honestly. Validators stake their SOL tokens to participate in the consensus process and earn rewards. This economic incentive helps to prevent malicious actors from attempting to disrupt the network.
\end{itemize}
 This unique approach allows Solana to achieve high scalability and transaction throughput. 
\paragraph{Algorand \cite{Gilad2017}-\cite{Chen2019a}} is a PoS algorithm that employs secret self-selection to randomly choose a leader and validators committee through cryptographic Sortition. The committee and leader selection are secured by privately computing the verifiable random function (VRF) using users' private keys and a seed generated in the previous block, without any communication among users. The selection process favors validators with the highest token balance (Algos) in their account, serving as a protection mechanism against Sybil attacks. As a result, the selected parties only discover their selection through the lottery when they propagate their winning tickets and their validation decision for the block. This renders it impractical for a malicious actor to corrupt the committee, influence their decision, or launch a DDoS attack against the members.

However, the committee's random selection process does not prevent the election of two-thirds malicious delegates. Once selected, the committee achieves consensus on the new block using a Byzantine agreement protocol called BA*, a variant of PBFT. BA* allows the participating members to reach consensus on a new block with low latency and without the possibility of forks (forks may occur with negligible probability). BA* guarantees consensus as long as an honest majority of \emph{n $>$ 2f/3} exists (assuming synchronous communication) and utilizes threshold signatures for efficiency and fault tolerance. This allows the protocol to handle large numbers of users with low latency and minimal risk of forks.

\paragraph{Thunderella \cite{Pass2018}} is a novel protocol built upon Pass and Shi's Sleepy and Snow White protocols, introducing a permissionless chain for failure recovery. Thunderella integrates two blockchains: a BFT chain representing the "fast path" and an underlying chain, considered a slow "fall-back" path, which can be any standard blockchain like Bitcoin or Ethereum. The fast path facilitates optimistic instant confirmation of transactions, while the synchronous slow chain ensures consistency and liveness.

The fast path is centralized, with a designated central authority, the Accelerator, serving as a leader responsible for transaction linearization. Concurrently, a validating committee comprising stakeholders is randomly elected using various approaches, such as utilizing all stakeholders as the committee (similar to Snow White or Algorand) or employing recent miners. The fast path confirms new transactions as long as the accelerator and $3/4$ of the committee are both online and behaving honestly; otherwise, the chain halts, awaiting recovery.

Periodically, Thunderella posts messages (referred to as alive messages) containing the hash and notarization of a checkpoint block to the underlying chain. In the event of a halt in the fast path due to a faulty accelerator or a dishonest committee, Thunderella nodes transition to the slow chain for recovery.

\paragraph{Casper Friendly Finality Gadget or CFFG \cite{Buterin2017}} is a protocol that Ethereum plans to utilize as a transitional method for transitioning from PoW to PoS (CASPER). Casper FFG represents a chain-based hybrid incorporating elements of both PoS and PoW. Blocks are mined using PoW, while PoS validators regularly verify the 100th block (known as the checkpoint) created by miners. In this setup, PoS validators do not confirm blocks or add transactions but ensure transaction finality.

To become active validators, participants lock Ethers (Ethereum's cryptocurrency) in Casper's smart contract running on the PoW chain. The verification and validation of new blocks are ensured by block validators selected based on their stake, with the voting power of each validator equivalent to the amount of their stake. Additionally, Byzantine behavior is prohibited through stake slashing.

\paragraph{Tendermint \cite{Buchman2016}} is a fault-tolerant protocol inspired by the PBFT SMR algorithm \cite{Castro2002} and the DLS algorithm \cite{Dwork1988}, designed for partially synchronous networks in both permissioned and permissionless settings. In a permissionless setting, it utilizes PoS as the underlying security mechanism.

Tendermint operates as a leader-based BFT protocol proceeding in rounds similar to PBFT's rounds. In each round, a new leader responsible for proposing a new block is elected through deterministic round-robin selection from a set of validators who have locked financial stakes. The frequency of a validator being chosen as a leader is proportional to their share of the total stake. If a validator misbehaves during their turn, they can be punished by having a portion of their deposited stake slashed. After a set duration, the stakes are unlocked and returned to the validator.

For block finality, Tendermint requires a supermajority (a quorum of over $2/3$) of all validators to validate the block. Tendermint can tolerate up to $1/3$ Byzantine validators, but if more validators disagree or become unresponsive, the network chooses to halt instead of proceeding with potentially incorrect transactions. Therefore, Tendermint prioritizes consistency over availability.

\paragraph{LibraBFT \cite{Baudet2019}} is a variant of the HotStuff consensus protocol, specifically utilizing chained HotStuff, and introducing several enhancements and improvements. In LibraBFT, the elected leader proposes a block (a set of transactions) to extend the longest chain of requests it knows. Nodes vote for the proposed block, unless it conflicts with a longer chain they already know. In such cases, they send their votes to the next leader to help it learn the longest chain.

In contrast to HotStuff, LibraBFT employs a different approach for leader election by randomizing the process using a VRF scheme. It utilizes aggregate signatures, eliminating the need for a complex threshold key setup, to preserve the identity of validators signing quorum certificates. Additionally, LibraBFT specifies a clear implementation of the Pacemaker mechanism introduced by HotStuff to ensure the advancement of rounds. The pacemaker of each validator keeps track of votes and time, triggering a leader election when a timeout occurs due to a faulty leader, lack of votes, or after a leader successfully proposes a block.

LibraBFT guarantees safety when at least \emph{2f+1} nodes are honest and provides liveness as long as there exists a global stabilization time (GST). These properties, coupled with rapid consensus, make it suitable for permissioned blockchains, such as the Libra blockchain.

\subsubsection{ DAG-based Protocols}

\paragraph{IOTA \cite{Popov2017}} operates on a unique hybrid consensus mechanism, mixing Proof-of-Work (PoW) with a custom Tangle structure. Transactions first undergo a lightweight PoW puzzle demanding computational effort to prevent spam and ensure network integrity, as transactions within the network are fee-less. Additionally, the new transaction should randomly approve two previous valid transactions, referred to as tips, by extending them and thereby increasing their initial weights.

In IOTA, the tip selection process involves choosing two previous valid transactions (tips) to reference when adding a new transaction to the Tangle. This is performed through a proof-of-work mechanism by the end-user or device. While the protocol does not strictly dictate the tip selection process, users often employ the recommended Markov Chain Monte Carlo (MCMC) weighted random walk. This process is biased toward transactions with higher weights, contributing to the prioritization of transactions with more confirmations in the Tangle.

During the network's initial stages, a central node called the Coordinator aids in achieving consensus by emitting milestones—special transactions that solidify the Tangle and serve as validation reference points. This ensures consensus in small networks with a sparse Tangle. However, the IOTA project plans to replace the centralized Coordinators with a new voting system called “Shimmer” \cite{IOTA} as part of the Coordicide project \cite{Popov2020}.

\paragraph{Avalanche Protocol} Avalanche stands out as a leaderless Byzantine fault tolerance protocol that operates on a metastable mechanism through network subsampling.
This process allows nodes to achieve consensus without relying on a designated leader, eliminating potential vulnerabilities associated with centralized control.
\begin{itemize}
    \item Random Sampling: Each node repeatedly samples a small random subset of the network, significantly reducing communication overhead and preventing any single node from dominating the consensus process.
    \item Query Rounds: Sampled nodes participate in multiple rounds of communication, exchanging information and collecting responses. This iterative approach allows for the emergence of a consensus decision even when faced with conflicting information or Byzantine behavior.
    \item Threshold Adoption: Upon reaching a predetermined threshold based on the majority of responses, nodes adopt the agreed-upon decision. This ensures that the network converges towards a consistent state even in the presence of faulty or malicious nodes.

\end{itemize}

Avalanche organizes transactions in a DAG. Each new transaction extends one or more transactions. When faced with conflicting transactions, Avalanche utilizes the DAG structure, relying on a transaction's progeny (all children transactions). Corrected nodes then vote positively on valid transactions based on the entire ancestry's validity.

Avalanche ensures strong probabilistic safety, even when faced with \emph{f $\leq$ n/3} Byzantine nodes. This is achieved through random sampling, minimizing the impact of individual Byzantine nodes. Nodes utilize probabilistic verification with weighted random samples for increased security, and transactions necessitate multiple confirmations from independent nodes to enhance fraud detection.

Additionally, Avalanche employs a financial mechanism, the AVA token, to safeguard the network against sybil attacks, ensuring open membership and enabling economic governance.

\paragraph{Hashgraph \cite{Baird2016}} operates as an asynchronous Byzantine fault-tolerant protocol, employing a DAG as its underlying data structure to store events, which encompass transactions and associated details. The protocol utilizes a combination of a voting algorithm and a gossip protocol to achieve consensus.

In the Hashgraph protocol, nodes engage in random gossiping, sharing their knowledge of known events with other nodes. This can involve either events originating from the gossiping node itself or those received from other nodes. This continuous gossiping process facilitates the widespread dissemination of transactions.

Each event within Hashgraph comprises two critical elements: a timestamp and two hashes.
\begin{itemize}
    \item The timestamp records the precise time the event was created, providing a chronological order to the events within the DAG.
    \item The two hashes reference two prior events: one from the gossip receiver and another generated by the sender. These references establish a clear lineage within the DAG and allow nodes to verify the validity of transactions.
\end{itemize}
Additionally, the gossiping node signs the shared event information, creating a verifiable audit trail and preventing malicious modifications. Consequently, each node in the network possesses a comprehensive record of the transaction history, along with details about nodes that previously received this information.

To ensure consensus on the order and validity of transactions, nodes participate in a virtual voting process. IN this phase, no votes are cast or exchanged. Instead, each node leverages its understanding of the DAG (constructed through gossip history) and the timestamps of events to calculate what other nodes should vote for. 
A transaction is considered valid and finalized only if it receives virtual validation from at least $2/3$ of the nodes in the network. This threshold ensures a high degree of confidence in the consensus decision, even in the presence of malicious actors or network failures.
Hashgraph operates under a closed membership model, implying that the total number of nodes in the network is known and fixed. This feature simplifies the virtual voting process by eliminating uncertainties about the voting quorum.
Furthermore, it enables efficient operation without the need for synchronized clocks or global knowledge, making the protocol suitable for geographically dispersed networks. By leveraging virtual voting and gossip-based information dissemination, Hashgraph achieves rapid consensus on the order and validity of transactions. However, 
Hashgraph faces limitations with a closed membership model, reliance on patented technology, and potential transparency issues. Scalability challenges and vulnerability risks from its gossip protocol reliance further contribute to considerations for widespread adoption.

\subsubsection{Federated BFT}
Ripple \cite{Schwartz2014} was the first the first implementation of a Federated Byzantine Agreement System (FBAS). The Federated Byzantine Agreement (FBA) approach redefines Byzantine Fault Tolerance (BFT) settings, introducing an open membership service based on a trust model. Unlike traditional BFT protocols, FBA protocols, exemplified by the Unique Node List (UNL) in Ripple and the quorum slice in Stellar, allow nodes to interact with a limited group of trusted peers, eliminating the need for a global unanimous agreement among network participants.

Ripple's consensus algorithm, the Ripple Protocol Consensus Algorithm (RPCA), functions in rounds where validators from a server's Unique Node List (UNL) strive for a supermajority consensus, typically set at least $80\%$. In each round, a designated server proposes a candidate set of transactions, and validators on its UNL individually vote on the proposal. The iterative process continues until a supermajority is achieved, indicating widespread agreement among validators. If consensus is not reached, the server identifies and blocks less-supported transactions, ensuring the reliability and security of the network.
However, if only less than $20\%$ of nodes in the network agree, a temporary network halt may occur. The network's safety and liveness depend on the proper server configuration and the intersection of correct nodes' UNLs. While Ripple suggests a minimum overlap requirement of $20\%$ of the UNL, RPCA guarantees safety and liveness under specific conditions, including a minimum overlap requirement of $40\%$ \cite{Armknecht2015}.

Stellar Consensus Protocol (SCP)\cite{Stellar} is based on Ripple protocol. It provides a first provably safe consensus, while assuming network transitivity and strong concreteness \cite{Mazieres2015}. Unlike Ripple's fixed Unique Node List (UNL) and supermajority voting, SCP utilizes flexible quorum slices, allowing nodes to define their own sets of trusted validators, enabling efficient and adaptable consensus. This, coupled with federated voting and a provably safe design, ensures network stability and Byzantine fault tolerance (up to $33\%$ Byzantine nodes). Furthermore, SCP promotes decentralization through open membership, empowering anyone to participate in the network.

\vspace{5pt}
\section{Execution Layer} \label{sect 6}
In this section, We will unpack the execution layer, examining its essential building blocks and their properties, before introducing the now-ubiquitous execution component driving state-of-the-art technologies.
\subsection{Components and properties}
DLT systems offer two primary avenues for translating agreements into code: smart contracts, which provide extensive flexibility for crafting custom logic, and built-in scripts, which offer a more structured and protocol-defined approach to rule execution.

\subsubsection{Execution environment}
\begin{figure}[b]
\centering
\includegraphics[width=8cm, height=4cm]{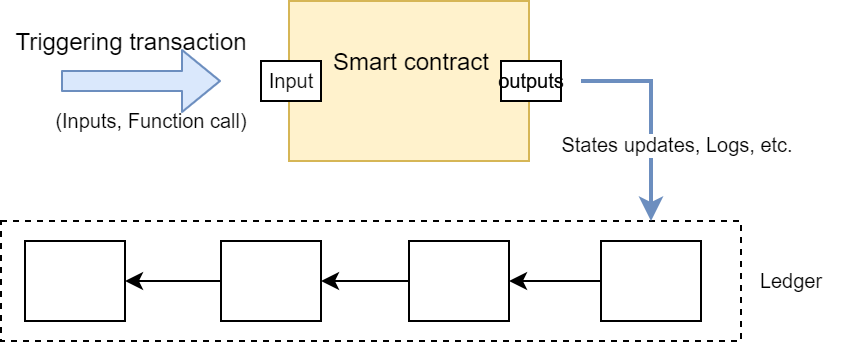}
\caption{An overview of the smart contract machine } \label{fig: sc}

\end{figure}

\paragraph{Smart Contract Model} In this paradigm, agreements between participants are encoded as self-executing programs that operate on a predefined set of states. Typically implemented in a dedicated language or using existing programming languages like Java or C++, these programs, known as smart contracts, are executed within a specialized environment such as a virtual machine or compiler. The execution involves processing the clauses specified in the triggering transaction, producing an output (Fig. \ref{fig: sc}), and often updating states. 
While smart contracts can facilitate native asset manipulation (e.g., tokens or cryptocurrency), their versatility extends beyond this function. They can be utilized to implement a wide range of logic and automate complex workflows, enabling applications beyond simple asset transfers.
Despite the term "self-executing," smart contracts require external triggering transactions to initiate their execution. While smart contracts enforce agreed-upon collaboration logic, they do not possess legal contract status.

\paragraph{Scripting Model} 
While smart contracts offer extensive programming flexibility, scripting model guides users within a predefined framework of rules and functions, ensuring adherence to specific usage patterns and protocol-level constraints.
This model leverages a predefined and limited set of rules established by the DLT protocol itself, restricting the possible scenarios for implementation. By restricting the scope of permissible logic, the scripting model reduces the potential attack surface and minimizes security vulnerabilities that complex smart contracts might introduce. The predefined rules eliminate the need for developing and deploying custom smart contracts, streamlining the process and reducing the overall complexity of the DLT system. Typically found in DLTs emphasizing the secure manipulation of built-in assets (e,g. Bitcoin), the scripting model focuses on providing a framework for executing predefined rules rather than supporting universal program execution.


\subsubsection{Turing Completeness}
In a general sense, an environment or programming language is deemed Turing-complete if it is computationally equivalent to a Turing machine \cite{Turing1936}. This means that a Turing-complete smart contract language or environment can execute any possible calculation within finite resources. Some DLTs support a Turing-complete execution environment, enabling users the flexibility to define intricate smart contracts. Conversely, certain DLTs employ Non-Turing complete execution environments, characterized by inherent limitations, such as the inability to have iteration structures with arbitrarily high upper bounds.
\subsubsection{Determinism}
Determinism is a crucial characteristic of the execution environment in DLT systems. Given that distributed programs, such as smart contracts, are executed across multiple nodes, deterministic behavior is imperative to produce consistent and identical outputs, avoiding discrepancies within the network.

To guarantee determinism within DLT systems, various approaches are employed:
\begin{itemize}
    \item Disabling non-deterministic features: Some DLTs opt for a conservative approach by simply disabling non-deterministic operations altogether. This ensures complete predictability but restricts the range of functionalities that can be implemented.
    \item Sandboxing and controlled environments: Certain DLTs employ sandboxes or other controlled environments for executing programs involving non-deterministic features. This allows for some flexibility while maintaining isolation and preventing unintended consequences.
    \item Deterministic alternatives: Developers strive to design deterministic alternatives for non-deterministic operations whenever possible. For instance, cryptographic hash functions can be used to generate deterministic pseudo-random numbers.

\end{itemize}

\subsubsection{Runtime Openness}
In the majority of DLTs, the execution environment or runtime is intentionally designed as an isolated component with no connections to external networks, such as the Internet. This isolation ensures security and immutability of the ledger, but it also limits the capabilities of DLT applications. However, there are scenarios where the need to access information from outside the DLT arises, such as weather forecasts, stock prices, or exchange rates. To accommodate this requirement, various design choices have been introduced, leading to three distinct approaches:
\begin{itemize}
    \item Isolated: Prohibiting interactions between the smart contract execution environment and external environments.
    \item Oracle-based: Allowing interactions with external environments through members of the network known as oracles. Oracles can be third parties or decentralized data feed services providing external data to the network.
    \item Open: Enabling the execution layer to connect directly to external environments.
\end{itemize}

\subsubsection{Interoperability}
Currently, DLT networks are intentionally siloed and isolated from each other. Interoperability, the ability to exchange data, assets, and transactions across different DLTs, emerges as a critical need to unlock the true power of this transformative technology. Given its significance, various solutions have been proposed to facilitate interoperability among different existing DLTs, falling into the following approaches:
\begin{itemize}
    \item Sidechain \cite{Back2014}: A blockchain operating in parallel with another chain (main chain) allowing the transfer of data (cryptocurrency) from the main chain to itself. Sidechains typically operate in either a one-way pegged or two-way pegged mode, with the former facilitating data movement to and from the main blockchain using locking mechanisms, and the latter allowing data movement only toward the sidechain.
    \item Multichain \cite{Greenspan2015}: A network of interconnected blockchains designed to facilitate seamless cross-chain communication and interaction. It features a central "major ledger" that governs and synchronizes transactions across various sub-ledgers, each representing a specific blockchain. This architecture enables users to securely swap assets, tokenize real-world objects, and build decentralized applications that function across diverse blockchain ecosystems.
    \item  Interoperability protocols: function as bridges between distinct DLTs, enabling seamless communication and exchange of data or assets. They often leverage smart contracts and other technical mechanisms to establish compatibility and facilitate cross-chain interactions.
    \item Interoperable DLT: New DLTs are being designed with interoperability as a core principle. These DLTs incorporate features and protocols specifically intended to facilitate seamless interaction with other DLT platforms.

\end{itemize}

\subsection{Execution layer: state of the art}
In this section, we present a comprehensive overview of the most prevalent execution environments implemented in both industry and academic literature, along with a discussion of their distinctive properties. Notably, our focus extends to the Ethereum Virtual Machine (EVM), given its widespread adoption across numerous existing DLTs. In a broader classification, current DLT-based smart contract platforms can be categorized into two primary groups: those compatible with the EVM-compatible and those not compatible with EVM.

\subsubsection{Execution environments}
\paragraph{Ethereum Virtual Machine (EVM)}
Smart contracts in Ethereum are written in high-level languages \cite{S-tikhomirov} such as Solidity, LLL, Viper, or Bamboo. These programs are compiled into low-level bytecode using an Ethereum compiler, and the resulting bytecode is stored in a dedicated account on the blockchain, effectively providing it with an address.

\begin{figure}[b]
\centering
\includegraphics[width=9cm,height=5cm]{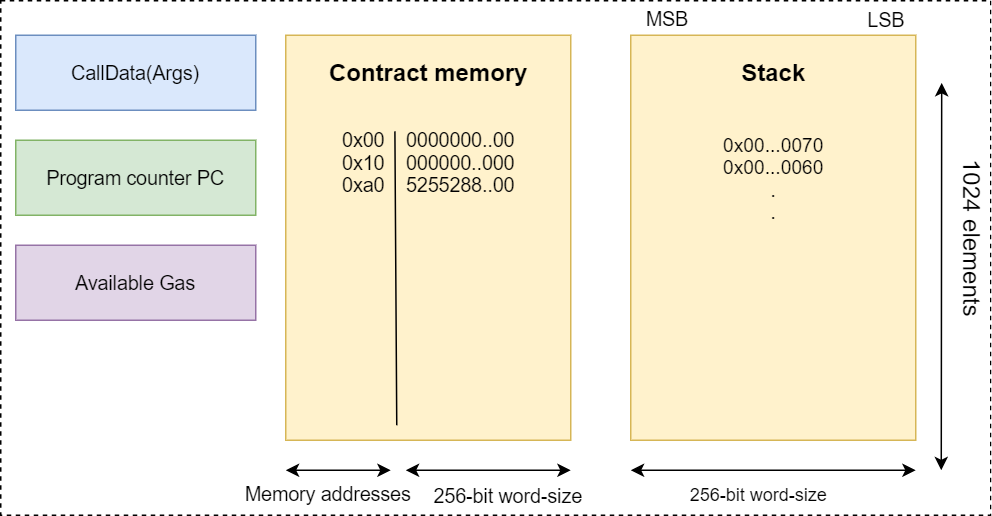}
\caption{The stack-based architecture of the EVM } 
\label{fig : evm2}
\end{figure}

The bytecode resides in the ledger (Fig. \ref{fig : evm2}) and is assigned an address for interaction. Interactions with a smart contract are facilitated through transactions, which carry inputs and specify the function to be called. The associated bytecode for the invoked function is simultaneously executed on the Ethereum Virtual Machines (EVMs) of all network nodes, processing the transaction's payload.

Upon successful termination of the bytecode execution, the smart contract's states are updated on the blockchain's state tree, capturing the outcomes of the executed code. This process ensures that all network nodes maintain a consistent view of the smart contract's state and its interaction history.

Operating as a stack-based virtual machine, the EVM efficiently processes bytecode and manages state updates. The stack, with a maximum size of $1024$ entries, employs a 256-bit register architecture, enabling simultaneous access and manipulation of the most recent $16$ items. The stack's dynamic nature facilitates the execution of complex operations within smart contracts. Complementing this stack, the EVM incorporates volatile memory, organized as a word-addressed byte array. Each byte is uniquely identified by its memory address, providing a flexible data structure for contract execution. In contrast, the EVM features persistent storage represented as a word-addressable word array. This storage, comprising $2^{256}$ slots, each holding $32$ bytes, operates as a non-volatile key-value mapping. Unlike volatile memory, the contents of storage persist across transactions, forming an essential component of the Ethereum blockchain's state. 

Furthermore, the EVM operates as a sandboxed runtime, creating an isolated environment for smart contracts execution. Each smart contract running within the EVM lacks access to the network, file system, or other processes running on the host computer. As a security-oriented virtual machine designed to execute potentially unsafe code, the EVM implements strict isolation measures. To counter Denial-of-Service (DoS) attacks, the EVM incorporates the gas system, where every computation within a program must be prepaid in a dedicated unit called gas, as per the protocol's definition. If the provided gas amount fails to cover the execution cost, the transaction is unsuccessful. However, it's important to note that the gas mechanism, while mitigating DoS attacks, can still be vulnerable if settings are not appropriately configured, as demonstrated by \cite{Chen2017} \cite{Perez2020}. Assuming adequate memory and gas, the EVM can be considered a Turing-complete machine, allowing the execution of a wide range of calculations.


\color{black}

\begin{figure}[b]
\centering
\includegraphics[width=8cm, height=4cm]{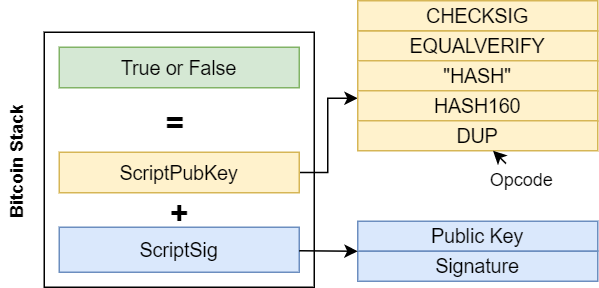}
\caption{Bitcoin loads and executes the locking and unlocking scripts onto the stack. If the supplied public key matches the public-key hash and the supplied signature matches the provided public key, the execution is correct (True). }\label{fig: btcstack3}

\end{figure}

\paragraph{Bitcoin Scripting}
Bitcoin employs a stack-based scripting engine. This engine operates on a stack-based architecture, utilizing a Forth-like language to define scripts that control how funds are transferred within the network. A Bitcoin script is a sequence of instructions, known as opcodes, that are loaded into a stack and executed sequentially (Fig. \ref{fig: btcstack3}). The script follows a push-pop stack approach, executing from left to right. A script is deemed valid if the top stack item is true (non-zero) upon completion of its execution.

Bitcoin utilizes two key scripts for handling transactions:
\begin{itemize}
    \item ScriptPubKey: This script functions as a locking script attached to the output of a transaction. It specifies the conditions that must be fulfilled for a recipient to redeem the corresponding funds. For instance, it might require a specific signature or a combination of signatures from multiple parties.
    \item ScriptSig: This script acts as the unlocking script. It serves as a proof that the recipient fulfills the conditions set by the ScriptPubKey, essentially unlocking the funds for transfer.
\end{itemize}


Bitcoin scripting deliberately lacks Turing completeness. Additionally, the execution time is constrained by the script's length, capped at $10$ kilobytes after the instruction pointer \cite{Script}. This limitation serves to mitigate denial-of-service attacks on nodes responsible for block validation. The language used in Bitcoin scripting is acknowledged as complex and limited for smart contract development. To address this, various projects have emerged, including Ivy \cite{IVY}, Simplicity \cite{Blockstream}, and BitML \cite{Stefanolande}. These projects offer high-level languages with enhanced features that compile into Bitcoin scripts. Furthermore, BALZaC \cite{Zhang2005a} and Miniscript provide alternatives—a high-level language based on formal models and a structured approach to writing (a subset of) Bitcoin Scripts, respectively. Notably, Rootstock (RSK) \cite{Mining2018} was proposed as an EVM-compatible two-way pegged sidechain with Ethereum, using a merge-mining process involving both Rootstock and Bitcoin. RSK's virtual machine, the Rootstock Virtual Machine (RVM), is based on the EVM. This means that smart contracts written for Ethereum are generally compatible with Rootstock with minimal modifications. This compatibility allows developers to build smart contracts with the security of the Bitcoin blockchain.

\paragraph{Stellar}
Unlike Ethereum and other platforms that rely on virtual machines and dedicated smart contract languages, Stellar \cite{Stellar} takes a distinct approach to smart contracts. Instead of executing general-purpose code, Stellar Smart Contracts (SSCs) are constructed from a series of interconnected transactions subject to specific constraints. 
Participants engaging with SSCs do not directly interact with on-chain code but instead agree to the conditions specified within transactions. These transactions are constructed using a predefined set of $13$ operations, each representing an individual command that can modify the Stellar ledger. Furthermore, various constraints can be applied to transactions, enhancing their functionality. Stellar supports built-in constraints such as Multisignature, Batching, Atomicity, Sequence, Time bounds, among others \cite{Stellara}.
SSCs are not Turing complete and developers can write SSCs in multiple programming languages (such as Python, $C\#$, Ruby, Scala, C++) using the Stellar SDK \cite{Developpers}.  \cite{Developpers}.

\paragraph{NXT and Ardor}

NXT prioritizes security with predefined smart contract templates, known as smart transactions \cite{Nxter} . These templates minimize code vulnerabilities, making NXT ideal for secure transactions like asset transfers and multi-signature accounts.

Ardor, as NXT's successor, adopts the Java Virtual Machine (JVM) for smart contracts, enabling Turing-complete code execution. This unlocks a broader range of computational capabilities, supporting more complex smart contracts. Ardor further introduces Lightweight Contracts, allowing developers to automate tasks without the full computational cost of Turing-complete contracts. These Java classes are executed executed by a subset of nodes selected to run the ContractRunner addon which facilitates the execution of smart contracts on the Ardor blockchain. Together, NXT and Ardor exemplify varied smart contract approaches, catering to different blockchain use cases.

\paragraph{NEO Virtual Machine (NeoVM)}
NEO introduces the lightweight NeoVM (NEO Virtual Machine) \cite{NeoVM20}, a virtual stack-based machine designed for processing smart contracts. NeoVM is designed to be language-agnostic, meaning that it supports multiple programming languages.  Languages such as $C\#$, Java, Python can be used to write smart contracts and NEO's compiler (NeoCompiler) translates the resulting source code (with limitations \cite{NeoVMa}) into a unified bytecode. This enables cross-platform programming. Notably, NeoVM provides an InteropService that facilitates communication between the virtual machine and the underlying blockchain infrastructure. This service allows smart contracts to interact with the blockchain, access data, and perform various operations.
Designed to be Turing-complete, NeoVM adopts a gas concept similar to Ethereum's, contributing to predictable resource management during contract execution.

\paragraph{EOS virtual machine}
The EOS Virtual Machine (EVM) primarily uses the WebAssembly language for smart contracts. WebAssembly is a portable binary format designed to provide a high-performance, secure, and platform-agnostic environment for executing code. Thus, smart contracts are typically written in languages like C++ or Rust and then compiled into WebAssembly bytecode for deployment on the EOSIO blockchain.
The compiled WebAssembly bytecode is deployed onto the EOSIO blockchain. EOS utilizes a resource model where users need to stake tokens to obtain resources (CPU and NET) to execute their smart contracts. EOS employs a unique resource model where users need to stake tokens to access resources like CPU and NET for executing their smart contracts. The staked tokens act as a form of rent, ensuring fair access to resources.

\paragraph{Cardano CCL}
The Cardano blockchain consists of two essential layers: the Cardano Settlement Layer (CSL) and the Cardano Computational Layer (CCL). The focus on the Cardano Computational Layer (CCL) lies in its role as a platform for decentralized applications (DApps) and smart contracts. Operating above the Settlement Layer, the CCL enables developers to create diverse applications, such as decentralized finance (DeFi) and identity verification solutions.
The CCL uses Plutus \cite{Plutus}, which is inspired from Hashkall, as a new smart contract language. Similarly, to the EVM, CCL utilizes a cost accounting model to prevent DoS attacks. Any changes resulting from the execution of the smart contract are reflected in the global Cardano ledger. The Cardano Settlement Layer (CSL) is responsible for maintaining this ledger, tracking the state changes brought about by the execution of smart contracts.

\paragraph{Zilliqa virtual machine}
Zilliqa \cite{Team2017} introduces a smart contract engine called Zilliqa Virtual Machine (ZVM). The ZVM adopts a sharding architecture that allows for parallel execution of smart contracts across different shards. This parallelization significantly increases transaction throughput and network scalability compared to sequential execution models
The ZVM primarily uses Scilla\cite{Teama} as its smart contract language. Scilla stands out as an intermediate-level programming language designed specifically for crafting safe smart contracts with formal verification. Its intermediate nature positions it as a dual-purpose tool: first, as a compilation target for high-level languages like Solidity, and second, as an autonomous programming framework in its own right.

Scilla is designed to facilitate the formal verification of smart contract programs, ensuring correctness and eliminating known vulnerabilities at the language level. Once compiled to Scilla, the program is interpreted using an interpreter (Scilla-runner), which takes the Scilla code, the current contract's state, and the message triggering the execution as inputs, mutating the smart contract states accordingly. Although Scilla helps write more secure and easily verifiable smart contracts, it lacks expressiveness as it is non-Turing-complete. Zilliqa adopts a gas protocol based on the computation complexity, storage usage, and network congestion involved in processing a smart contract.

\paragraph{Java Virtual Machine}
Hyperledger Fabric takes a distinctive approach by leveraging the Java Virtual Machine (JVM) and nodeJs \cite{Fabric20} runtime as smart contract environments. Consequently, smart contracts, referred to as Chaincodes, can be written in Java, JavaScript, TypeScript, and, Go or any other language compatible with the supported runtimes. Fabric Chaincodes are executed within Docker containers to ensure execution isolation from the peer, providing an additional layer of security. Similarly, Corda R3 opts for the JVM as its smart contract execution environment, without containerization, and utilizes Kotlin and Java as the primary languages for smart contracts.

\paragraph{Stratis CLR}
Stratis leverages the Microsoft .NET framework, and specifically, the Common Language Runtime (CLR) as an execution environment for its smart contracts. Developers write smart contracts in languages supported by the .NET framework, such as $C\#$ or $F\#$ and The smart contract code is compiled into Intermediate Language (IL) code \cite{Startisacademy}. IL is a low-level, platform-independent representation of the smart contract logic. This design choice allows Stratis to support CIL, enabling the use of theoretically any language that can be translated into CIL for writing smart contracts. Stratis also employs the "gas" mechanism for paid execution, similar to Ethereum's gas model, to manage resource consumption and prevent denial-of-service attacks.

\paragraph{NEM}
NEM is a Java-based blockchain platform and and features a native P2P cryptocurrency (XEM). It encompasses both private and public blockchains, offering key value-added features such as ease of deployment, extensive customization, high performance, and robust security. NEM provides additional features such as the creation of custom digital assets (mosaics), identity verification through namespaces, and document timestamping using apostille.


\paragraph{MOVE for Libra}
Move is a bytecode language designed for direct execution in Move's Virtual Machine (VM). Its distinctive feature is the ability to define custom resource types with semantics inspired by linear logic, reminiscent of Rust. In Move, a resource can only be moved between program storage locations and cannot be copied or implicitly discarded, similar to Rust's ownership system. This uniqueness aligns with Rust, where values can only be assigned to one name at a time, making them inaccessible under the previous name after reassignment.

Move's transaction script introduces flexibility by supporting both one-off and reusable behaviors. Smart contract functions can be executed multiple times, providing a broader range of capabilities compared to Ethereum, which is limited to invoking a single smart contract method for reusable behaviors. Moreover, Move's executable format is a typed bytecode that is higher-level than assembly yet lower-level than a source language. The bytecode undergoes on-chain checks for resource, type, and memory safety by a bytecode verifier before being executed directly by a bytecode interpreter. This approach allows Move to offer safety guarantees typically associated with a source language without adding the source compiler to the trusted computing base or incurring the cost of compilation on the critical path for transaction execution.
While Move was originally designed for the abandoned Libra's blockchain, Move is still actively used in the Diem  project \footnote{https://www.diem.com/en-us/}

\paragraph{Solana Runtime}
 Departing from conventional virtual machine reliance, Solana adopts an operating system-inspired model, enabling direct program execution on validators' machines and eliminating the interpretational overhead associated with VMs, resulting in unparalleled speed and efficiency. This architecture relies on Rust-powered programs compiled into bytecode, state-holding accounts, signed transactions, and decentralized validators. The execution process involves transaction submission, validation, account lookup, direct program execution, and consensus. Solana's optimized mechanism, featuring parallel processing through Sealevel, predictable fees, and upgradeable programs, underscores its commitment to performance, scalability, and user-friendly interactions, positioning it as a highly efficient and innovative blockchain platform.
 
\subsubsection{Interoperability}
The lack of communication between isolated DLTs has posed a notable obstacle to the advancement of the blockchain ecosystem. As a result, several suggestions have surfaced to overcome this challenge. In the following section, we underscore key strategies implemented at the execution layer to address this issue.

\paragraph{Sidechains}
Various sidechains have been proposed in the DLT ecosystem. Rootstock \cite{Mining2018} serves as a sidechain of Bitcoin, featuring an integrated Ethereum virtual machine known as RVM. The Rootstock chain is connected to the Bitcoin (BTC) blockchain through a two-way peg mechanism. This innovative approach facilitates seamless transfers between Bitcoin (BTC) and SBTC (Rootstock's native currency). This connection is established by utilizing Bitcoin scripts, allowing for interoperability between the two blockchains. In this process, users can send their BTC to the Rootstock chain, locking it up in a special smart contract. In return, they receive an equivalent amount of SBTC (Rootstock Bitcoin) on the Rootstock chain.
Similarly, Counterparty \cite{Counterparty} utilizes a sidechain architecture built on top of the Bitcoin blockchain. To facilitate asset movement onto the Counterparty sidechain, users "lock" their Bitcoin by sending them to a designated address, effectively making them unavailable on the main chain.
Drivechain \cite{Drivechain} proposes a mechanism for transferring BTC between the Bitcoin blockchain and sidechains. In contrast to most DLTs where the sidechain is a separate project, Cardano introduces Cardano KMZ as an integral part of its ecosystem. Cardano KMZ is a protocol facilitating the movement of assets from its two-layer CSL to the CCL (Cardano Computation Layer) or other blockchains supporting the Cardano KMZ protocol. Another noteworthy sidechain project is Plasma \cite{Poon2017}, which aims to create hierarchical trees of sidechains (or child blockchains) using smart contracts on the root chain (Ethereum). Plasma enhances Ethereum's scalability by shifting transactions to sidechains operated by individuals or a group of validators rather than the entire underlying network. Currently, Plasma is actively developed and utilized by projects such as OmiseGo \cite{Poon2017a}, focusing on building a peer-to-peer decentralized exchange, and Loom \cite{Loom}, providing tools for constructing high-performance DApps while operating on the Ethereum network.

\paragraph{Interoperability Protocols}
Interledger (ITL) \cite{Thomas2015a}-\cite{Siris2019} is a standardized protocol developed by the World Wide Web Consortium for facilitating payments across different ledgers. It consists of a network of untrusted connectors that link various ledgers and employ escrow transactions (conditional locks of funds) to facilitate transfers between accounts on different ledgers. Additionally, Atomic swap \cite{Buterin2016} allows the trading of digital assets across unrelated blockchains. Atomic swaps use Hashed Time-Lock Contracts (HTLC) \cite{Poon2016} to coordinate operations, such as trading digital assets, on different chains. These operations are triggered by the revelation of a specific hash preimage. Alternatively, the Hyperledger project proposes the Hyperledger Labs Blockchain Integration Framework \cite{Catus}, a communication model enabling permissioned blockchain ecosystems to exchange on-chain data independently of the platform (e.g., Hyperledger Fabric, Quorum) without the need for intermediaries. The BTCRelay \cite{Network} project serves as a bridge between Bitcoin and Ethereum, implementing a BTCRelay smart contract on Ethereum that acts as a Bitcoin SPV (Simplified Payment Verification) node. It stores Bitcoin block headers provided by external parties known as Relayers, allowing other Ethereum contracts to verify transactions on the Bitcoin network.

\paragraph{Multi-chains}
Polkadot \cite{Wood2016} constitutes a network of interconnected chains, featuring a central connector called the Relay chain and multiple linked ledgers known as $Parachains$. The Relay chain finalizes transactions, facilitates cross-chain transactions \cite{Polkadot}, and shares states. To link the Relay chain with other networks like Ethereum or Bitcoin, Polkadot introduces bridge Parachains \cite{Parachain}, enabling two-way compatibility. Similarly, COSMOS \cite{Kwon2016} is composed of "Zones," which are blockchain networks interconnected through a central hub known as the Cosmos Hub Network.  Each Zone operates by maintaining its own state through validators that secure the blockchain and contribute to the consensus algorithm. Validators in a Zone operate independently with their own validator set. Meanwhile, intercommunication between Zones is facilitated by the Inter-Blockchain Communication (IBC) protocol \cite{IBC}, enabling the secure transfer of assets and information between Zones. Each Zone manages its state using a state machine, and the network as a whole benefits from the interoperability provided by the IBC protocol, allowing for a seamless exchange of value across different Zones in the Cosmos ecosystem. A key distinction between Polkadot and Cosmos lies in child chain sovereignty. Cosmos zones are independent chains built using the Cosmos SDK without sharing the same underlying environment. In contrast, in Polkadot, Parachains are dependent on and bound by the root chain's governance model, technical design choices, and limitations.

\paragraph{Interoperable chains}
Gravity Hub is a blockchain designed with the capability to communicate with other blockchains like Waves\cite{Waves} or Ethereum. Gravity Hub nodes can, for instance, fetch block headers from the Ethereum network and transmit them to the Waves Platform, providing proof of a specific transaction on Ethereum. Another DLT showcasing built-in interoperability is Wanchain \cite{Wanchain}. Wanchain is a blockchain platform focused on enabling interoperability between different blockchains. Initially centered on Ethereum, it now extends its cross-chain capabilities to include Bitcoin, EOS, and projects like AION. Wanchain facilitates the seamless transfer of assets across these blockchains, contributing to the development of a decentralized financial infrastructure. Wanchain proposes to interconnect different blockchains through a decentralized bridge infrastructure utilizing secure multi-party computation (sMPC) and threshold key sharing. This approach ensures the security and privacy of cross-chain transactions by distributing key functions and enhancing overall blockchain interoperability. \cite{Siris2019a} presents further details on interoperability solutions, including other projects like AIO, Blocknet, ARK or others.

\subsubsection{Determinism}
The Achilles' heel of many DLTs lies in their vulnerability to non-determinism, where seemingly identical transactions can produce inconsistent results due to factors like execution order or environmental variables. This undermines data integrity and consensus, jeopardizing the very foundation of trust and reliability in DLTs. To tackle this challenge, three main approaches have emerged (table \ref{tab:approachesdeterm-table}):
\begin{itemize}
    \item Determinism by design: This strategy eliminates non-determinism at the core, exemplified by Ethereum's Ethereum Virtual Machine (EVM). By excluding non-deterministic operations like floating-point arithmetic and external randomness sources, the EVM ensures consistent transaction execution across nodes. Solidity, the primary programming language for Ethereum smart contracts, also enforces determinism by design. However, recognizing the importance of randomness, the RANDAO \cite{Randaow} project proposes a decentralized autonomous organization (DAO) for registering random data on the Ethereum blockchain.
    \item Deterministic environments: Recognizing the need for occasional randomness, some projects embrace existing runtime environments like Java Virtual Machine (JVM) or Google's V8 engine, but modify them to enforce determinism. Examples include: Multichain \cite{Greenspan2015}, \cite{Corda}, and Stratis \cite{Startisacademy}.
    \item Determinism by endorsement: Introduced by Hyperledger Fabric, this novel approach leverages the network's endorsement process to achieve consensus on determinism itself. Each transaction undergoes simulation and execution by designated "endorsement" nodes. If any node produces a divergent result, the transaction is deemed invalid and rejected, preventing inconsistent outcomes from entering the ledger. Chaincode, the native smart contract language for Hyperledger Fabric, is specifically designed to facilitate deterministic execution under this model.
\end{itemize}

\begin{table*}[ht!]
\centering
\label{tab:approachesdeterm-table}
\caption{Approaches to Addressing Non-determinism in DLTs with Examples}
\begin{adjustbox}{max width=\linewidth}
\begin{tabular}{|p{2cm}|p{13cm}|}
\hline
\textbf{Approach} & \textbf{Blockchain Projects (Examples)} \\ \hline
\textbf{Determinism by Design} & Ethereum (EVM, Solidity), Hyperledger Besu, Qtum, Tezos, Constellation Network, Elrond, Chainlink, Zcash, Narcissus Protocol, Diem, Cardano, Solana, Avalanche, NEAR Protocol, Cosmos \\ \hline
\textbf{Deterministic Environments} & Multichain, Corda, Stratis, Hyperledger Fabric (Chaincode), DFINITY, Hedera Hashgraph, EOSIO, Rchain, Hyperledger Burrow, Tezos Interledger Protocol (TIP), Polkadot, ICON, Neo3, Fabric 2.0, Hashgraph Consensus Service \\ \hline
\textbf{Determinism by Endorsement} & Hyperledger Fabric (endorsement nodes), Ripple, Stellar, Libra, EOS, Algorand, VMware Blockchain, R3 Corda (Consensus service), Quorum, Hyperledger Indy, POA Network, Byzantine Fault Tolerance (BFT) projects (Tendermint, Hyperledger Sawtooth), Ripple Consensus Protocol (RCP) \\ \hline
\end{tabular}
\end{adjustbox}
\end{table*}

\subsection{Environment Openness}
In many DLTs, oracles play a pivotal role in acquiring data from external sources. Essentially, an oracle serves as a smart contract maintained by an operator, facilitating interaction with the external world. Various data feeds are deployed for smart contract systems like Ethereum, including Town Crier \cite{Zhang2016}, Oraclize.it \cite{Provable}, Band Protocol (BAND), Tellor (TRB), API3 (API3), DIA (DIA), Witnet (WIT), and Uma (UMA). Oraclize.it  \cite{Provable} relies on the reputation of the service provider, while Town Crier incorporates the concept of enclave hardware root of trust \cite{GlobalPlatform2017}. Alternatively, oracles like Gnosis \cite{Team2017a} and Augur \cite{Peterson2018} utilize prediction markets \cite{Abramowicz2006}.
For MakerDAO \cite{Dai}, a decentralized lending platform on the Ethereum blockchain, ensuring both reliable price data and decentralization for its assets is a priority. To achieve this, MakerDAO adopts a multi-tiered oracle system. At its core is the Medianizer \cite{Maker}, which aggregates data from $14$ independent price feeds, thereby ensuring accurate pricing for Ethereum. This approach aligns with the principle employed by ChainLink \cite{Tschorsch2015a}, another decentralized oracle solution that gathers data from diverse sources, contributing to a robust and decentralized data acquisition mechanism within the blockchain ecosystem.
In contrast to most DLTs, Fabric's Chaincode can interact with external sources like online APIs \cite{HyperledgerComposer}. However, if different endorsers receive divergent answers from the API, the endorsement policy fails, preventing the transaction from occurring. Other DLTs, such as Aeternity \cite{Aeternity}, integrate an oracle into the blockchain consensus mechanism \cite{Blog}, eliminating the need for a third party.

\subsection{Execution Layer: Discussion}
Despite the promising benefits of smart contracts, past implementations have unveiled critical security and performance pitfalls. Several recent studies have reported security issues in smart contracts \cite{Jiang, Kalra2018, Parizi2018, Hewa2021, Aggarwal2021, Hewa2021a, Ante2021}, including:

\paragraph{Immutability vs. Smart Contract Security}
Due to the immutable nature of most DLTs, patching and correcting detected bugs and security vulnerabilities is challenging, leading to potential fund losses, as demonstrated by the Ethereum DAO project \cite{Bloga}. To address this issue, various automated tools like SolidityCheck \cite{Zhang}, Securify \cite{Tsankov2018}, and ChainSecurity \cite{ChainSecurity2018} have been proposed to assist in writing secure code and analyzing bytecodes on different platforms. Additionally, ÆGIS \cite{FerreiraTorres2020} is a dynamic analysis tool designed to protect smart contracts from exploitation during runtime.

\paragraph{Insecure Languages for Writing Safe Smart Contracts}
The inherent risks in certain languages used for writing smart contracts have led to the proposal of new security-oriented languages, such as Flint \cite{Schrans2019} and SOLIDITYX \cite{SOLIDITYX}. Moreover, in \cite{Pettersson}, dependent types from the IDRIS language \cite{Brady2013} are employed to write provable smart contracts for Ethereum. Several projects are actively adopting formal verification, involving mathematical proofs to demonstrate that a given contract satisfies specific safety properties. For example, Scilla is designed to be amenable to formal verification, Tezos uses the Coq Proof Assistant for facilitating formal verification of smart contracts, and KEVM \cite{KEVM} introduces a complete semantics of the Ethereum virtual machine.

\paragraph{Untrusted Execution Environment}
To enhance the security of the execution environment, particularly in private blockchain platforms where execution outcomes are susceptible to tampering, some DLT platforms, such as Sawtooth Lake or Fabric \cite{Brandenburger2018}, execute smart contracts in Trusted Execution Environments (TEEs), such as Intel Software Guard Extensions (SGX). Although the setup of TEEs is complex, they play a crucial role in improving the privacy and security of data. TEEs securely store sensitive data, such as encryption keys, without leakage and provide evidence of the correct execution of the contract.

\paragraph{Smart Contracts Upgradability}
An ongoing concern in the blockchain space is the upgradability of smart contracts. Due to the immutability of smart contracts on blockchains, upgrading them poses a challenge. Current recommendations advise adopting an upgradable design pattern, as outlined in \cite{NickJohnson} and \cite{ManuelAraoz}, which involves deploying contracts alongside another dispatcher contract. Some platforms, such as Kadena \cite{Kadena}, propose solutions for implementing upgradable smart contracts.

\paragraph{Lack of Interoperability}
Despite considerable efforts to foster interoperability across chains, challenges persist, and accomplishments remain incomplete. This situation is attributed to two main factors. Firstly, existing interoperability projects are predominantly Ethereum-centric, necessitating more robust endeavors to facilitate interoperability with other DLTs. Secondly, the absence of a universal and unified standard hinders interoperability across different DLTs. It is noteworthy that ongoing standardization initiatives, like those by the Enterprise Ethereum Alliance and the GS1 initiative \cite{GS1}, aim to address this standardization gap.

\paragraph{Privacy}
Many smart contract environments lack privacy, exposing contract states. Addressing this concern, Hawk \cite{Kosba2016} was proposed as a framework for constructing privacy-preserving smart contracts using cryptographic primitives like zero-knowledge proofs. Additionally, Ekiden \cite{Cheng2018} introduced a solution based on executing smart contracts in a trusted execution environment. Enigma \cite{Zyskind2015a} facilitates private smart contracts through the use of distributed hash-tables (DHT) and multi-party computations (MPC). Notably, the implementation of zero-knowledge proof techniques to enable private transactions or smart contracts has gained attention. However, their adoption comes at a significant cost \cite{Unterweger2018} and introduces latency. Moreover, several privacy-preserving solutions may require a trusted party, potentially compromising the decentralized nature of smart contracts.

\paragraph{Layer 2 (L2) Scaling Solutions}
Rollups are scaling solutions that use the underlying blockchain layer (L1) to store transaction data, while the actual transaction processing and computation occur on the rollup itself. Rollups periodically post specific data on L1 (e.g., Ethereum), such as state roots or compressed transaction data, enabling anyone to verify the validity of the rollup state and transition to a new state.  This data availability on the main chain allows anyone to transition the rollup into a new state and prove the validity of the transition through validity or fraud-proof issuance.
Here's a breakdown of the key components of the rollup process:
\begin{enumerate}
    \item Off-chain Execution:
    \begin{itemize}
        \item Prover(s): Verify transaction validity and generate cryptographic proofs.
        \item Verifier(s): Check the proofs on the mainnet, ensuring the validity of the batch without needing to process individual transactions.
    \end{itemize}
    
    \item Data Availability:
    \begin{itemize}
        \item  Validity Rollups: Only proofs are submitted, requiring trust in the prover(s).
        \item Fraud Proofs: Transaction data is also stored on-chain, enabling anyone to challenge fraudulent transactions.
        \item Optimistic Rollups: Assume all transactions are valid unless challenged, offering faster finality but relying on fraud detection mechanisms.
    \end{itemize}
    
    \item State Commitment:
    \begin{itemize}
        \item Rollups maintain a compressed state snapshot on the mainnet, representing the current state of the off-chain ledger.
        \item State updates: Merkle trees or other efficient methods track changes in the off-chain state.
        \item Withdrawal: Users can withdraw their assets from the rollup back to the mainnet.
    \end{itemize}
    
    \item Security: 
    \begin{itemize}
        \item  Fraud proofs: Anyone can challenge invalid transactions in fraud-proof rollups.
        \item Validity proofs: Verifier contracts on the mainnet ensure the validity of proofs in validity rollups.
    \end{itemize}
\end{enumerate}

Rollups generally fall into two categories: Optimistic Rollups (e.g., Optimism, Arbitrum) and ZK-Rollups (e.g., dYdX, Loopring, ZK Sync) \cite{thibault2022blockchain}. ZK-Rollups offer significant scalability advantages over Optimistic Rollups due to their data compression capabilities. For example, while Optimism posts data after every transaction, dYdX only posts data reflecting account balances, resulting in a $1/5th$ L1 footprint and an estimated $10x$ higher throughput. This translates to lower fees for ZK-Rollups. However, ZK-Rollups may have higher computational costs during proof generation and may have limitations in smart contract functionality compared to Optimistic Rollups.

\paragraph{Modular Blockchains, A New Design Paradigm for Enhanced Scalability and Security}

\begin{figure}[t]
\centering
\includegraphics[width=9cm, height=6cm]{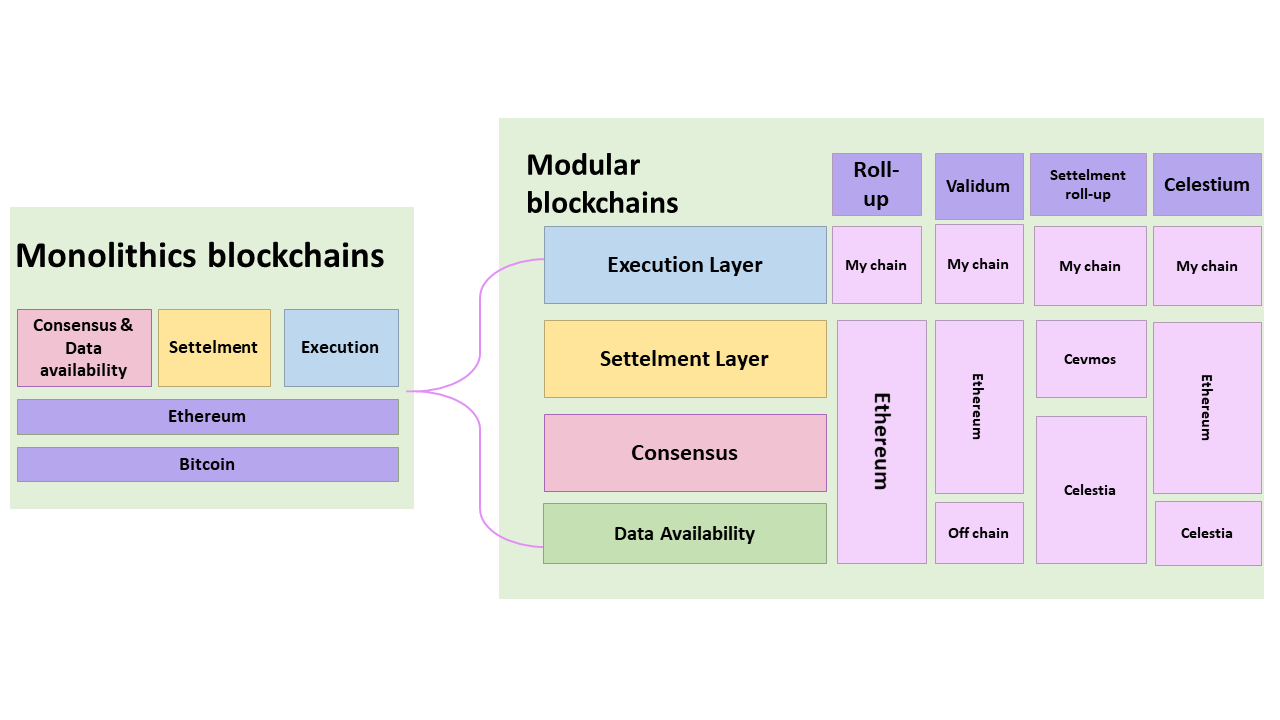}
\caption{Modular versus monolithic blockchains} \label{fig: modular}
\centering
\end{figure}

Many blockchain networks encounter challenges related to transaction volume, high fees, and optimization. A recent approach to achieve high scalability with robust security involves breaking down the blockchain into multiple components that can be scaled independently. In a traditional monolithic blockchain, the base consensus layer handles data availability, settlement, and execution. However, settlement and execution are typically coupled, limiting the system's overall capacity by design. In contrast, the modular blockchain paradigm separates the responsibilities, with the base consensus layer focusing solely on data availability, making transaction data available without executing it (see Figure \ref{fig: modular}). This design alleviates the burden on the base consensus layer, allowing it to concentrate on ensuring data availability. While technologies like Ethereum scaling rollups and Avalanche subnets incorporate modular components, many public blockchains remain designed as monolithic entities.
Modularity has become a prominent trend in the blockchain ecosystem, with the concept introduced by the Celestia project\footnote{https://www.projectcelestia.com/}. In a modular architecture, one of the network components (execution, consensus, or data availability) is decoupled (see Figure \ref{fig: modular}), enabling projects to deploy their blockchains without the complexities of establishing a new consensus network.
Several projects have embraced modularity:
\begin{itemize}
    \item Celestia: The first modular blockchain focusing on Consensus and Data Availability. It allows easy creation of individual blockchains, enabling developers to concentrate on DApp development.
    \item Celestiums: Integrates Celestia and Ethereum, serving as an Ethereum Layer 2 chain utilizing Celestia for data availability and Ethereum for settlement and dispute resolution.
    \item MEL (Fuel Labs): Prioritizes building the fastest modular execution layer (Modular Execution Level or MEL) on Celestia. MEL is designed as a fraud-provable computing system for modular blockchains, and Fuel v1 was launched as the first optimistic roll-up for scaling Ethereum.
\end{itemize}

The modular design offers several advantages over monolithic design:
\begin{itemize}
    \item  Scalability and Speed: Multi-level distribution allows modular blockchains to implement scalability mechanisms, significantly increasing throughput without compromising decentralization and security.
    \item \textbf Flexibility: Modularity reduces the cost of deploying smart contracts and facilitates experimentation with different technologies and environments. Notably, SwaySwap, a decentralized exchange similar to Uniswap, operates efficiently at the modular level.
\end{itemize}

\color{black}
\vspace{5pt}
\section{Application layer}  \label{sect 7}
In this section, we present a concise introduction to the components and attributes outlined by our DCEA framework at the application layer. Furthermore, we provide an overview of the current state-of-the-art within this context.

\subsection{components and properties}
\paragraph{Integrability}
DLTs go beyond mere data storage and transaction processing to offer real value by seamlessly integrating with existing technologies and systems. This focus on integrability is crucial for user experience and adoption. We propose integrability as a qualitative property, allowing the establishment of a "Level of Integrability" to assess if a DLT can easily integrate with Web, mobile, and other existing systems without requiring major overhauls?. This scale ranges from “High” to “Low” providing insights into the integrability of a DLT ecosystem:
\begin{itemize}
    \item High: Indicates strong integrability with other technologies, particularly web and programming technologies.
    \item Low: Suggests a lack of official integrability tools or limited availability with restricted capabilities in the DLT ecosystem.
    \item Medium: Represents an intermediate level between the two extremes.
\end{itemize}

\paragraph{DApp Orientation and DLT's Purpose}
Decentralized Applications, or DApps, represent a groundbreaking type of software that breaks away from the centralized model of traditional applications. Operating autonomously on decentralized networks, primarily blockchains, they bring resilience, transparency, and user empowerment to the forefront. In contrast to regular apps relying on centralized servers, DApps spread their architecture across multiple nodes, preventing any single entity from having control or the ability to censor.
However, not all applications utilizing blockchain technology qualify as true DApps. Simply storing data or relying on timestamps while keeping core logic outside the blockchain doesn't suffice. A genuine DApp leverages the full potential of a DLT by running core functions such as business logic and state transitions directly on the chain. 
Recognizing the important role DApps play in shaping the future of technology, some DLTs prioritize their development by becoming "DApp-oriented." These specialized networks go beyond simply providing blockchain infrastructure and actively cater to the needs of DApp creators.

\paragraph{Wallets and Identity Management}
Wallets play a crucial role in the application layer, serving as a central component for managing users' cryptographic identities. In the majority of DLTs, identities and ownership are established through public/private key pairs, making wallets a primary entry point to the network. Wallets are in charge of handling all the complex cryptographic tasks linked to creating or managing a user's cryptographic credentials, as well as signing transactions. They essentially act as a secure gateway, ensuring the safety and confidentiality of digital assets by managing cryptographic keys and authentication.

\subsection{Application Layer: State of the Art}
Considering the varied methodologies embraced by DLTs at the application layer, we present a summary of the application layer in several widely recognized DLTs.

\paragraph{Integrability}
DLTs typically bring about a layer of integration that acts as a bridge between external entities and their data and execution layer. Notable DLTs such as Ethereum \cite{Buterin2014a}, NEO \cite{NeoVM20}, and EOS \cite{EOSChart20} offer a rich toolset for integration. Ethereum provides a robust JSON-RPC API with strong support for JavaScript. In fact, Web3.js \cite{GitHubAPI20}, a JavaScript library, facilitates interaction with Ethereum-compatible nodes over JSON-RPC. For smooth integration into legacy systems, the Camel-web3j connector \cite{Camel-web3j} The Camel-web3j connector is used to integrate Apache Camel with the web3j library for interacting with Ethereum. Infura \cite{Infura} offers remote Ethereum nodes that developers and users can access through APIs to interact with the Ethereum blockchain. Metamask \cite{Metamask}, a popular cryptocurrency wallet, utilizes Infura's nodes by default to connect to the Ethereum network. This eliminates the need for users to run their own node, simplifying wallet functionality.
Similarly, EOS offers an extensive set of tools and features, simplifying integration with external systems. EOS provides various APIs, such as EOSIO RPC API, with implementations in different languages like EosJs \cite{EosJava20}, Py Eos \cite{PYEos20}, Scala Eos wrapper \cite{Wrapper20}, and Eos Java \cite{EosJava20}. These tools empower developers to interact with EOS across various programming platforms.
Business-to-business (B2B) focused DLTs like Hyperledger Fabric \cite{Vukolic2016} or the Corda platform \cite{Corda} address integrability challenges by providing robust integration SDKs. For instance, Fabric offers a Fabric SDK that simplifies the development and integration of NodeJs and Java \cite{Fabric20} applications within the Hyperledger Fabric blockchain framework. 
In contrast, other systems like Bitcoin, Litecoin, Dogecoin, or similar, offer limited integrability, as they were not designed to communicate with other systems. Bitcoin offers basic RPC features, and multiple unofficial implementations exist in various languages (e.g., BitcoinJ\cite{xiao2019java}, pybtc\cite{wickert2021python}), but lacks the comprehensive integrability features found in more business-oriented DLTs.

It is noteworthy that RPC can be a potential vulnerability vector susceptible to exploitation for launching Denial-of-Service attacks, particularly when there are RPC security issues \cite{li1995security} or inadequate server configurations. Unprotected JSON-RPC endpoints pose a security risk, as attackers may exploit them to transfer cryptocurrencies to accounts controlled by the attackers or to acquire admin privileges over a node.

\paragraph{ DApp orientation and DLT's purpose}
Bitcoin and its counterparts, like Zcash, Litecoin, and others, lead the charge in a digital cash revolution. These projects share a common bold goal: to challenge traditional finance and free value from the tight control of centralized authorities making them Cryptocurrency-oriented. Other DLTs aim to provide additional functionalities beyond cryptocurrency transactions. For instance, storage-oriented DLTs like Sia Network \footnote{https://sia.tech/}, Storj \footnote{https://www.storj.io/}, FileCoin \footnote{https://filecoin.io/}, and Ipfs manage data storage in addition to a cryptocurrency. Similarly, service-oriented DLTs offer specific services that consume the inherent token, such as "Steemit" \footnote{https://steem.com/}for a social network or Namecoin for decentralized DNS. On the contrary, several DLTs are DApp-oriented, enabling developers to create diverse applications. Examples include Ethereum, EOS, Stellar, TRON, among others, which provide a more flexible development environment for building decentralized applications (DApps) with built-in tokens. For a comprehensive overview of the current blockchain DApps landscape, refer to the study by Wu et al. \cite{Wu2019}.

b-1) Decentralized Finance (DeFi): DeFi applications leverage smart contracts to facilitate a range of functionalities such as margin trading, derivatives, stablecoins, and lending/borrowing. By incorporating smart contract capabilities and utilizing data oracles like Band Protocol, DeFi platforms achieve fully permissionless, enduring, and scalable operations. A notable example is Aave \footnote{https://aave.com/}, an open-source, non-custodial liquidity protocol that enables users to earn interest on deposits and borrow assets.
Additionally, Uniswap, a decentralized exchange, allows users to trade cryptocurrencies directly with each other without the need for a central intermediary. MakerDAO, a decentralized stablecoin platform, uses smart contracts to maintain the value of its stablecoin, DAI, pegged to the US dollar. Compound, another lending/borrowing platform, allows users to earn interest on their crypto holdings and borrow assets against their collateral.

b-2) NFTs and Asset Tokenization:  Tokenization \cite{Das2022UnderstandingEcosystem} \cite{wang2021sok} involves the digital representation of real-world assets as tokens traded on a blockchain platform and managed by smart contracts.  The tokenization enables to capitalize on traditional blockchain advantages such as indisputable ownership, transparency, trustless transactions, and an openly accessible ledger of records. Non-Fungible Tokens (NFTs), a notable application of tokenization, have garnered significant attention in the blockchain realm. They represent unique and indivisible digital assets on a blockchain, with each NFT being distinct and often associated with digital art, collectibles, virtual real estate, or other unique digital items.

\color{black}
b-3) Prediction Markets: Leveraging smart contracts and data oracles, prediction markets on blockchain platforms enable the incorporation of real-world, open-internet data. This includes information on market movements, weather conditions, sports results, and more. The use of smart contracts and oracles enhances the transparency and reliability of these prediction markets. Developers and end-users can create niche betting or prediction platforms, where the outcome of events is automatically determined and payouts are executed based on predefined rules encoded in smart contracts. This eliminates the need for centralized authorities in overseeing and settling predictions, offering a decentralized and trustless environment for participants. 

\paragraph{Wallets and Identity Management}
Bitcoin, as the original cryptocurrency, utilizes a simple private key system for user access and management, with wallets, whether software, hardware, or paper, ensuring secure storage and transfer capabilities. Ethereum's wallets act as gateways to its ecosystem, managing Ether (ETH) and ERC-20 tokens (or others), connecting to DApps, and enabling transaction signing. Notable options like MetaMask, MyEtherWallet, and Ledger hardware wallets enhance user interaction within the Ethereum network. 
Ethereum has introduced Account abstraction (AA). By decoupling transaction signing from the traditional private key model, AA introduces a layer of programmability and flexibility that opens up exciting possibilities for wallets. AA enhances wallet security by reducing direct user management of private keys, implementing advanced recovery mechanisms like multi-signature schemes, and allowing programmable permissions for added security against unauthorized transactions. 
Solana, emphasizing speed, offers wallets such as Phantom and Solflare, prioritizing rapid transaction processing, integration with dApps, and support for staking and managing Solana Programmatic NFTs (pNFTs). Hyperledger Fabric, designed for enterprise use, focuses on permissioned networks with access control, integrating wallets with identity management systems based on PKI to ensure authorized access and secure transactions.

\subsection{Application Layer: Discussion}
The concept of decentralized applications (DApps) takes full advantage of the unique characteristics of DLTs but also inherits some of their limitations. According to a recent report \cite{DAPP}, DApp projects in 2019 faced ongoing challenges, including poor user attraction and retention due to complex user experiences or the perceived uselessness of their services. As a result, a significant number (estimated at $1300$) of DApps were abandoned in 2019 \cite{DAPP}. Successful DApps, such as CryptoKitties (an online game built on Ethereum) or EIDOS (an EOS token), highlighted a critical limitation of public DLTs, namely, scalability. The popularity of these DApps led to unprecedented congestion on their underlying chains, causing thousands of unvalidated transactions. To address this, developers are increasingly exploring the use of L2, rapid sidechains like Lightning, Raiden, or Loom Network, which offer faster transaction processing compared to the main chains (e.g., Ethereum).

\vspace{5pt}
\section{Evaluation and discussion } \label{sect 8}
In this section, we undertake a comparative analysis and evaluation of DLTs. The analysis is conducted at two levels; First, we compare and evaluate, at a high level, the chosen DLTs based on the properties outlined by our framework; Second, we compare and evaluate multiple consensus protocols against the criteria introduced in the section \ref{sect 5}.

\subsection{A Comparative evaluation of Blockchain and blockchain-like system based on DCEA framework}
Tables \ref{tab:my-table7} and \ref{tab:my-table9} provide a comprehensive overview of a diverse and substantial selection of Distributed Ledger Technologies (DLTs) from both industry implementations and recent research contributions. The comparative analysis encompasses four key dimensions: the composition of the four-component DCEA framework, operational scope, level of decentralization, and the higher taxon classification. In this subsection, our attention is directed towards a detailed examination of governance and conflict resolution methodologies, with a concurrent evaluation of decentralization. These properties assume an important role in determining the categorization of a system as a blockchain or not.

\subsubsection{Decentralization}
Decentralization is a fundamental element in the design of DLT. Our assessment delves beyond the surface, scrutinizing the topology of nodes, the cost and distribution of running full nodes, and the mechanisms governing decision-making. A truly decentralized DLT resists the control of any singular entity, whether physical or logical. This multi-layered analysis dissects the network's backbone, ensuring power is distributed rather than concentrated across nodes and participants. A DLT is deemed decentralized if it avoids physical or logical control by a singular entity, with consideration given to the aforementioned factors.  The $44$ DLTs, as listed in Table \ref{tab:my-table7}, underwent assessment on a three-step scale:
\begin{itemize}
    \item centralized : in this model, a single entity or a small group of entities control all aspects of the system. They possess full decision-making power, maintain the ledger, and dictate the rules and processes for transaction validation and consensus.
    \item semi-decentralized : This model introduces some elements of decentralization but still retains significant control in the hands of specific entities.
    \item  fully decentralized : This model aims to distribute power and control among all participants in the network, eliminating the need for a central authority.
Decisions are made collectively through consensus mechanisms, and no single entity has exclusive control over the system.
\end{itemize}
The pursuit of enhanced performance and scalability in DLT projects often compels developers to grapple with the fundamental tension between decentralization and centralization. Our comparative analysis reveals striking examples of this compromise, where projects like Ripple, Stellar, and Libra integrate centralized elements to optimize network efficiency, albeit at the expense of absolute decentralization. In Ripple's case, a pre-defined "starter list" of trusted nodes, chosen by the founding team, serves as the initial validator set. While users can theoretically update this list, concerns about divergent pathways and compromised security (as noted by Armknecht \cite{Armknecht2015}) deter most users from doing so. This static list effectively cedes operational and decisional power to the starter nodes, creating a de facto centralized system heavily influenced by the Ripple company, which also holds a significant stake in the network's token (XRP). This centralized control raises concerns about potential censorship, manipulation, and vulnerability to single points of failure. Similarly, Stellar's network exhibits dependence on two core validators managed by the Stellar Foundation. \cite{Kim2019}  highlights that deleting these nodes could trigger a network-wide collapse, further accentuating the centralized nature of its governance and operational structure.
These examples illustrate the complexities and trade-offs inherent in the decentralization vs. performance dilemma for DLT projects. While centralized elements can offer undeniable benefits in terms of speed and stability, they simultaneously introduce weaknesses in trust, security, and resilience. The long-term viability and societal impact of these projects hinge on finding innovative solutions that bridge this gap and pave the way for scalable, yet truly decentralized, DLT frameworks. Conversely, most Proof-of-Stake (PoS)-based DLTs exhibit decentralization. Nonetheless, PoS faces criticism for potentially favoring entities with a larger token stake, leading to centralized validation in instances of unfair token distribution. \cite{Nguyen} demonstrates the significant impact of the ratio between block reward and total network stake on the decentralization of PoS networks. IOTA serves as an example of a semi-decentralized DLT, utilizing Coordinator (COO) nodes run by the IOTA foundation to safeguard the network from 34\% attacks. Transactions cannot be confirmed unless approved by the Coordinator through milestones. Delegated Proof-of-Stake (DPoS)-based DLTs face criticism for susceptibility to validation centralization due to the potential collusion of elected validators. 


The case of EOS exemplifies this tension. Despite its claim to democratic governance through staking-based voting, concerns persist regarding centralization tendencies within its ecosystem. Studies have uncovered correlations in votes for different candidates, suggesting potential collusion among a limited group of influential actors \cite{Coindesk}. Additionally, the extreme concentration of EOS tokens, with the top 100 holders possessing over $75.13\%$ of the total supply \cite{Voting}, raises worries about the undue influence they may wield in validator selection and network governance. These findings align with broader concerns expressed by researchers such as Micali, who caution against poorly designed incentive mechanisms exacerbating centralization within blockchains. Kwon et al. echo these sentiments, emphasizing the inherent challenges in achieving robust decentralization in permissionless blockchain systems \cite{Kwon2019}.

\subsubsection{Governance}
In our assessment, we prioritized the examination of the decentralized governance structures implemented in the chosen Distributed Ledger Technologies (DLTs). Diverse decision-making approaches are employed across different projects to modify DLT protocol parameters and upgrade network rules, reflecting varied political forms of governance. Networks such as Bitcoin and Ethereum adopt an anarchic governance model. In these systems, a anybody can initiate an improvement proposal to address issues or to change protocol settings (e.g., increasing block size). The proposal undergoes public discussion, and upon garnering sufficient favorable peer review, it is implemented into the project's codebase. 
Depending on the technical enhancements, deploying it might need either a soft fork or a hard fork. A soft fork is compatible with the existing system and doesn't require all network nodes to accept it, whereas a hard fork requires nodes to update their software. For instance, Bitcoin's SegWit upgrade was a soft fork, and Ethereum's shift from Proof-of-Work to Proof-of-Stake was a hard fork. The risk with a hard fork is that it may lead to a split in the network, as upgraded nodes disconnect and reject nodes with a different protocol version that didn't undergo the hard fork.
Bitcoin employs on-chain governance through Version Bits voting \cite{Bips} for Soft-fork implementations, measuring miner support. 
Projects like Qtum aim to avoid significant disruptions, such as hard forks, for minor changes like block size and gas parameters. They achieve this by including built-in features that allow participants to collectively decide on system adjustments through dedicated smart contracts, as seen in the Decentralized Governance Protocol. Similarly, Tezos utilizes the Tezos governance protocol, enabling the network to decide on protocol upgrades. In enterprise-grade DLTs such as Hyperledger Fabric designed primarily for private or consortium contexts networks governance is managed by a predefined governing body. For example, a technical or administrative committee makes decisions, halts the network \cite{Fabrica}, and implements upgrades.

\begin{table*}[ht]
\caption{Applications of blockchain and blockchain-like systems}
\label{tab:comparaison-table}
\colorbox{mycolor}{%
\begin{adjustbox}{max width=\linewidth}
\begin{tabular}{|l|l|l|l|l|l|l|l|l|l|l|l|l|l|l|l|l|}
\hline
 &
\begin{sideways}\parbox{2cm}{ \cellcolor{TopRow}Cryptocurrency and payment} \end{sideways} &
\begin{sideways}\parbox{2cm}{ \cellcolor{TopRow}Digital identity} \end{sideways} &
\begin{sideways}\parbox{2cm}{ \cellcolor{TopRow}IoT} \end{sideways} &
\begin{sideways}\parbox{2cm}{ \cellcolor{TopRow}Healthcare} \end{sideways} &
\begin{sideways}\parbox{2cm}{ \cellcolor{TopRow}Logistics} \end{sideways} &
\begin{sideways}\parbox{2cm}{ \cellcolor{TopRow}Smart city} \end{sideways} &
\begin{sideways}\parbox{2cm}{ \cellcolor{TopRow}Telecom} \end{sideways} &
\begin{sideways}\parbox{2cm}{ \cellcolor{TopRow}Smart grid} \end{sideways} &
\begin{sideways}\parbox{2cm}{ \cellcolor{TopRow}AI} \end{sideways} &
\begin{sideways}\parbox{2cm}{ \cellcolor{TopRow}Prediction markets} \end{sideways} &
\begin{sideways}\parbox{2cm}{ \cellcolor{TopRow}Security} \end{sideways} &
\begin{sideways}\parbox{2cm}{ \cellcolor{TopRow}E-government} \end{sideways} &
\begin{sideways}\parbox{2cm}{ \cellcolor{TopRow}Banking} \end{sideways} &
\begin{sideways}\parbox{2cm}{ \cellcolor{TopRow}Finance} \end{sideways} &
\begin{sideways}\parbox{2cm}{ \cellcolor{TopRow}Banking} \end{sideways} &
\begin{sideways}\parbox{2cm}{ \cellcolor{TopRow}Games and E-sports} \end{sideways} \\ \hline
\textbf{Blockchain} &
  \cite{nakamoto} \cite{Buterin2014} &
  \cite{Wang2019a}&
  \cite{Novo2018a} \cite{GIULIO2015}&
  \cite{Liang2018}&
  \cite{Shipchain} &
  \cite{Loss2019} &
 \cite{Weiss2019}&
 \cite{Khalid2020} &
  \cite{Harris2019} \cite{bellajgbtrust2023} &
 \cite{Peterson2019} &
  \cite{Noyes2016}  \cite{Ouaddah2017a} &
  - &
  - &
  -&
  -&
 \cite{Min2019}\\ \hline
  
\textbf{Blockchain-like} &
  \cite{EosJs} &
\cite{Dorri2017a} &
\cite{Peterson2016}&
  \cite{Antipova2018} &
 \cite{Casino2019} &
 \cite{Loss2019} &
 \cite{Haris2020} &
  \cite{Dekhane2019} &
  \cite{Mamoshina2018} &
  - &
- &
 \cite{Antipova2018a} &
 \cite{Casino2019}&
  \cite{Liu2019a} &
\cite{Dozier2019a} &
 -\\ \hline
\end{tabular}
\end{adjustbox}}
\end{table*}


\subsubsection{Conflict Resolution}

An intriguing aspect within the selected DLTs lies in their methodologies for transaction ordering and conflict resolution. In systems employing Nakamoto consensus protocols, miners autonomously arrange transactions in blocks, validate them, and include them in the chain of blocks. In the event of conflicting chains, nodes shift toward the longest chain. DPoS protocols also follow the longest chain rule when a block fails to receive the majority (\emph{2/3 +1}) of votes from block producers. Nevertheless, a recent study \cite{Brown-Cohen2019} questions the safety of applying the "longest-chain PoS" rule to PoS protocols. Conversely, Nano's DAG relys on balance-weighted voting such that the network reaches consensus through individual node decisions and voting power.

Certain networks, such as Hyperledger Fabric, utilize a specialized ordering service and dedicated mechanisms. The Ordering Service sequences transactions into blocks, ensuring a specific processing order and contributing to conflict prevention in the ledger. Hyperledger Fabric, employs Multi-version Concurrency Control (MVCC) to allow parallel transactions without conflicts. Each ledger key maintains a version history, preventing data inconsistencies. Additionally, an Endorsement Policy mandates validations from specific peers before submitting transactions, preventing invalid or conflicting entries. 
Similarly, Corda employs the notary service \cite{Cordaa} to order transactions and detect conflicts utilizing multiple consensus mechanisms like RAFT, PBFT, or custom implementations. \cite{Cordaa}. Once the notary service validates a transaction and adds it to the ledger, it becomes final and binding for all participating nodes. This provides certainty and immutability to the transaction data.
Transaction ordering in IOTA is partially ensured by senders and weights. Senders choose two previous confirmed transactions (called "tips") to attach their new transaction to, creating a DAG structure. While some suggest the tips selection can be random from confirmed transactions, the IOTA Foundation recommends using a Markov Chain Monte Carlo (MCMC) weighted random walk. This method prioritizes attaching to heavier branches, which helps them grow faster and become the dominant valid tangle. Hashgraph aims to ensure "ordering fairness" \cite{Hedra} through its gossip-about-gossip protocol. This concept ensures that transactions on a blockchain or distributed ledger are processed and committed in the same order that they were received by the network nodes. Recognizing the significance of correct and fair ordering, Asayag et al. proposed Helix \cite{Asayag2018} which leverages an "in-protocol randomness" mechanism to elect validators from a larger pool and determine which messages should be included in a block. This randomness helps prevent manipulation and ensures fair ordering of transactions. Moreover, Kelkar et al. introduced Aequitas \cite{Kelkar2020} which uses a lottery-based approach to select validators and assign transaction slots within a block. This lottery is designed to be provably fair and tamper-proof.

\subsubsection{Application Scenarios for Blockchain and Blockchain-like Systems}

In recent years, the utilization of DLTs proliferated across diverse domains. The examples presented in Tables \ref{tab:comparaison-table} and \ref{tab:taxonomy_survey} highlight numerous sector-specific applications. While there may be some overlap in their application domains, it is clear that these technologies address distinct business scenarios with unique requirements. Broadly speaking, blockchain systems are well-suited for global decentralized C2C models, emphasizing high user autonomy. In contrast, blockchain-like platforms find greater utility in B2B use-cases within a single organization or consortium, especially in corporate settings where controlled governance and restricted data access are crucial. For instance, a consortium of financial actors may opt for a blockchain-like system to limit the sharing of financial transaction details to relevant parties. Conversely, blockchain systems excel in egalitarian networks, providing considerable freedom to end-users, such as enhanced transparency, decentralized control, and heightened security. A prime example is the utilization of cryptocurrencies like Bitcoin and Ethereum. In these blockchain networks, participants benefit from a more equitable distribution of authority and access, fostering a trustless environment where transactions are verifiable and immutable.

\subsection{A comparative analysis of consensus protocols} \label{subsection consensus analysis}
\begin{figure}[t]
\centering
\includegraphics[width=9cm, height=6cm]{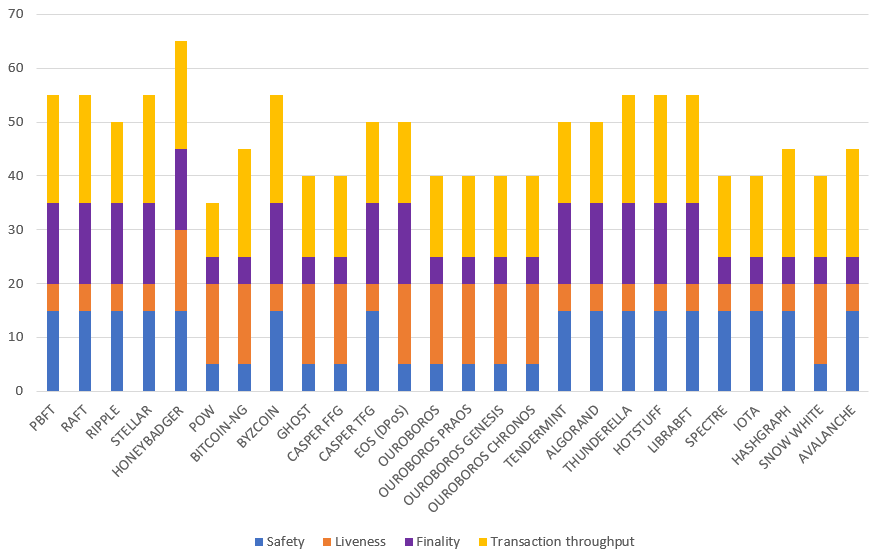}
\caption{Visual comparison of consensus protocols in terms of safety, liveness and finality } \label{fig: scale}
\centering
\end{figure}

For a better comparison of DLTs, it is pertinent to compare and contrast the consensus mechanisms separately as they have a direct influence on other aspects and properties of the DLTs (e.g. centralization, membership, scalability).
To achieve a thorough comparison, we adopt the metrics defined in section \ref{sect 5}, and we consider the evaluative scale presented in Table \ref{tab:my-table5}. Table \ref{tab:my-table6} and Figure \ref{fig: scale}, summarize the result of our analysis and evaluation. 
\begin{table}[b]
\caption{The evaluative scale used to compare the consensus protocols}
\begin{adjustbox}{max width=\linewidth}
\label{tab:my-table5}

\begin{tabular}{|l|l|l|l|l|}
\hline
\textbf{ \cellcolor{TopRow} Liveness}           & Strong              & Weak                    & \multicolumn{2}{l|}{}             \\ \hline
\textbf{ \cellcolor{TopRow} Safety}             & Strong              & Weak                    & \multicolumn{2}{l|}{}             \\ \hline
\textbf{ \cellcolor{TopRow} Transaction   throughput} & Very   high (\textgreater{}1000 tps) & High({[}1000,100 tps{]}) & \multicolumn{2}{l|}{Low(\textless{}100   tps)} \\ \hline
\textbf{ \cellcolor{TopRow} Finality}           & Absolute            & Probabilistic           & \multicolumn{2}{l|}{}             \\ \hline
\textbf{ \cellcolor{TopRow} Network   Model}    & Synchronous         & Partially   synchronous & \multicolumn{2}{l|}{  Asynchronous} \\ \hline
\textbf{ \cellcolor{TopRow} Adversarial   mode} & Strongly   adaptive & Middly   adaptive       & Adaptive     & Non   adaptive     \\ \hline
\textbf{\cellcolor{TopRow} Identity   model}   & Permissioned        & Permissionless          & \multicolumn{2}{l|}{}             \\ \hline
\end{tabular}
\end{adjustbox}
\end{table}

\begin{table*}[ht]
\caption{a comparison of consensus protocols}
\label{tab:my-table6}
\begin{adjustbox}{max width=\linewidth}
\begin{tabular}{|l|l|l|l|l|l|l|l|l|l|l|}
\hline
 &
  \textbf{ \cellcolor{TopRow}Network   Model} &
  \textbf{ \cellcolor{TopRow}Adversarial   model
  } &
  \textbf{ \cellcolor{TopRow}Adversary   mode} &
  \textbf{ \cellcolor{TopRow} \vtop{\hbox{\strut Fault}\hbox{\strut tolerance}}  } &
  \textbf{ \cellcolor{TopRow}Identity   Model} &
  \textbf{ \cellcolor{TopRow}Safety} &
  \textbf{ \cellcolor{TopRow}Liveness} &
  \textbf{ \cellcolor{TopRow}Finality} &
  \textbf{ \cellcolor{TopRow} \vtop{\hbox{\strut transaction}\hbox{\strut throughput}}
     } &
  \textbf{ \cellcolor{TopRow}Type*} \\ \hline
\textbf{PBFT} &
  Asynchronous &
  Threshold   Adversary  &
  NA &
  f\textless{}n/3 &
  Permissioned &
  Strong &
  Weak &
  Absolute &
  Very high &
  Blockchain-like \\ \hline
\textbf{RAFT} &
  Asynchronous &
  Crash-failure &
  NA &
  f\textless{}n/2 &
  Permissioned &
  Strong &
  Weak &
  Absolute &
  Very high &
  Blockchain-like  \\ \hline
\textbf{RIPPLE} &
  Asynchronous &
  Threshold   Adversary  &
  Adaptive &
  f\textless{}n/5 &
  Permissionless &
  Strong &
  Weak &
  Absolute &
  High &
  Blockchain-like \\ \hline
\textbf{STELLAR} &
  Asynchronous &
  Threshold   Adversary  &
  Adaptive &
  f\textless{}n/3 &
  Permissionless &
  Strong &
  Weak &
  Absolute &
  Very High &
  Blockchain-like \\ \hline
\textbf{HONEYBADGER} &
  Asynchronous &
  Threshold   Adversary  &
  Non   adaptive &
  f\textless{}n/3 &
  Permissioned &
  Strong &
  Strong &
  Absolute &
  Very high &
  Blockchain-like \\ \hline
\textbf{POW} &
  Partially-synchronous &
  Threshold   Adversary  &
  Strongly   adaptive &
  f\textless{}n/2 &
  Permissionless &
  Weak &
  Strong &
  Probabilistic &
  Low &
  Blockchain \\ \hline
\textbf{BITCOIN-NG} &
  Partially-synchronous &
  Threshold   Adversary  &
  Strongly   adaptive &
  f\textless{}n/2 &
  Permissionless &
  Weak &
  Strong &
  Probabilistic &
  Very high &
  Blockchain \\ \hline
\textbf{BYZCOIN} &
  Partially-synchronous &
  Threshold   Adversary  &
  Non   adaptive &
  f\textless n/3 &
  Permissionless &
  Strong &
  Weak &
  Absolute &
  Very high &
  Blockchain \\ \hline
\textbf{GHOST} &
  Partially-synchronous &
  Threshold   Adversary  &
  Non   adaptive &
  f\textless{}n/2 &
  Permissionless &
  Weak &
  Strong &
  Probabilistic &
  High &
  Blockchain \\ \hline
\textbf{CASPER   FFG} &
  Asynchronous &
  Stake   Threshold Adversary &
  Non   adaptive &
  f\textless{}n/3 &
  Permissionless &
  Weak &
  Strong &
  Probabilistic &
  High &
  Blockchain \\ \hline
\textbf{CASPER   TFG} &
  Asynchronous &
  Stake   Threshold Adversary &
  Non   adaptive &
  f\textless{}n/3 &
  Permissionless &
  Strong &
  Weak &
  Absolute &
  High &
  Blockchain \\ \hline
\textbf{DPoS (EOS)} &
  Partially-synchronous &
  Stake   Adversary  &
  Non   adaptive &
  f\textless{}n/3 &
  Permissionless &
  Weak &
  Strong &
  Absolute &
  High &
  Blockchain-like \\ \hline
\textbf{OUROBOROS} &
  synchronous &
  Stake   Threshold Adversary &
  Middly   adaptive &
  f\textless{}n/3 &
  Permissionless &
  Weak &
  Strong &
  Probabilistic &
  High &
  Blockchain \\ \hline
\textbf{\vtop{\hbox{\strut OUROBOROS}\hbox{\strut PRAOS}}   } &
  Partially-synchronous &
  Stake   Threshold Adversary &
  Strongly   adaptive &
  f\textless{}n/3 &
  Permissionless &
  weak &
  Strong &
  Probabilistic &
  High &
  Blockchain \\ \hline
\textbf{\vtop{\hbox{\strut OUROBOROS}\hbox{\strut GENESIS}}   } &
  Partially-synchronous &
  Stake   Threshold Adversary &
  Strongly   adaptive &
  f\textless{}n/3 &
  Permissionless &
  Weak &
  Strong &
  Probabilistic &
  High &
  Blockchain \\ \hline
\textbf{\vtop{\hbox{\strut OUROBOROS}\hbox{\strut CHRONOS}}   } &
  Partially   synchronous &
  Stake   Threshold Adversary &
  Strongly   adaptive &
  f\textless{}n/3 &
  Permissionless &
  weak &
  strong &
  Probabilistic &
  High &
  Blockchain \\ \hline
\textbf{TENDERMINT} &
  Partially   synchronous &
  Stake   Threshold Adversary &
  Non   adaptive &
  f\textless{}n/3 &
  Permissioned &
  Strong &
  Weak &
  Absolute &
  High &
  Blockchain-like \\ \hline
\textbf{ALGORAND} &
  Partially   synchronous &
  Stake   Threshold Adversary &
  Strongly   adaptive &
  f\textless{}n/3 &
  Permissionless &
  Strong &
  Weak &
  Absolute &
  High &
  Blockchain \\ \hline
\textbf{THUNDERELLA} &
  Synchronous &
  Stake   Threshold Adversary &
  Middly   adaptive &
  f\textless{}n/3 &
  Permissionless &
  Strong &
  Weak &
  Absolute   (Fast Path) &
  Very high &
  Blockchain-like \\ \hline
\textbf{HOTSTUFF} &
  Partially-synchronous &
  Threshold   Adversary  &
  Adaptive &
  f\textless{}n/3 &
  Permissionless &
  Strong &
  Weak &
  Absolute &
  Very high &
  Blockchain-like \\ \hline
\textbf{LIBRABFT} &
  Partially   synchronous &
  Threshold Adversary  &
  Adaptive &
  f\textless{}n/3 &
  Permissionless &
  Strong &
  Weak &
  Absolute &
  Very high &
  Blockchain-like \\ \hline
\textbf{SPECTRE} &
  Partially   synchronous &
  Threshold Adversary  &
  Non   adaptive &
  f\textless{}n/3 &
  Permissionless &
  Strong &
  Weak &
  Probabilistic &
  High &
  Blockchain-like \\ \hline
\textbf{IOTA} &
  Partially   synchronous &
  Threshold Adversary  &
  Non   adaptive &
  f\textless{}n/3 &
  Permissionless &
  Strong &
  Weak &
  Probabilistic &
  High &
  Blockchain-like \\ \hline
\textbf{HASHGRAPH} &
  Asynchronous &
  Threshold Adversary  &
  Non   adaptive &
  f\textless{}n/3 &
  Permissioned &
  Strong &
  Weak &
  Probabilistic &
  Very high &
  Blockchain-like \\ \hline
\textbf{SNOW   WHITE} &
  Asynchronous &
  Stake   Threshold Adversary &
  Strongly   adaptive &
  f\textless{}n/3 &
  Permissionless &
  Weak &
  Strong &
  Probabilistic &
  High &
  Blockchain \\ \hline
\textbf{AVALANCHE} &
  Partially   synchronous &
  Stake   Threshold Adversary &
  Non   adaptive &
  f\textless{}n/3 &
  Permissionless &
  Strong &
 Weak  &
  Probabilistic &
  Very high &
  Blockchain \\ \hline
\end{tabular}
\end{adjustbox}
\begin{tablenotes}
        \scriptsize
         \item[1]  \emph{f} : is the faulty nodes or actors and \emph{n}: is the total number of nodes that are coming to consensus.
        \item[2]* We present the DLT type according to the analysis performed and summarized in table VII.

    \end{tablenotes}
\end{table*}
 

\begin{table*}[htbp!]
\caption{Comparative Analysis Table for Selected DLTs} \label{tab:my-table7}
\includegraphics[]{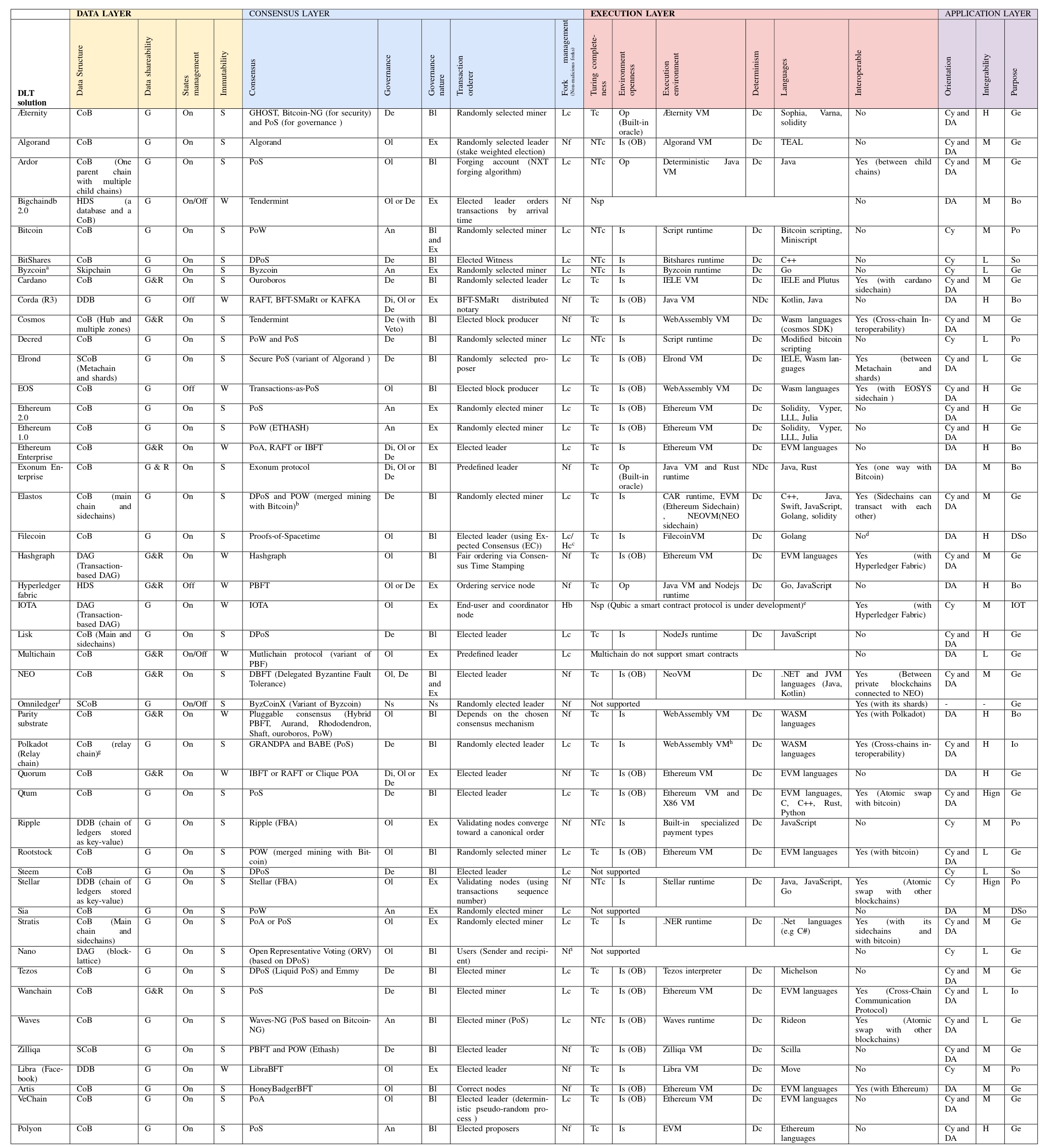}
\begin{tablenotes}
        \scriptsize
         \item LEGEND : \\  CoB: Chain of blocks, SCoB: Sharded Chain of blocks, DDB: Distributed database, HDS: Hybrid data structure
        \item G: Global, S: Strong, R: Restricted, W: Weak, Cy: Cryptocurrency, DA: DApps, Ol: Oligarchic, De: Democratic, An: Anarchic, Di: Dictatorship
        \item Bl: Built-in, Ex: External, Ns: Not specified, Nsp: Not supported, Nf: No forks, Lc: Longest chain, Hb: Heaviest branch, Hc: Heaviest chain,  
        \item Op: Open, Is: Isolated, OB: Oracle-based, Dc: Deterministic, NDc: Non-Deterministic, H: High, M: Medium, L: Low, Ge: General.
        \item Bo: Business-oriented,  Po: Payment-oriented, So: Service-oriented, DSo: Decentralized storage-oriented,  IoT: IoT-oriented, Io: Interoperability-oriented
        \\
        \item[a] Based on the project repository on \url{https://github.com/dedis/cothority/tree/master/byzcoin}
        \item[b] Elastos sidechains can theoretically use any consensus mechanism
        \item[c] Filecoin incentivizes miners with greater storage capacity
        \item[d] Different software implementations of the Filecoin protocol should be able to work together seamlessly.
        \item[e] \url{https://qubic.iota.org}
        \item[f] Based on the implementation available on \url{https://github.com/dedis/student_18_byzcoin}
        \item[g] Unlike the main chain, known as the Relay Chain, which has a standardized structure, parachains can be customized with a wide range of data models and representations.
        \item [h] Parachains Runtime logic provides isolation between parachains. If one chain experiences an issue or vulnerability, it doesn't automatically affect the others.
        \item [i] In Nano, a single network-wide DAG is shared by all participants and a balance-weighted voting system is used to handle conflicting transactions.
    \end{tablenotes}
\end{table*}

An analysis of the results reveals important observations about the strengths and limitations of the analyzed protocols. Nakamoto consensus protocols face criticism for their performance limitations and susceptibility to centralization \cite{Gervais2014}. Bitcoin's proof-of-work, a well-known representative, is notorious for its energy consumption and significant latencies \cite{Karame2012}. Pass and Shi \cite{Pass2017b} show that, in order for Nakamoto consensus protocols to maintain security, the block interval must be set as a constant factor larger than the network's maximum delay. Consequently, the inherent design of these protocols imposes a limit on the potential network's throughput, resulting in a scalability bottleneck.

Despite these limitations, Nakamoto consensus protocols exhibit interesting properties. They demonstrate resilience against significant Byzantine minorities (\emph{n $>$ f/2}) with anonymous open membership. Moreover, they do not necessitate extensive message exchange between nodes to reach an agreement, allowing them to scale efficiently to a large number of participants. These characteristics make them widely employed in global cryptocurrency networks and well-suited for public permissioned networks, where implementing economic incentives is feasible as a safeguard mechanism to ensure liveness and network security.

PoS is praised as a better alternative to Nakamoto protocols, as it is energy-efficient and allows unlimited open membership with equivalent fault tolerance (\emph{n $>$ f/2+1}). However, it is vulnerable to numerous security threats such as Short and Long-range attacks \cite{Deirmentzoglou2019} and the nothing-at-stake attack. These issues are mitigated by some PoS protocols, such as Ouroboros, which employs only one designated leader in each round. Moreover, PoS protocols face the weak subjectivity issue, where a node joining the network for the first time or after a long absence has to rely on other nodes to synchronize the correct ledger. This dependence undermines the trustless nature of blockchains entirely.

Algorand is a PoS protocol designed to scale independently of the network’s size \cite{Gilad2017a}. It achieves this scalability by utilizing the verifiable random function (VRF) to randomly select private delegates in a representative way without the need for coordination between nodes. However, similar to other PoS protocols, Algorand is secure only against a $1/3$ adversary bound. Additionally, it results in an inherently slower block production rate compared to protocols with probabilistic finality, such as Snow White or Ouroboros, due to the requirement of multiple rounds.

On the other hand, BFT protocols are efficient and well-suited for small or midsize permissioned networks. However, they involve extensive communication exchanges between nodes and demand accurate knowledge of membership. Take PBFT, for example, which incurs quadratic communication overhead, making it challenging to scale in terms of the number of participants. Additionally, these protocols are vulnerable to Sybil attacks, rendering them unsuitable for public DLTs but well-suited for private DLTs.

Despite these limitations, some protocols aim to leverage the advantages of PBFT to construct highly performant open consensus protocols for public blockchains while mitigating its inherent drawbacks. Tendermint \cite{Kwon}, for example, combines BFT and DPoS. It prevents Sybil attacks and offers open membership based on proof of stake.
%
%
\\
Safety and liveness are crucial properties for DLTs. Figure \ref{fig: scale} provides an expressive overview of the safety and liveness of the analyzed consensus mechanisms. We observe that PoW-based protocols sacrifice safety (forking can happen) for strong liveness. These protocols provide probabilistic safety as the network converges toward a canonical chain using the probabilistic Nakamoto’s longest chain fork choice rule (or a similar rule) coupled with economic incentives. Similarly, many PoS-based protocols favor safety over liveness. For instance, Algorand and Casper FFG ensure safety and liveness if dishonest participants control less than $1/3$ of the deposited stake. Ouroboros is proven to achieve safety and liveness in synchronous settings, under the assumption of an honest majority of all stake in the system, even if a significant portion of participants is offline \cite{Badertscher2018}. Algorand achieves safety with a “weak synchrony” assumption, whereas to achieve liveness, Algorand assumes strong synchrony.
%
%
Other protocols such as BFT protocol and its variants \cite{Lamport1982}, \cite{Pease1980}, PBFT implementations  and FBA favor fault tolerance and termination over safety. This means that in case of accidental fork, the network halts waiting for recovery (restoration of the consensus). In the Stellar Federated Byzantine Agreement, nodes choose their quorum slice (set of trusted nodes) according to their trust relationship. Thus, the safety and security of the protocol is highly dependent on the structure of the quorum slices. However, \cite{Kim2019} shows that the Stellar system is significantly centralized and proved that FBA is not better than PBFT in terms of safety and liveness.

Another important aspect to consider when analyzing consensus protocols is finality. Protocols such as Snow white \cite{Daian2019}, Ouroboros \cite{Kiayias2017a}, Casper FFG \cite{Buterin2017}, and PoW achieve probabilistic finality. DLTs adopting these probabilistic protocols define rules to avoid the risk of double spending by urging users to wait for a given delay –usually expressed number of blocks— before considering the transaction as final. For instance, in Bitcoin, it is widely advised to wait at least for 6 confirmation blocks (around 60 minutes) before accepting the validated payment. The reason behind, is that after 6 confirmation the probability of reverting the transaction decreases to $0.02428\%$  [\cite{nakamoto} page 8] assuming that an adversary controls less than $10\%$ of the overall hash-power which is very unlikely. In contrast, multiple consensus mechanisms such as PBFT-based protocol (e.g Tendermint), Ripple, Stellar, DPoS-based protocols, some PoS protocols (e.g. Algorand), Thunderella’s fast path and others achieve absolute finality if the validating majority is honest. Moreover, DLTs networks adopt measures to secure finality. For instance, EOS, a DPoS DLT, utilizes the concept of Last irreversible block (LIB) to improve finality and honest nodes wait for 330 confirmation blocks (less than 2 seconds) before considering a transaction irreversible.

When considering scalability, a crucial property for DLTs, the situation is continually improving. Recently proposed protocols like Avalanche \cite{Rocket2019} ($3400$ Transactions Per Second or tps) and Algorand \cite{Gilad2017a} (around 1000 tps) outperform classical blockchain consensus protocols like PoW \cite{Charts} (about $4$ tps), while providing a comparable level of security and better finality. Permissioned protocols such as PBFT can achieve much higher throughput, for example, $15,000$ tps if the number of validating peers is under 16 \cite{Miller2016}. However, the throughput falls to under 5000 tps when the number of validating peers is $64$ \cite{Miller2016}. HoneyBadgerBFT attains roughly equal performance of between $10,000$ and $15,000$ tps \cite{Miller2016}, a performance that comes with higher latency (around 6 minutes in a network of 104 nodes \cite{Miller2016}). On the other hand, Tendermint can process thousands of transactions per second with very low latency (one-second block latency). However, similar to other PBFT-like protocols, the scalability of Tendermint in large-scale networks is questionable, as demonstrated in \cite{Buchman2016a} with only a maximum of $64$ nodes.


\vspace{5pt}

\begin{table*}[htbp!]
\caption{ A comparison of selected DLTs}
\label{tab:my-table9}
\centering
\begin{adjustbox}{max width=\textwidth}
\begin{tabular}{|l|l|l|l|}
\hline
\rowcolor{TopRow}
\textbf{DLT}  & Scope                & Decentralization Level & Taxon           \\ \hline

Æternity               & Public                          & Decentralized          & Blockchain      \\ \hline
Algorand               & Public                          & Decentralized          & Blockchain      \\ \hline
Ardor                  & Public                          & Decentralized          & Blockchain      \\ \hline
Bigchaindb 2.0         & Private                         & Semi-centralized       & Blockchain-like \\ \hline
Bitcoin                & Public                          & Decentralized          & Blockchain      \\ \hline
BitShares              & Public                          & Decentralized          & Blockchain      \\ \hline
Byzcoin              & Public                          & Decentralized          & Blockchain      \\ \hline
Cardano                & Public                          & Decentralized          & Blockchain      \\ \hline
Corda (R3)             & Private or consortium           & Semi-Decentralized     & Blockchain-like \\ \hline
Cosmos             & Cosmos hub is public, zones can be public or private & Decentralized                                        & Blockchain      \\ \hline
Decred                 & Public                          & Decentralized          & Blockchain      \\ \hline
Elrond                 & Public                          & Decentralized          & Blockchain      \\ \hline
EOS                    & Public                          & Semi-Decentralized     & Blockchain-like \\ \hline
Ethereum 1.0 and 2.0        & Public                          & Decentralized          & Blockchain      \\ \hline
Ethereum Enterprise    & Private or consortium           & Decentralized          & Blockchain      \\ \hline
Exonum Enterprise  & Public or private                                    & Semi-Decentralized                                   & Blockchain-like \\ \hline
Elastos                & Public                          & Decentralized          & Blockchain      \\ \hline
Filecoin               & Public                          & Decentralized          & Blockchain      \\ \hline
Hashgraph              & Public                          & Centralized            & Blockchain-like \\ \hline
Hyperledger fabric & Private or consortium                                & Semi-Decentralized                                   & Blockchain-like \\ \hline
IOTA                   & Public                          & Semi-Decentralized     & Blockchain-like \\ \hline
Lisk                   & Public                          & Decentralized          & Blockchain      \\ \hline
Multichain             & Private or consortium           & Semi-Decentralized     & Blockchain      \\ \hline
NEO                    & Public or private               & Semi-Decentralized     & Blockchain      \\ \hline
Omniledger            & Public                          & Decentralized                      & Blockchain      \\ \hline
Parity substrate       & Private or consortium           & Semi-Decentralized     & Blockchain      \\ \hline
Polkadot (Relay chain) & Public                          & Decentralized          & Blockchain      \\ \hline
Quorum                 & Private                         & Semi-Decentralized     & Blockchain      \\ \hline
Qtum                  & Public                          & Decentralized          & Blockchain      \\ \hline
Ripple                 & Public                          & Centralized            & Blockchain-like \\ \hline
Rootstock              & Public                          & Decentralized          & Blockchain      \\ \hline
Steem                  & Public                          & Decentralized          & Blockchain      \\ \hline
Stellar                & Public                          & Centralized            & Blockchain-like \\ \hline
Sia                    & Public                          & Decentralized          & Blockchain      \\ \hline
Stratis            & Main chain is public, seidechains are private        & Main chain is decentralized, the BAAS is centralized & Blockchain-like \\ \hline
Nano                   & Public                          & Semi-decentralized     & Blockchain-like \\ \hline
Tezos                  & Public                          & Decentralized          & Blockchain      \\ \hline
Wanchain               & Public                          & Decentralized          & Blockchain      \\ \hline
Waves                  & Public, private or permissioned & Decentralized          & Blockchain      \\ \hline
Zilliqa                & Public                          & Decentralized          & Blockchain      \\ \hline
Libra (Facebook)       & Public                          & Centralized            & Blockchain-like \\ \hline
Artis                  & Public                          & Decentralized          & Blockchain      \\ \hline
VeChain                & Public                          & Semi-Decentralized     & Blockchain      \\ \hline
Red Belly              & Public                          & Semi-Decentralized     & Blockchain      \\ \hline
\end{tabular}
\end{adjustbox}
\end{table*}



\subsection{Zero-Knowledge Rollups and zkEVM: A Comparative Overview}

The Zero-knowledge Virtual Machine (zkVM) is a emerging technology currently in the early stages of development, designed to enhance rollup capabilities through the utilization of zero-knowledge proofs. At its core, zkVM introduces innovative features, including the execution of smart contracts within the rollup in a manner that prioritizes both security and privacy. Unlike traditional transaction verification methods, zkVM goes beyond individual transactions and validates the entire computation within smart contracts. This approach significantly reduces the volume of on-chain data, contributing to improved efficiency.

One of the primary technical aspects of zkVM lies in its ability to execute smart contracts securely and privately. By leveraging advanced cryptographic techniques associated with zero-knowledge proofs, zkVM can conceal sensitive details of smart contract execution, including transaction amounts and asset types. This heightened level of privacy protection aligns with the growing demand for secure and confidential transaction processing on blockchain networks.

The potential technical advantages of zkVM are noteworthy:
\begin{itemize}
    \item Increased Scalability: zkVM's offloading of smart contract execution to the rollup has the potential to significantly amplify transaction throughput, addressing scalability concerns associated with traditional on-chain execution.
    \item Enhanced Privacy Mechanisms: Leveraging zero-knowledge proofs, zkVM ensures robust privacy guarantees for smart contract execution, mitigating concerns related to data exposure.
    \item Facilitation of Complex Use Cases: zkVM's ability to validate entire computations within smart contracts opens doors to new use cases that require not only heightened privacy but also on-chain verification of intricate and computationally intensive processes.
\end{itemize}

However, it is important to underscore the challenges associated with zkVM's early developmental phase:
\begin{itemize}
    \item Technical Complexity: Implementing and verifying zkVM proofs necessitates the application of advanced cryptographic techniques, introducing a level of technical complexity that requires careful consideration.
    \item Limited Ecosystem: The current support for zkVM is constrained, with ongoing efforts in the development of necessary tools and applications. The ecosystem is evolving, and broader adoption is contingent on continued maturation.
\end{itemize}

As zkVM progresses through its developmental stages, these technical considerations will likely shape its trajectory, determining its viability and potential impact within the broader blockchain landscape. In table \ref{tab:zktable} we provide a comprehensive comparison of some well-known zk protocols within the Ethereum ecosystem, highlighting their distinctive features, applications, and trade-offs. zkSync and zkPorter both utilize zk-SNARKs, emphasizing scalability and privacy, with efficient proof generation and strong privacy guarantees. However, they exhibit limitations in smart contract functionality when compared to Optimism. Hermez and zkTube, employing PLONK, share similar focuses on scalability and privacy, boasting fast verification and versatile proof systems. Nevertheless, akin to zkSync and zkPorter, they present constraints in smart contract functionality compared to Optimism. StarkWare, relying on STARK, prioritizes scalability and security, delivering highly secure proofs and fast verification processes. However, it places less emphasis on privacy compared to zk-SNARKs. zk-rollups on Arbitrum leverage various zk-SNARKs and PLONK implementations to address scalability and privacy concerns, capitalizing on existing Arbitrum infrastructure. Despite the flexibility, users face the challenge of navigating through multiple implementations, introducing complexity in selecting the most suitable one. Aztec Protocol, utilizing zk-SNARKs, stands out for its focus on privacy-preserving DeFi, ensuring strong privacy for financial transactions. Nevertheless, its smart contract functionality is somewhat limited for non-DeFi applications. Each protocol exhibits a unique set of advantages and disadvantages, catering to specific use cases and user priorities within the Ethereum ecosystem.

\subsection{DApps Attractiveness: A Comparative Overview}

\begin{figure}[t]
\centering
\includegraphics[width=7cm, height=4cm]{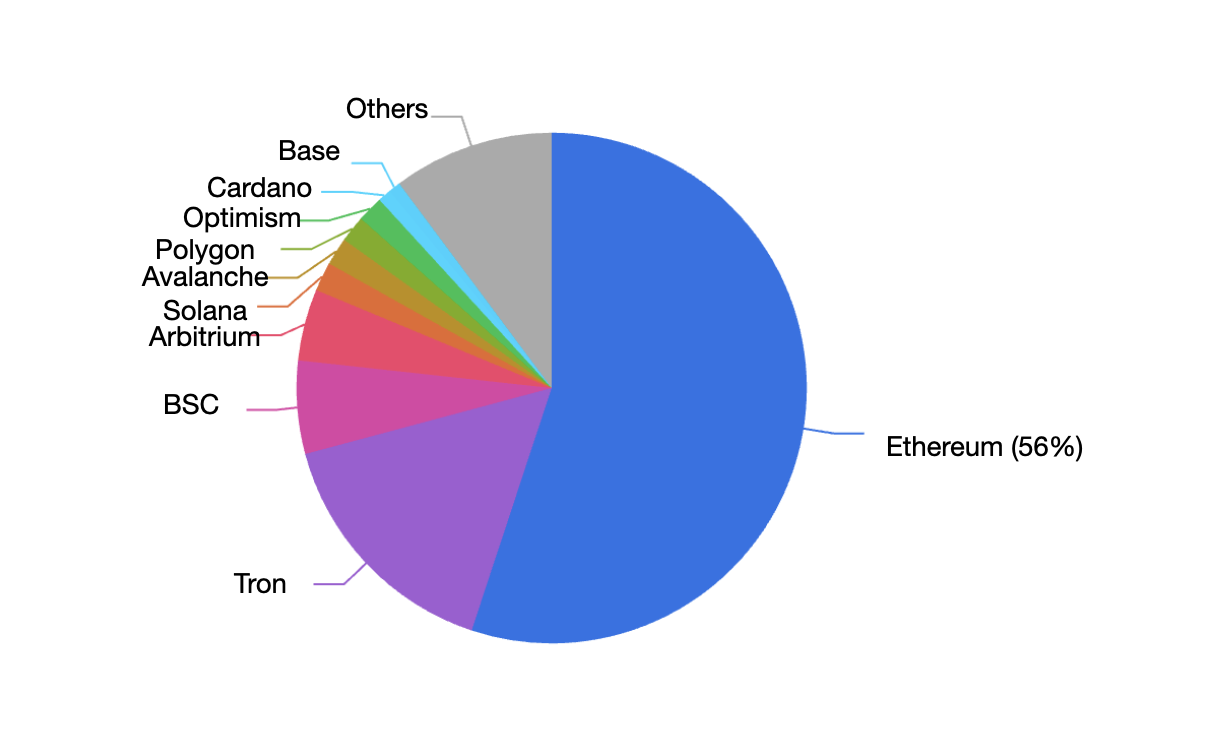}
\caption{TVL in the most prominent DApp-oriented blockchains (data source: DeFiLama)}
\label{fig:tvl}
\centering
\end{figure}

In evaluating the appeal of specific blockchain network solutions for DApp builders, the Total Value Locked (TVL) metric serves as a crucial benchmark within the cryptocurrency sector. TVL quantifies the total U.S. dollar value of digital assets locked or staked through decentralized finance (DeFi) platforms or decentralized applications (DApps). Figure \ref{fig:tvl} illustrates that Ethereum stands as the predominant network for DApp development, controlling more than half of the total value locked across blockchains. Unsurprisingly, alternative Layer-1 (L1) chains, particularly those compatible with Ethereum Virtual Machine (EVM), have proven attractive to developers. The top five contenders—Arbitrum, Optimism, Base, Polygon, and Era— collectively controlling about $10\%$ of the TVL. This phenomenon can be attributed to the appeal of EVM-compatible L1 chains like Polygon and Avalanche, which leverage their compatibility to entice Ethereum users seeking improved transaction speeds, reduced fees, and potentially higher returns. This establishes a mutually beneficial relationship between the L1 chain and its user base. Conversely, non-EVM projects such as Solana, Cardano, and Bitcoin exhibit lower TVL percentages, with Solana at $2\%$, Cardano at $1\%$, and Bitcoin at less than $1\%$. The divergence in TVL may be explained by the ease of DApp development on EVM-compatible platforms compared to their non-compatible counterparts.

\begin{table*}[ht]
\centering
\caption{Comparison of zk Protocols in the Ethereum Ecosystem}
\label{tab:zktable}
\begin{adjustbox}{max width=\linewidth}
\begin{tabular}{|c|c|c|c|c|}
\hline
\textbf{Protocol} & \textbf{ZK Protocol Used} & \textbf{Focus} & \textbf{Advantages} & \textbf{Disadvantages} \\
\hline

zkSync & zk-SNARKs & Scalability, privacy & Highly efficient proofs, strong privacy guarantees & Limited smart contract functionality compared to Optimism \\
\hline
Hermez & PLONK & Scalability, privacy & Fast verification, versatile proof system & Limited smart contract functionality compared to Optimism \\
\hline
StarkWare & STARK & Scalability, security & Highly secure proofs, fast verification & Less focus on privacy compared to zk-SNARKs \\
\hline
zkPorter & zk-SNARKs & Scalability, privacy & Efficient proof generation, strong privacy guarantees & Less mature than other protocols \\
\hline
zkTube & PLONK & Scalability, privacy & Fast verification, versatile proof system & Limited smart contract functionality compared to Optimism \\
\hline
zk-rollups on Arbitrum & Various zk-SNARKs and PLONK implementations & Scalability, privacy & Leverages existing Arbitrum infrastructure & Multiple implementations, complexity in choosing the right one \\
\hline
Aztec Protocol & zk-SNARKs & Privacy-preserving DeFi & Strong privacy for financial transactions & Limited smart contract functionality for non-DeFi applications \\
\hline
\end{tabular}
\end{adjustbox}
\end{table*}

\section{Lessons Learned and Tutorials for Designing New DLT Solutions} \label{sect 9}
While preparing this survey, we found multiple challenges that directly impact the performance of DLT. Here, we summarize the most crucial challenges across the four layers.

\paragraph{Managing Blockchain Size} When exploring different blockchain projects, a common problem that surfaces is what we call "blockchain bloat". Blockchain bloat is a phenomenon where the size of a blockchain becomes excessively large, posing various challenges. It results from factors like large block sizes, increased transaction volume, data redundancy, and the immutability of blockchain data. This condition has significant consequences, including increased storage requirements, scalability limitations, and concerns about decentralization. Several solutions are being explored to address blockchain bloat, such as state pruning, sidechains, sharding, and Layer-2 solutions. Tackling this challenge is crucial for ensuring the sustainability and scalability of blockchain technology in the long run.

Ensuring the security of smart contracts presents a significant challenge in the DLT domain, especially in public blockchains with associated financial risks. 
Several strategies have been recommended to enhance smart contract security, encompassing the implementation of code analyzers, the adoption of secure smart contract libraries like OpenZeppelin \cite{OpenZeppelin}, formal verification techniques, and the establishment of coding best practices. Due to the lack of a clear upgrading process for vulnerable smart contracts in the majority of DLTs, designers often focus on providing secure architectural and design approaches, upgradability patterns, and detailed best practice guidelines. These measures aim to assist developers in writing secure smart contracts, avoiding well-known vulnerabilities, and steering clear of security pitfalls. Furthermore, for risk mitigation, formal verification proves to be an efficient means of addressing security concerns and ensuring better smart contracting. Interestingly, bug bounty programs have proven to be an effective and cost-efficient way to enhance smart contract security.

\paragraph{Upgrading Consensus Mechanisms} There is a plethora of consensus protocols in the literature, with continuous research efforts aimed at producing new ones. However, when designing a new DLT, one should be aware of the upgradability pitfall. For various reasons, such as the emergence of a high-performing protocol or a new improvement, a project may need to shift or upgrade its underlying consensus protocol. In public blockchains, the protocol upgrade can come with its risks. For instance, Ethereum's plan to shift from PoW to PoS will have financial implications for its current miners, which may lead to opposition and a potential network split. Thus, it is crucial to choose the most suitable mechanism at the beginning of the project.
Moreover, the decision to switch from one protocol to another must be fault-tolerant and should be secured through built-in mechanisms (e.g., Ethereum difficulty bomb \cite{Age}).

\paragraph{Accessible applications} The complexity of the actual end user experience is a common observation among most DLTs.  This is explained by the fact that the inherent design does not consider to provide a good user experience but rather focuses on the internal machinery. To remedy that, the current solutions (e.g. wallets plugins) are not only non-intuitive for a new user but they are also challenging to use and maintain. A DLT designer should consider providing gateways, APIs and SDKs, and other enabling solutions that will allow developers to design more user friendly DApps and allow a seamless interaction between both product and end-user. An accessible blockchain design should enable average users to interact with the hosted DApps without prior knowledge or the need to synchronize the whole ledger, in order to lower the entry level of blockchain use.

\vspace{5pt}
\section {BLOCKCHAIN CHALLENGES AND FUTURE RESEARCH DIRECTIONS} \label{sect 11}
 DLTs (blockchain and blockchain-like), despite their immense potential, face significant challenges that need to be addressed for their wider adoption and mainstream success. Here, we present a non-exhaustive list of the key challenges:

\paragraph{Limited Scalability} This problem is primarily due to the challenge of finding a perfect balance among decentralization, security, and scalability. A well-known paradigm, called the scalability trilemma \cite{altarawneh2020buterin}, posits that it is impossible to build a system with all the aforementioned characteristics. However, multiple approaches to scaling distributed protocols are presented, such as off-chain processing, sharding (dividing a whole blockchain into multiple shards), and overhead reduction (Segwit, MAST, etc.). Nevertheless, these solutions themselves raise new challenges and security concerns. For example, the work in \cite{Yu2020} highlights the remaining challenges of sharding mechanisms, such as intra-consensus safety, cross-shared communication, and more. It is worth noting that although the theoretical propositions are not fully sharded, the blockchain is still operational.

\paragraph{Formal Verification} Recent costly bugs in smart contracts have highlighted the critical role of formal verification in ensuring the correctness and security of these programs. While significant research has explored formal verification techniques for smart contracts, achieving promising results \cite{Murray2019}-\cite{Liu2019}, existing approaches exhibit limitations in handling complex contracts. Current methods often struggle with contracts featuring intricate control flow, extensive state transitions, or interactions with external oracles, hindering their practical applicability in real-world scenarios. Additionally, the computational resources required for verifying complex contracts can be substantial, further limiting their scalability.

\paragraph{Data immutability and integrity in DLTs}
In private or consortium DLTs with permissioned access, ensuring a high level of data immutability is challenging compared to the assurance provided by public blockchains. This difficulty raises concerns about the suitability of DLTs in such environments where immutability is a crucial aspect. While there are some partial solutions like Exonum (which anchors data on the Bitcoin blockchain) or Kadena, the issue is far from being fully resolved.
 
\paragraph{Data availability problem} The efforts of scaling blockchain face significant challenge: guaranteeing data availability. In simpler terms, all nodes on the network, not just the block producers, need to be confident that the data in each new block is complete and hasn't been tampered with. Traditionally, verifying this meant downloading the entire block, which is inefficient and impractical for large blockchains. To address this hurdle, most scaling projects utilize data availability proofs. These mechanisms allow nodes to confirm, with near certainty, that all block data is present, even if they only download a tiny fraction of the block itself. 

\color{black}

\paragraph{Decentralized governance mechanisms}
Implementing effective decentralized governance poses a multifaceted challenge due to the absence of a central authority. Decentralized Autonomous Organizations (DAOs) initially appeared promising for transparent and decentralized management, but practical implementations revealed crucial limitations. One significant drawback is the lack of robust mechanisms for establishing user reputations within the system, affecting trust assessment and governance quality. Additionally, the anonymous nature of DAOs renders them vulnerable to Sibyl attacks, allowing malicious actors to manipulate voting processes and undermine governance integrity. The regulatory landscape surrounding DAOs remains unclear in many jurisdictions, creating uncertainty and hindering widespread adoption. Looking ahead, exploring alternative governance models beyond token-based voting, implementing decentralized dispute resolution mechanisms, and assessing the broader societal and economic implications of decentralized governance practices are essential considerations for overcoming these challenges.

\paragraph{Quantum resistance} Many cryptographic algorithms used by different DLTs are not quantum-resistant. For instance, the use of Schnorr or ECDSA (used by Bitcoin and Ethereum and others) for signing transactions is under threat. Aware of this problem, a few researchers have attempted to advance efficient solutions. Notably, \cite{Sun2018} reported the experimental realization of a quantum-safe blockchain. However, as research remains limited, more effort should be placed on the adoption of alternative cryptographic signature schemes (e.g. XMSS, hash ladder signatures, and SPHINCS) to replace the classical schemes and build a secure, resilient post-quantum DLT protocol.

\paragraph{Smart Contracts for IoT} Despite the proposition of a few projects dedicated to IoT (e.g. IOTA, Vechain \cite{Schenker2019}), there is no platform providing a smart contracts environment tailored to the special IoT requirements (e.g. lightweight execution runtime.). Most solutions propose hybrid DLTs where the IoT objects rely on external resources to run smart contracts.

\paragraph{Useful Proof-of-work} Proof-of-work (PoW) is one of the secure consensus mechanisms, but it is severely criticized for being wasteful. \cite{Ball2017} and \cite{Lihu2020} proposed to build new useful Proof-of-Work protocols solving useful calculation.

\paragraph{Sustainable Liquidity pools} One of the biggest problems that DeFi protocols face is the difficulty of sustainably attracting long-lasting liquidity. Most of these protocols distribute an important proportion of their native tokens into the liquidity mining incentives. This usually attracts investors and accelerate the growth of DeFi projects quickly. However, the vast majority of liquidity is unloyal and moves to new projects that offer better financial incentives which creates a huge selling pressure for the native token and thus drops its value. 
To mitigate this financial risk new DeFI projects, considered as being part of the next generation DeFi (DeFi 2.0), try to attract long-lasting liquidity without depending on the never-ending cycle of subsidising the users with liquidity mining rewards. To reach that goal more effort should be placed to design new protocols creating sustainable liquidity through a decentralized market-making and enabling a quick bootstrapping phase and attracting initial capital to a new chain or L2.

\paragraph{Upgrading Runtimes Without Forks}
While hard forking is a prevalent approach for upgrading public blockchains, it proves to be inefficient and error-prone in large-scale networks. The challenges stem from the considerable offline and online coordination needed to prevent network splits. Emerging solutions, such as those seen in projects like Polkadot, are investigating alternative methods. One promising avenue involves leveraging portable technologies like Wasm on-chain, allowing nodes to autonomously adopt upgraded logic at a predefined block height, eliminating the need for external intervention.

\paragraph{Low-Latency Byzantine Agreement Protocols Using RDMA} Remote Direct Memory Access (RDMA) is a technology that enables networked computers to exchange data in shared memory to improve throughput and performance. The shared memory model has been widely researched for key/value stores, databases and distributed file systems. However, the leverage of RDMA for consensus mechanisms especially for BFT has been neglected. Further research is needed to integrate blockchain with the shared memory and RDMA technologies and to propose BFT protocols for low-latency and RDMA-enabled consensus algorithms.

\paragraph{Emerging Trends in Zero-Knowledge Proof (ZKP) Protocols}
Zero-Knowledge Proof (ZKP) protocols have evolved significantly in recent years, playing a vital role in enhancing privacy and security across various domains.

Non-Interactive Zero-Knowledge Proofs (NIZKPs) \cite{chi2023privacy}: Traditionally, ZKPs involve interaction between the prover and verifier. However, there is a growing interest in Non-Interactive Zero-Knowledge Proofs (NIZKPs), where the prover can generate a proof that can be verified without interaction. Recent research has focused on constructing NIZKPs based on various cryptographic assumptions, including the use of bilinear pairings, hash functions, and algebraic structures. NIZKPs are not only theoretically intriguing but also practical for decentralized systems where participants may not be online simultaneously.

Lattice-Based Zero-Knowledge Proofs: Lattice-based cryptography, considered a post-quantum alternative, is being explored in the context of ZKPs for its resistance to quantum attacks. The challenge lies in designing efficient lattice-based ZKPs without sacrificing performance. Ongoing research aims to strike a balance between security and practicality in these constructions. 
Integration with Homomorphic Encryption: Combining ZKPs with homomorphic encryption allows computations to be performed on encrypted data without decryption, providing an additional layer of privacy and security. Research in this area explores ways to integrate ZKPs with homomorphic encryption schemes, enabling secure and private computations on encrypted data while proving the validity of the computations. This integration has potential applications in secure multi-party computation and privacy-preserving data analytics. Succinct Arguments of Knowledge: Efficiency remains a key concern in ZKP protocols, and succinct arguments of knowledge aim to reduce the size of proofs and improve verification speed without compromising security. Techniques like zk-SNARKs (Zero-Knowledge Succinct Non-Interactive Arguments of Knowledge) have demonstrated significant success in achieving succinctness. Ongoing research focuses on refining zk-SNARK constructions, exploring new mathematical frameworks, and addressing limitations to further enhance their efficiency and applicability.

\vspace{5pt}
\section{CONCLUSION} \label{sect:conclusion}

This document introduces an extensive examination of contemporary DLTs through a thorough and multi-layered state-of-the-art analysis. We introduce a conceptual and referential framework aimed at enhancing comprehension and categorization of DLTs. Our specific focus is on defining clear boundaries and reducing ambiguity between blockchain and other DLT systems, referred to as "blockchain-like". 
Moreover, we employ the provided framework as a guide to examine the different design-choices adopted by various DLTs across four layers: data structure, execution, consensus, and application layers. The application of our reference framework results in the creation of a novel taxonomy for classifying DLTs, broadly categorizing them into two groups: blockchain and blockchain-like systems.
Additionally, we conduct a qualitative and comparative analysis of numerous existing DLTs, with a dedicated examination of the most significant and recent consensus mechanisms. This survey aims to provide valuable insights for designers of new DLT systems and consensus mechanisms, along with assisting decision-makers. It offers a clear and detailed overview of recent contributions at each layer, highlighting trade-offs and limitations resulting from different design choices.
Furthermore, the analysis can aid in advocating the selection of a DLT solution for building decentralized systems. Finally, we identify several important open issues that merit attention in future research.

In conclusion, our work contributes to advancing the understanding of DLTs, their classifications, and the nuances among various systems. We hope this survey serves as a valuable resource for both academia and industry, guiding the development of robust and efficient decentralized technologies.

\bibliographystyle{IEEEtran}




\end{document}